\documentclass[letterpaper,11pt]{article}

\usepackage{jheppub}
\usepackage{booktabs} 
\usepackage{graphicx} 
\usepackage{siunitx}
\usepackage{hyperref}
\usepackage[T1]{fontenc}
\usepackage{cleveref}

\usepackage{tikz}

\usepackage{bm}
\usepackage{bbm}
\usepackage{xspace}

\usepackage{amsmath,latexsym}
\usepackage{pstricks}
\usepackage{color}

\usepackage{tikz}

\begin{document}

	\title{\texorpdfstring{Vector Superradiance without Separability: \\ Instability Rates from the Worldline Effective Theory}{Vector Superradiance without Separability: Instability Rates from the Worldline Effective Theory}}
	
	\author[a]{Keegan Cove,}
	\author[a,b]{Jingping Li,}
	\author[a]{Riccardo Penco}
	\affiliation[a]{Department of Physics, Carnegie Mellon University, Pittsburgh, PA 15213}
	\affiliation[b]{Cimulate, Salesforce}
	
	\emailAdd{kcove@andrew.cmu.edu}
	\emailAdd{jingping.li@salesforce.com}
	\emailAdd{rpenco@andrew.cmu.edu}

	\vspace{0.3cm}

	\abstract{We use a point-particle effective field theory to study the superradiant instability rates of an ultralight massive vector around a slowly rotating black hole. Earlier effective-theory treatments reported an apparent $\mathcal{O}(1)$ discrepancy with the rates obtained by solving the Proca equation on a fixed black-hole background, e.g. for the mode with $j=1$ and $\ell=2$. We show that there is in fact no discrepancy: once the leading-order states in the degenerate $\{\ell=0,\ell=2\}$ subspace are correctly identified, the effective theory---implemented here with the graviton integrated out exactly, and expanded in $\alpha$ only afterwards---reproduces \emph{all} of the $j=1$ Proca instability rates exactly, and it does so \emph{without} ever invoking the FKKS ansatz on which the full-theory calculation rests. Since it never relies on separability, the same approach applies, at least in principle, to rotating backgrounds and bodies that lack the hidden symmetries underlying that ansatz. The single ingredient beyond the far zone is a conservative worldline contact term---the static monopole response of the horizon in the parity-even $j=1$ Proca sector---which we fix by matching to the near zone. The same term shifts the fine-structure splitting of the bound states into exact agreement with the published Proca spectrum. In addition to providing a conceptually straightforward model of vector superradiance and resolving an apparent tension in the literature, we present this work to illustrate potential technical subtleties involving dissipative effects in point particle effective theories.}
	\maketitle

\section{Introduction} \label{sec:Introduction}

Superradiance is a phenomenon whereby the modes of a bosonic field scattering off a rotating object are amplified at the expense of the object's rotational energy.
Interest in this effect has recently been revived by the realization that it offers a purely gravitational probe of ultralight bosons. Any such particle must be weakly interacting to have gone undetected thus far, but it would still interact gravitationally.
This is where the mass becomes essential: a massive field admits gravitational bound states around a black hole (BH), and a bound quantum that has been amplified once remains in place to be amplified again. Superradiant scattering thereby turns into a superradiant \emph{instability}, in which the occupation number of a bound state---and with it the energy and angular momentum extracted from the BH---grows exponentially. Massless particles, by contrast, can only interact with BHs via asymptotic scattering states\footnote{ 
While light can orbit a BH in the photosphere, this orbit is unstable, precluding long-term amplification. We focus on giving the field mass to enable multiple interactions, but it can also be achieved by changing the external boundary conditions with a mirror~\cite{Press_1972}.}, limiting their ability to extract meaningful quantities of energy.
With gravitational-wave astronomy now providing direct access to BH data, the boson clouds grown by this instability could be detected---if not individually, then statistically, through the angular momentum they drain from the BH population~\cite{Arvanitaki_2011,Baumann:2018vus,Hannuksela:2018izj,Zhang:2018kib,Berti:2019wnn,Zhang:2019eid,Baumann:2022pkl,Tomaselli:2024dbw}.

Two complementary strategies are available to compute superradiant instability rates. One may treat this problem using the worldline effective field theory (EFT) approach, which models the BH as a point particle and encodes its short-distance structure in a series of operators localized on the worldline.
The power of this technique is its portability: once a few Wilson coefficients have been fixed by matching, the same effective theory can be carried over to more complicated scenarios---most notably binary inspirals, where worldline methods have proven very useful~\cite{Goldberger:2004jt}. Alternatively, one may tackle this problem by the more traditional means of solving the wave equation on the Kerr background using the method of matched asymptotic expansions~\cite{Rosa:2011my,Baryakhtar:2017ngi,Baumann:2019eav}. The implementation of the effective theory in this paper occupies an intermediate position between these two strategies: we integrate out the graviton exactly---the equation for the mode functions is formulated in the tree-level metric resummed by the worldline mass---and only then expand it in powers of $\alpha$, organizing the calculation by ordinary (degenerate) perturbation theory. A fully diagrammatic treatment, in which the same potentials are built up from Feynman diagrams involving potential gravitons~\cite{Goldberger:2004jt} and the power counting is manifest at every step, is an interesting problem that we leave for future work. Wherever both methods apply they must of course agree, and whether they actually do is the question that motivates this paper.

Our main result is that the effective theory and the full theory in fact agree \emph{exactly}---for every $j=0$ and $j=1$ instability rate, including the $\alpha^{11}$ rate of the $\ell=2$ mode that has long been a source of tension. Taken at face value, the effective theory overestimates this particular rate by a factor of three, and an earlier treatment~\cite{Endlich:2016jgc} quoted a value larger still [eq.~(95) there, $\Delta\Gamma\simeq\tfrac{1280}{59049}(GM\mu)^{11}\Omega$].\footnote{The factor of three is the basis subtlety resolved in \S\ref{sec:diagonal-fix} and \S\ref{sec:Perturbative-corrections}. The residual factor of two relative to eq.~(95) of~\cite{Endlich:2016jgc} is not a matter of convention---there, as here, the reported rate is the occupation-number growth rate $\Gamma_\mathrm{em}-\Gamma_\mathrm{abs}=2\,\mathrm{Im}\,\omega$---but originates in the Wilson coefficients matched there, which are twice the values obtained in \S\ref{sec: results}; see footnote~\ref{fn:endlich-factor2}.} The resolution has two complementary faces. On one hand, the state that the full theory labels $\ell=2$ is not the naive spherical-harmonic mode but a definite admixture of the degenerate $\ell=0$ and $\ell=2$ states; adopting this basis removes the factor of three (\S\ref{sec:diagonal-fix}). On the other hand, the same rate follows from first principles---by degenerate perturbation theory closed off with a single piece of static near-zone information, the monopole response of the horizon in the parity-even $j=1$ sector---without any change of basis performed by hand (\S\ref{sec:Perturbative-corrections}). Rather than demanding detailed information about the ultraviolet, then, the effective theory needs only this one number, fixed by matching, to reproduce the full-theory result.

It is worth stressing what the effective theory does \emph{not} require. The full-theory rates rest on the FKKS ansatz~\cite{Frolov:2018ezx}, a separation of variables tied to the hidden symmetries of the Kerr--NUT--(A)dS spacetimes---and one whose ability to capture all three polarizations of the Proca field was established only after the fact~\cite{Dolan:2018dqv}. Our treatment never invokes it. In \S\ref{sec:diagonal-fix} we do \emph{connect} to it, attributing the factor-of-three discrepancy to a change of basis: the separated variables of the full theory single out an eigenstate that already differs from the naive $\ell=2$ bound state at the order that matters on the worldline. In \S\ref{sec:Perturbative-corrections}, by contrast, the very same admixture---the small $\ell=0$ component acquired by the $\ell=2$ bound state---emerges as an \emph{output} of degenerate perturbation theory and near-zone matching, with no basis change imposed by hand. That the two routes agree is therefore a genuine check rather than a tautology, and it indicates that the effective theory stays predictive even where no FKKS-type separability is available---\textit{e.g.}, for backgrounds deformed away from Kerr--NUT--(A)dS, or for rotating bodies other than BHs.

This near-zone information has a natural interpretation in the language of static response and Love numbers. Within the parity-even $j=1$ sector, the response to a static $\ell=2$ perturbation vanishes---mirroring the well-known vanishing of black-hole Love numbers in four dimensions~\cite{Hui:2020xxx,Rodriguez:2026iot}---whereas the response to a static $\ell=0$ perturbation does not. This nonvanishing static Proca susceptibility of the Schwarzschild horizon is physical: it shifts the bound-state fine structure at order $\alpha^4\mu$ and is exactly what closes the $\ell=2$ instability rate. The scalar case is markedly different: there the analogous response vanishes and the fine structure is a purely far-zone effect.

The rest of this paper is organized as follows: in \S\ref{sec: field expansion}, we outline the full problem of massive vector bound states around BHs and establish the relevant scales and power counting in preparation for the EFT setup. In \S\ref{sec: eft}, we review the EFT framework introduced in previous work and scrutinize each component of the EFT calculation in detail. In \S\ref{sec: results}, we carry out the matching procedure in detail and summarize the resulting EFT instability rates, relegating a $j=0$ cross-check to Appendix~\ref{sec:j0-appendix}. In \S\ref{sec:diagonal-fix}, we trace the apparent discrepancy with the full theory to a change of basis, connecting the effective theory to the FKKS ansatz on which the full-theory calculation relies. In \S\ref{sec:Perturbative-corrections}, we rederive the very same result from first principles---through degenerate perturbation theory and near-zone matching---without performing that change of basis by hand, providing an independent and complementary route to the exact rate. Finally, \S\ref{sec: discussion} concludes the paper with an outlook on the future of EFTs of superradiance. 

\vspace{6pt}
\noindent \emph{Conventions:} Throughout this paper, we work in units such that $\hbar = c = 1$ and adopt a `mostly plus' metric signature. We will also use capital Latin letters $(I, J, K, ...)$ for spatial indices and lowercase Greek letters $(\nu,\rho,\sigma,...)$ for space-time indices.

\section{Perturbative Expansion for Massive Vectors on a Schwarzschild Background} \label{sec: field expansion}

In this section, we start with a systematic discussion of the Proca equation on a Schwarzschild
background and carefully establish the power-counting scheme we will use throughout this paper. Note that although superradiance requires a spinning BH, we can avoid using the Kerr metric in our regimes of interest by taking the angular momentum to be small. Superradiance is of course stronger the faster the BH spins, but working at small spin simplifies the analysis considerably without obscuring any of the conceptual points we wish to make. BH interactions with the field will be addressed in \S\ref{sec: eft}.

\subsection{Proca equation on a Schwarzschild background}

The propagation of a vector field $A_{\nu}$ with mass $\mu$  on a fixed curved spacetime metric $g_{\lambda\rho}$ is described by the action 
\begin{equation}
    S_\mathrm{v}[g,A]=\int d^{4}x\sqrt{- g}\left( -\frac{1}{4}F_{\lambda\nu}F^{\lambda\nu}-\frac{1}{2}\mu^{2}A_{\nu}A^{\nu}\right) ,\label{eq:procaac}
\end{equation}
with the conventional field strength tensor $F_{\lambda\nu}=\partial_{\lambda}A_{\nu}-\partial_{\nu}A_{\lambda}$.
The equation of motion for $A_{\nu}$ is the Proca equation $\nabla_{\lambda}F^{\lambda\nu}-\mu^{2}A^{\nu}=0$,
which is equivalent to the combination of a vector wave equation and a Lorenz condition,
\begin{equation}
\Box A^{\nu}=\mu^{2}A^{\nu},\qquad\qquad \nabla_{\nu}A^{\nu}=0.\label{eq:procaeq}
\end{equation}
Then, the physical degrees of freedom are the spatial
components of the vector field $A_{I}$ and their derivatives $E^{I} = F^{0I}$,
$B^{I}=\frac{1}{2}\epsilon^{IJK}F_{JK}$. The Lorenz condition implies that the temporal component of the field is not dynamical. The equations \eqref{eq:procaeq} can be expanded as
\begin{subequations} \label{eq: proca eqns in components}
\begin{equation}
\left(\partial_{t}+\Gamma_{\lambda t}^{\lambda}\right)A^{t}=-\left(\partial_{I}+\Gamma_{\lambda I}^{\lambda}\right)A^{I},
\end{equation}
\begin{equation}
\partial^{\rho}\partial_{\rho}A^{I}-g^{\sigma\lambda}\Gamma_{\lambda\sigma}^{\rho}\partial_{\rho}A^{I}+2\Gamma_{J\lambda}^{I}\partial^{\lambda}A^{J}+g^{\lambda\sigma}\partial_{J}\Gamma_{\lambda\sigma}^{I}A^{J}+2\Gamma_{t\lambda}^{I}\partial^{\lambda}A^{t}+g^{\lambda\sigma}\partial_{t}\Gamma_{\lambda\sigma}^{I}A^{t}-\mu^{2}A^{I}=0 \ , \label{eq: 2nd proca equation spelled out}
\end{equation}
\end{subequations}
where in \eqref{eq: 2nd proca equation spelled out} we have exploited the fact that the background metric satisfies the vacuum equation $R_{\lambda\nu}=0$ to trade the terms quadratic in $\Gamma$ for linear ones.

Equations \eqref{eq: proca eqns in components} are valid for any vacuum solution of the Einstein equations. In the remainder of this section, we will focus on the Schwarzschild solution, in which case it becomes convenient to work in spherical coordinates and separate the metric and connection coefficients as
\begin{align}
	g=\eta+h \ ,\ \qquad \qquad \quad  \Gamma=\Gamma_{\eta}+\bar{\Gamma} \ ,
\end{align}
where $\Gamma_{\eta}$ denotes flat connections in spherical
coordinates, while $\bar{\Gamma}$ depends on the Schwarzschild radius $r_\mathrm{s}$ and encodes the departure from Minkowski spacetime due to gravity. The only nonzero components of $h^{\mu\nu}$ and $\bar \Gamma^\lambda_{\mu\nu}$ are
\begin{gather}
	h^{tt}=1-\left(1-\frac{r_\mathrm{s}}{r}\right)^{-1},\quad  h^{rr}=-\frac{r_\mathrm{s}}{r} \ ,\\
    \bar{\Gamma}_{tt}^{r} =\frac{r_\mathrm{s}(r-r_\mathrm{s})}{2r^{3}},\quad  \bar{\Gamma}_{tr}^{t}=\frac{r_\mathrm{s}}{2r(r-r_\mathrm{s})},\quad \bar{\Gamma}_{rr}^{r}=-\frac{r_\mathrm{s}}{2r(r-r_\mathrm{s})}, \quad \bar{\Gamma}_{\theta\theta}^{r}=r_\mathrm{s}, \quad  \bar{\Gamma}_{\varphi\varphi}^{r}=r_\mathrm{s}\sin^{2}\theta.
\end{gather}

Taking advantage of the time translation invariance of the Schwarzschild solution, 
we can expand the vector field in creation and annihilation operators with mode functions that are time independent, 
\begin{align}\label{eq:field-expansion}
\hat{A}^{\nu} & (t, \vec x)=\sum_{\beta}\frac{1}{\sqrt{2E_{\beta}}}\left( \hat{a}_{\beta}f_{\beta}^{\nu}(\vec{x})e^{-iE_{\beta}t}+\hat{a}_{\beta}^{\dagger}f_{\beta}^{\nu*}(\vec{x})e^{iE_{\beta}t}\right) \ ,
\end{align}
where the collective index $\beta$ stands for both discrete and continuous eigenvalues. Then, \eqref{eq: proca eqns in components} reduce to the following equations for the various components of the mode functions:
\begin{subequations} \label{eq:vec_eoms}
\begin{gather}
i E f^{t}= \vec \nabla \cdot \vec f ,\label{eq:constr} \\
\begin{split}
\left(\nabla^{2}\vec{f}\right)^{r}-\mu^{2}f^{r}+\left(1-\frac{r_\mathrm{s}}{r}\right)^{-1}E^{2}f^{r}&+\frac{4r_\mathrm{s}}{r^{3}}f^{r}-\frac{r_\mathrm{s}}{r^{2}}\partial_{r}f^{r}-\frac{r_\mathrm{s}}{r}\partial_{r}^{2}f^{r} \\
&+\frac{3r_\mathrm{s}\cot\theta}{r^{2}}f^{\theta}+\frac{3r_\mathrm{s}}{r^{2}}\partial_{\theta}f^{\theta}+\frac{3r_\mathrm{s}}{r^{2}}\partial_{\varphi}f^{\varphi}=0 , 
\end{split} \label{eq:vecr}\\
\left(\nabla^{2}\vec{f}\right)^{\theta}-\mu^{2}f^{\theta}+\left(1-\frac{r_\mathrm{s}}{r}\right)^{-1}E^{2}f^{\theta}-\frac{3r_\mathrm{s}}{r^{2}}\partial_{r}f^{\theta}-\frac{r_\mathrm{s}}{r}\partial_{r}^{2}f^{\theta}=0 , \label{eq:vecth}\\
\left(\nabla^{2}\vec{f}\right)^{\varphi}-\mu^{2}f^{\varphi}+\left(1-\frac{r_\mathrm{s}}{r}\right)^{-1}E^{2}f^{\varphi}-\frac{3r_\mathrm{s}}{r^{2}}\partial_{r}f^{\varphi}-\frac{r_\mathrm{s}}{r}\partial_{r}^{2}f^{\varphi}=0 , \label{eq:vecph}
\end{gather}
\end{subequations}
where we have streamlined the notation by suppressing the collective index $\beta$, and the Laplacian $\nabla^2$ and the divergence $\vec \nabla \cdot$ are defined in terms of the flat connection $\Gamma_\eta$.\footnote{Note also that the terms in the second line of \eqref{eq:vecr} 
	can be combined into a covariant divergence on the 2D sphere. This
	allows us to rewrite them in terms of the Newman-Penrose covariant
	derivatives in Appendix~\ref{sec:Degeneracies}. } 
 The constraint \eqref{eq:constr} determines the time component of the mode functions in terms of the spatial ones and, interestingly, is independent of $r_\mathrm{s}$. The spatial components, instead, should solve \eqref{eq:vecr}--\eqref{eq:vecph} with
appropriate boundary conditions on the horizon and at infinity. Unfortunately, these equations are prohibitively difficult to solve exactly, and therefore we need to resort to some approximations.

\subsection{Perturbative expansion and power counting for bound states} \label{sec: bound states}

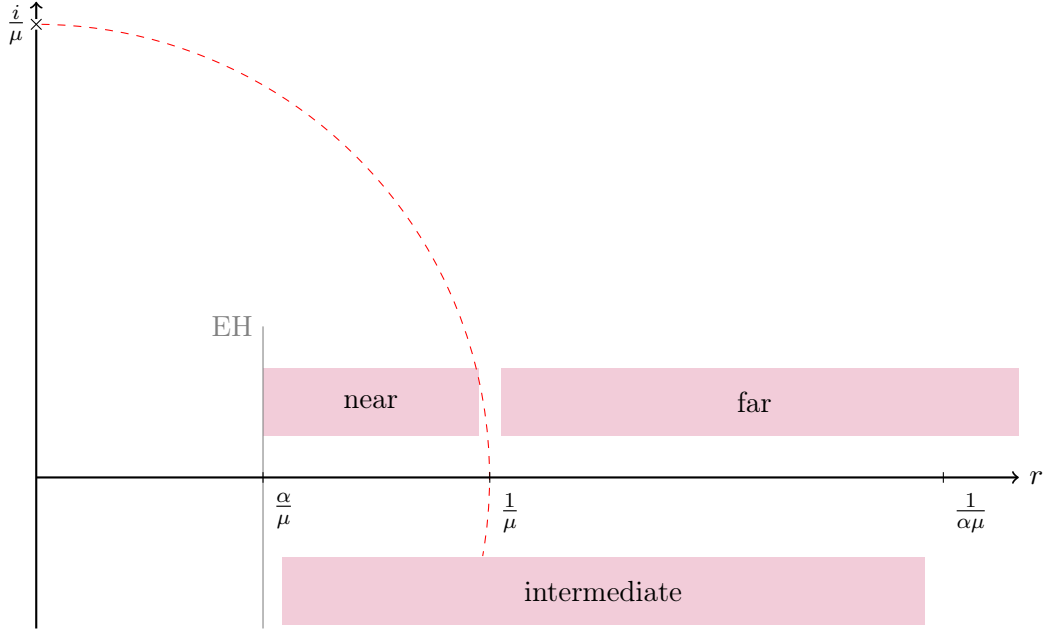
\begin{figure}
    \centering
\begin{tikzpicture}

\fill[purple!20!white] (3,0.55) rectangle (5.85,1.45);
\draw (4.425,1) node{near};
\fill[purple!20!white] (6.15,0.55) rectangle (13,1.45);
\draw (9.5,1) node{far};
\fill[purple!20!white] (3.25,-1.05) rectangle (11.75,-1.95);
\draw (7.5,-1.5) node{intermediate};

\draw[dashed,red] (2pt,6) arc (89.3:-10:6);
\draw[gray] (3,-2) -- (3,2) node[anchor = east]{EH};

\draw[thick,->](0,0) -- (13,0) node [anchor = west]{$r$};
\draw (6,2pt) -- (6,-2pt) node [anchor = north west]{$\frac{1}{\mu}$};
\draw (12,2pt) -- (12,-2pt) node [anchor = north west]{$\frac{1}{\alpha \mu}$};
\draw (3,2pt) -- (3,-2pt) node [anchor = north west]{$\frac{\alpha}{\mu}$};

\draw[thick] (0,-2)--(0,6cm-2pt);
\draw[thick,->] (0,6cm+2pt)--(0,6.3);
\draw (0,6) node [anchor = east]{$\frac{i}{\mu}$};
\draw (-2pt,6cm-2pt)--(2pt,6cm+2pt);
\draw (-2pt,6cm+2pt)--(2pt,6cm-2pt);

\end{tikzpicture}
    \caption{
    The radial Proca equation features imaginary poles close to $\pm i$ times the Compton wavelength.
    Consequently, the near zone (centered at the event horizon, EH) and the far zone (centered at infinity) both end near the Compton wavelength with no overlap, necessitating an intermediate zone to bridge between them in the matched asymptotic expansion. Although the bound-state problem of \S\ref{sec: bound states} is analyzed in the far zone, the near zone re-enters through static contact terms on the worldline, which we fix by matching in \S\ref{sec:Perturbative-corrections}.}
    \label{fig:zones}
\end{figure}

Due to the attractive nature of gravitational interactions, \eqref{eq:vec_eoms} admit an infinite number of solutions with $E < \mu$ that are regular on the horizon and vanish at spatial infinity. These mode functions describe the bound states of spin-1 particles and the BH. Their equations feature two length scales: the Schwarzschild radius $r_\mathrm{s}$ and the Compton wavelength $1/\mu$. In this paper, we will be interested exclusively in the situation where $r_\mathrm{s} \ll 1/\mu$, in which case the system admits a naturally small parameter,
\begin{align}
	 \alpha \equiv G M \mu = \frac{r_\mathrm{s} \mu}{2} \ll 1,
\end{align}
which captures the weakness of the interactions between the BH and the spin 1 particles described by the vector field. We can then use perturbation theory to calculate all observable quantities as an expansion in powers of the small parameter~$\alpha$. However, when it comes to the mode functions, this expansion cannot be implemented uniformly across the entire region outside the horizon. Therefore, the strategy commonly adopted (see \textit{e.g}.~\cite{Rosa:2011my,Baumann:2019eav}) is to subdivide the range $r_\mathrm{s} \leqslant r < \infty$ into two or more overlapping regions where the $\alpha$ expansion can be implemented uniformly, and then smoothly match these solutions where they overlap, as in Figure~\ref{fig:zones}. This procedure is possible only for discrete values of $E$ corresponding to the allowed energies of bound states.

The outermost region is known as the \emph{far field zone}, or simply the \emph{far zone}, and covers values of $r \gg r_\mathrm{s}$. The equations of motion (\ref{eq:vecr}\textendash\ref{eq:vecph}) can be multiplied by $(1-r_\mathrm{s}/r)/(2\mu)$ and cast in the form of an eigenvalue equation, organized in powers of $r_\mathrm{s}/r$:
\begin{align} \label{eq: eqs for spatial components of mode functions}
	\frac{E^2 - \mu^2}{2\mu} \ f^I = - \frac{1}{2\mu}\left( \nabla^2 \vec f \right)^I - \frac{\alpha}{r} \ f^I + {V^I}_J f^J \ ,
\end{align}
where on the right-hand side we have singled out the two terms that are most dominant at very large $r$ and collected the remaining terms into the differential operator ${V^I}_J$, with the indices $I$ and $J$ running over $r, \theta, \varphi$. 

As a first approximation, we can neglect ${V^I}_J$ and expand the solution to \eqref{eq: eqs for spatial components of mode functions} in terms of ``pure-orbital'' vector harmonics~\cite{Thorne:1980ru}, 
\begin{align} \label{eq: expansion in pure orbital vector harmonics}
	\vec f (r, \theta, \varphi) =  R (r) \vec Y_{\ell, \, j m} (\theta, \varphi) .
\end{align}
The pure-orbital vector harmonics are eigenfunctions of the Laplacian operator on the 2-sphere with eigenvalue $- \ell (\ell +1)$ for integers $\ell \geqslant 0$, determining the orbital angular momentum. The other two eigenvalues, integers $j$ and $m$, determine the total angular momentum; their possible values are $j = \ell -1, \ell, \ell+1 \geqslant 0$ and $|m| \leqslant j$. This approximation and ansatz reduce \eqref{eq: eqs for spatial components of mode functions} to the Schr\"odinger equation for the hydrogen atom for $R$, so we introduce the overtone number $n \geqslant \ell + 1$ and take the leading-order radial solution to be $R=R_{n\ell}$~\cite{Sakurai1993Modern}, given by
\begin{equation}\label{eq:radial-mode}
    R_{n\ell}(r) = \sqrt{\left(\frac{2\alpha\mu}{n}\right)^3 \frac{(n-\ell-1)!}{2n(n+\ell)!}} e^{-\rho/2} \rho^\ell L^{2\ell+1}_{n-\ell-1}(\rho).
\end{equation}
Above, $\rho = 2\alpha\mu r/n$ and $L^{2\ell+1}_{n-\ell-1}(\rho)$ is a generalized Laguerre polynomial~\cite{abramowitz+stegun} of degree $n-\ell-1$\footnote{In Sakurai~\cite{Sakurai1993Modern}, the same polynomial is instead denoted $-L^{2\ell+1}_{n+\ell}(\rho)/(n+\ell)!$. This is because Sakurai's definition of $L_n(x)$ is $n!$ times that in Abramowitz and Stegun, and Sakurai defines $L^m_n(x) = d^m L_n(x)/dx^m$ instead of the result $L_n^{m}(x) = (-1)^m d^m L_{n+m}(x)/dx^m$.}.
Bound states are thus labeled by the complete set of integer eigenvalues $(n, j, m, \ell)$.

In this solution, the only length scale that appears on the right-hand side of \eqref{eq: eqs for spatial components of mode functions} is $1/(\alpha \mu)$, defining the Bohr radius for this problem and indicating the power-counting rule for this region:
\begin{align}
    \nabla \ \sim \  \frac{1}{r} \ \sim \ \alpha \mu \, .
\end{align}
This characteristic length scale is parametrically larger than $1/\mu$, so it indeed describes a particle in the far zone. The energy eigenvalues depend on the single energy scale $\alpha\mu$ as  
\begin{align} \label{eq: eigenvalue}
	E^2 - \mu^2 \simeq - \frac{(\alpha \mu)^2}{n^2}.
\end{align}

Based on these scaling rules, the perturbation ${V^I}_Jf^J$ in \eqref{eq: eqs for spatial components of mode functions} consists of terms scaling like $\alpha^4 \mu$ and $\alpha^6 \mu$. The former is the leading perturbation to the system $V^{(1)} \sim \alpha^4\mu$, and its components in the standard $(r,\theta,\varphi)$ basis satisfy
\begin{multline}\label{eq:pert}
    {V^{(1)I}}_Jf^J=\frac{r_\mathrm{s}}{2\mu r} \left(
    \left(\nabla^2 \vec{f}\right)^I + \left(\partial_r^2 + \frac{3}{r} \partial_r\right) f^I \right.\\
    \left.- \frac{1}{r^2} \left( 2r\partial_rf^r + 4f^r + 3r\partial_\theta f^\theta + 3r\cot\theta f^\theta + 3r\partial_\varphi f^\varphi \right)  \delta^I_r\right).
\end{multline}
It is worth emphasizing that, whereas the flat Laplacian $\nabla^2$ operates on the vector $\vec{f}$, all other derivatives operate solely on the components.
Using standard time-independent perturbation theory, we can solve \eqref{eq: eqs for spatial components of mode functions} to calculate the mode functions in the far zone order by order in $\alpha$. We will return to these corrections in \S\ref{sec:Perturbative-corrections}.

In order to calculate the spectrum of bound states at higher order in $\alpha$, including the instability rates associated with the imaginary parts of the energy eigenvalues, it becomes necessary to go beyond the far zone and consider matching with regions closer to the BH horizon. In these regions, modes with the same parity mix with each other and the ansatz \eqref{eq: expansion in pure orbital vector harmonics} no longer diagonalizes the equation for the mode functions. The more sophisticated FKKS ansatz~\cite{Frolov:2018ezx} that is also suitable for studying Kerr BHs is required, although modes can still be classified using quantum numbers $\left\{ n, j, m, \ell \right\}$. This asymptotic matching procedure yields the following result for the instability rates at leading order in $\alpha$:\footnote{Within the source material, $\Gamma$ is defined as the imaginary part of the frequency, $\omega = E + i \Gamma$, which is \textit{half} the decay rate; \eqref{eq:Baumann-result} is two times the result of~\cite{Baumann:2019eav} to put it on par with our results.}
\begin{gather}
\Gamma_{njm\ell}=4\left(1+\sqrt{1-\tilde{a}^2}\right)C_{n\ell j}\left(\prod_{k=1}^{j}\bigl(k^2 + (m^2-k^2)\tilde{a}^2\bigr)\right)(m\Omega-\mu)\alpha^{2\ell+2j+5}, \label{eq:Baumann-result}\\
C_{n\ell j}=\frac{2^{2\ell+2j+1}(n+\ell)!}{n^{2\ell+4}(n-\ell-1)!}\left(\frac{\ell!}{(\ell+j)!(\ell+j+1)!}\right)^{2}\left(1+\frac{2(1+\ell-j)(1-\ell+j)}{\ell+j}\right)^{2} , \nonumber
\end{gather}
where $\tilde{a} = J/(GM^2)$ (with $J$ the angular momentum of the black hole) is the dimensionless spin parameter. Taking the limit $\tilde{a}^2 \ll 1$ for a slowly rotating black hole and restricting our attention to modes with $j=m=1$ and $\ell = 0,1,2$, the general result \eqref{eq:Baumann-result} reduces to
\begin{subequations}\label{eq: instability rates for j=1 bound states}
\begin{align} 
	\Gamma_{n,1,1,0} &= \frac{16}{n^3} \alpha^7 \Omega  \\
    \Gamma_{n,1,1,1} &= \frac{64(n^2-1)}{9n^5}\alpha^9 \Omega \\
    \Gamma_{n,1,1,2} &= \frac{16 (n^2-1)(n^2-4)}{81 n^7}\alpha^{11} \Omega, \label{eq: instability rates for j=1 bound states l=2}
\end{align}
\end{subequations}
with $\Omega \gg \mu$ the angular velocity of the BH\footnote{
These results for $\Gamma_{1,1,1,0}$ and $\Gamma_{2,1,1,1}$ match the rates derived in~\cite{Baryakhtar:2017ngi} with $a_*\mu \approx 4 \alpha \Omega$}. In this paper, we will show that the conventional point-particle EFT, applied naively, appears to miss the rate \eqref{eq: instability rates for j=1 bound states l=2} by a factor of three, but that this shortfall is an artifact of the leading-order basis: once the degenerate $\ell=0$ and $\ell=2$ states are correctly identified, the EFT reproduces \eqref{eq: instability rates for j=1 bound states l=2} exactly. Along the way, we also correct an earlier result in the EFT literature~\cite{Endlich:2016jgc}.

\subsection{Perturbative expansion and power counting for relativistic asymptotic states} \label{sec:asymptotics}

In what follows, we will also be interested in the probability $P_\mathrm{abs}$ of absorption of asymptotic scattering states with particle momentum $\vec k$, where the particle energy is $\omega_k \equiv \sqrt{\mu^2 + k^2} >\mu$ and with de Broglie wavelength large compared to the size of the BH, i.e. $r_\mathrm{s} \ll 1/k$. 
In this case, there is an additional dimensionful parameter, $k$, and thus the perturbative structure becomes richer. The situation simplifies considerably, however, if we choose to work in the relativistic regime,
\begin{align}
	r_\mathrm{s} \ll \frac{1}{k} \ll \frac{1}{\mu} ,
\end{align}
where the vector modes can be treated as approximately massless, and observable quantities admit at leading order an expansion in powers of a single small parameter---the product $k r_\mathrm{s}$. In this limit, we treat spacetime as flat, so the wave equation \eqref{eq: eqs for spatial components of mode functions} reduces to
\begin{align} 
	\frac{k^2}{2} \vec f \simeq - \frac{1}{2} \nabla^2 \vec f  \ ,
\end{align}
and the mode functions are 
\begin{align}
	\vec  f_{\vec{k}, \lambda} (\vec x) = \vec \epsilon \, (\vec k, \lambda) \frac{e^{i \vec k \cdot \vec x}}{\sqrt{2 \omega_k}} \ , 
\end{align}
with $\lambda$ the helicity, an eigenvalue of the operator $\vec J \cdot \vec P / \vert \vec P \vert$. To express solutions in terms of angular momentum, rather than labeling with the eigenvalues $\vec k$ of momentum $\vec P$, it is more convenient to use spherical helicity states with labels $\left\{\omega, j, m, \lambda\right\}$, where $\omega = \omega_k$ and the angular momentum numbers $j$ and $m$ are as they appear in \eqref{eq: expansion in pure orbital vector harmonics}, diagonalizing the Hamiltonian $H$, total angular momentum $\vert \vec J\vert^2$ and $J_z$, and helicity. (For free particles, the total angular momentum is conserved and hence the appropriate quantum number to adopt.)

The ultrarelativistic limit takes $\mu \to 0$, which is a delicate limit for a vector field. A massive vector has three polarizations, whereas a massless vector has just two. The limit can be taken continuously at the level of the action using the Stueckelberg trick~\cite{Hinterbichler:2011tt}, which consists of introducing a massless scalar field with the replacement $A_\nu \to A_\nu + {\mu}^{-1} \partial_\nu \phi$. The Lagrangian of $A_\nu$ and $\phi$ has a gauge symmetry, $A_\nu \to A_\nu + \partial_\nu \lambda$, $\phi \to \phi - \mu \lambda$, so gauge-fixing can remove the scalar field whenever $\mu \ne 0$, rendering the modified action equivalent to the Proca action $S_\mathrm{v}$. In the limit $\mu \to 0$, the longitudinal mode decouples and becomes a scalar degree of freedom, preserving the number of physical polarizations:
\begin{equation}\label{eq:decoupling}
    S_\mathrm{v}[A,g] \, \stackrel{\mu \to 0}{\longrightarrow} \int d^{4}x\, \sqrt{-g}\left( -\frac{1}{4}F_{\lambda\nu}F^{\lambda\nu}-\frac{1}{2}\partial_{\nu}\phi \partial^{\nu} \phi \right). 
\end{equation}

Consequently, a complete description of the massless regime requires considering both scalar and vector absorption. The known absorption probabilities for a Schwarzschild BH are~\cite{Page:1976df}:
\begin{align}
    P_\mathrm{abs}^\mathrm{scalar} &= \frac{16}{9} (GM)^4 \omega^3 (\omega - m \Omega), \label{eq:Pabs-scalar}\\
    P_\mathrm{abs}^\mathrm{vector} &= \frac{64}{9} (GM)^4 \omega^3 (\omega - m \Omega), \label{eq:Pabs-vector}
\end{align}
for $j = 1$ ($\ell = 1$ for the scalar) and each helicity of the massless vector. From these, the absorption cross sections can be derived. Note that these probabilities are proportional to $\omega - m \Omega$. When this quantity is negative, the system exhibits superradiant amplification.

\section{The Worldline EFT Approach} \label{sec: eft}

 Rather than describing the BH as a fixed curved metric in the full general relativistic theory, we can model it with an effective theory that treats it as a point particle~\cite{Goldberger:2004jt,Porto:2005ac}. Such an approach should be able to describe all interactions between the vector and the BH that take place at scales much larger than $r_\mathrm{s}$. These include the behavior of bound states with Bohr radius $1/(\mu \alpha) \gg r_\mathrm{s}$, and the scattering and absorption of waves with a wavelength large compared to $r_\mathrm{s}$. Therefore, we can construct the EFT via matching with known results from scattering calculations and make predictions for the bound-state instability rates. In this section we detail the construction and derive the general formalism, which applies equally to bound states and to asymptotic scattering states; the application to bound-state instability rates is the subject of \S\ref{sec: results}.

\subsection{Introduction to the worldline EFT}

Following EFT principles, we need to include all possible couplings to the external field consistent with the symmetries, and their form can be determined by considering the pattern of spontaneous symmetry breaking. In addition to the usual Poincar\'e symmetry, taking the point particle to be spherically symmetric introduces an internal $\operatorname{SO}(3)$ symmetry. The presence of the point particle spontaneously breaks boosts and spatial translations, and the spacetime rotations and internal rotations together are broken to their diagonal $\operatorname{SO}(3)$ subgroup because the ground state is invariant under a rotation of the particle coupled with the same rotation of spacetime. The Goldstone bosons resulting from the broken translations can be identified with the vierbein ${e_\mu}^a$, and those from the broken spacetime rotations are the Euler angles $\xi^a$ after being projected onto the BH worldline. There are no Goldstone bosons associated with the broken boosts. The Euler angles enter the Lagrangian in the form of the relativistic angular velocity, which includes Lorentz transformations parameterized by the Euler angles~\cite{Delacretaz:2014oxa}.

The relativistic angular velocity is~\cite{Endlich:2015mke}
\begin{equation}
    \Omega_I = - \frac{1}{2}\epsilon_{IJK} \left( \eta^{ab}{\Lambda_a}^J \frac{d}{d\tau} {\Lambda_b}^K + \frac{dx^\nu}{d\tau}{\omega_\nu}^{ab} {\Lambda_a}^J{\Lambda_b}^K \right),
\end{equation}
where $\tau$ is the proper time defined by $d\tau=\sqrt{-g_{\lambda\nu}dx^{\lambda}dx^{\nu}}$, $x^\nu$ is the worldline of the particle, ${\omega_\nu}^{ab}$ is the spin connection, and ${\Lambda_a}^b = {B_a}^c {R_c}^b$ is a Lorentz transformation taking a comoving frame into an inertial one, which can always be represented as a rotation $R$ followed by a boost $B$; ${R_I}^J$ is a typical $\operatorname{SO}(3)$ matrix and ${R_0}^a = \delta_0^a$. In the nonrelativistic limit, $\tau\to t$ and ${B_a}^c \to \delta_a^c$. Then, the angular velocity as a function of the Euler angles $\xi^a$ becomes
\begin{equation}
    \Omega_I(\xi) = \frac{1}{2} \epsilon_{IJK} R^{LJ}(\xi) \frac{d}{dt} {R_L}^K(\xi).
\end{equation}
By using the exponential parameterization of the rotation matrix $R$, it is easy to see that this is equivalent to the angular velocity of the body.

Putting this together, we construct a general action for a dynamical, isolated point particle EFT with the worldline sourcing the gravitational potential~\cite{Goldberger:2004jt}.
The dynamics are described by an invariant relativistic particle action
on the worldline 
\begin{equation}
S_\mathrm{pp}[g,x,\xi]=-\int d\tau\, M(\Omega^{2})
\end{equation}
where gravity is consistently included by the definition of the proper
time $d\tau$, and $M(\Omega^{2})$ gives the mass of the spinning object which includes contributions from rotational energy~\cite{Porto:2005ac, Endlich:2015mke,Delacretaz:2014oxa}. This dependence is related to the Hansen-Regge
function of the spinning top~\cite{Hanson:1974qy}. For our present purpose,
where the object starts with stationary initial conditions and the backreaction is assumed negligible, the details are suppressed, and we use $S_\mathrm{pp}[g,x,\xi]$ alongside the Einstein-Hilbert action $S_\mathrm{EH}[g]$ to determine the leading-order deviations of $g$ in $r_\mathrm{s}/r$ from Minkowski spacetime.

The complete action is $S_\mathrm{EH}[g]+S_\mathrm{pp}[g,x,\xi]+S_\mathrm{v}[g,A]+S_\mathrm{int}[g, x, \xi, A, X]$,
where the third term is defined in \eqref{eq:procaac} and the last describes the internal dynamics of the particle via its microscopic degrees of freedom $X$ and interactions with the macroscopic degrees of freedom~\cite{Goldberger:2005cd}. First, we use $S_\mathrm{EH}$ and $S_\mathrm{pp}$ to determine $g_{\lambda\nu}$, $x^\nu$, and $\xi^a$, neglecting the backreaction from the vector field and the internal degrees of freedom. Taking the non-relativistic limit for the motion of the BH, we can choose an inertial frame where $x^\nu = (t,\vec{0})$ and $\partial_t\xi^a = \Omega \delta^a_3$ where $\Omega$ is the angular velocity about the $z$-axis\footnote{The description is similar to the case of an extreme mass ratio system at zeroth order where the backreaction is completely neglected~\cite{Cheung:2023lnj}.}~\cite{Goldberger:2004jt}. Similarly, we can expand the metric as $g_{\lambda\nu} = \eta_{\lambda\nu}+h_{\lambda\nu}$, use the BH as a fixed gravitational source and define the metric to any desired order in ${r_\mathrm{s}}/{r}$ and with the Kerr spin parameter neglected. To obtain the normal modes of the vector field as in the previous section, we neglect the interaction in $S_\mathrm{int}$. For bound state modes, we insert $\eta + h$ as a background metric in $S_\mathrm{v}$, and for asymptotically free scattering states, we can simply use $\eta$. Substituting this background into $S_\mathrm{v}$ is equivalent to integrating out the graviton at tree level to all orders in $GM$: the ladder of potential-graviton exchanges and the nonlinear graviton vertices that a diagrammatic calculation would assemble one by one~\cite{Goldberger:2004jt} are resummed into the Schwarzschild metric from the outset~\cite{Neill:2013wsa}. (Graviton loops and the backreaction of the vector cloud are negligible in the test-field limit considered here, so this tree-level resummation is exact for our purposes.) The expansion in powers of $r_\mathrm{s}/r$ carried out in \S\ref{sec:Perturbative-corrections} then acts on this exact result. We comment on the prospect of a fully diagrammatic derivation in \S\ref{sec: discussion}.

The resummation just described captures only the potential sourced by the worldline mass; the couplings that encode the finite size of the horizon are not generated by integrating out the graviton and must be supplied as worldline operators with matched Wilson coefficients. These come in two kinds. The static contact term introduced in \S\ref{sec:Perturbative-corrections} is conservative despite its near-zone origin: it shifts and mixes the bound-state levels but produces no absorption by itself, and enters the instability rates only by selecting the correct leading-order states. Absorption and superradiant amplification are instead governed by dissipative couplings of the field to the microscopic degrees of freedom of the BH, to which we now turn. We choose not to keep track of the internal degrees of freedom in $S_\mathrm{int}$, instead modeling their interaction with the vector fields stochastically with couplings to composite operators $\mathcal{O}$. We already fixed $g_{\lambda\nu}$, $x^\nu$, and $\xi^a$, so $S_\mathrm{int}[A,X]$ is used to model the dissipation of energy from the vector field to excite modes in the microscopic sector. These degrees of freedom represent the short distance dynamics (e.g. quasinormal modes)
that we lose track of at long distances\footnote{While we specialize to the small spin limit here, the spinning black 	holes and their dissipative dynamics are also systematically described in the worldline formalism~\cite{Porto:2007qi}.}.

In the coset construction, the unbroken subgroup is linearly realized, requiring the degrees of freedom $A_\nu$ and $\mathcal{O}$ to be representations of $\operatorname{SO}(3)$. For the composite operators, defined in the corotating frame of the BH, the overall rank of the tensor
representation corresponds to the total angular momentum quantum
number $j$. We anticipate that the dominant modes have $j=1$, which
include the three polarization states of vector bosons (corresponding to $(\frac{1}{2},\frac{1}{2})$, $(1,0)$, and $(0,1)$ representations of the Lorentz group). Therefore our interaction terms will involve fields coupled to composite operators of the form $\mathcal{O}^I$ to yield $\operatorname{SO}(3)$ scalars.

For $A_\nu$, we need to consider all vector representations of $\operatorname{SO}(3)$ that can be constructed from the field ${A}_I$ and its derivatives (recall that $A_0$ is determined by a constraint, so only $A_I$ is dynamical). Then, the coset construction dictates that the Goldstone modes associated with broken rotations are included via rotation matrices that dress the vectors as $\tilde{A}_{I}\equiv {R_{I}}^J A_{J}$~\cite{Endlich:2016jgc}, where $\tilde{A}_{I}$ is the field measured in the corotating frame of the BH. We can use the equations of motion $\nabla^2 A_I = \partial_t^2 A_I + \mu^2 A_I$ to replace vector Laplacians with the constant factor $\mu^2 - E^2$, which means the only independent derivative operators we can use for our field degrees of freedom are those that are trace-free. Thus, there are only three independent fields: $\tilde A_I$, $\epsilon_{IJK}\partial^J\tilde A^K$, and $\partial_I \partial^J \tilde A_J$, which can be reorganized as $\tilde{A}_{I}$, $\tilde{B}_I$, and $\tilde{E}_{I}$. The benefit of these field definitions is that $B_I$ and $E_I$ are now gauge-invariant, each field is even or odd under parity transformations, and $E_I$ is even under time-reversal. The leading order interaction is linear in the fields, with dispersion included through $j=1$ representations of $\operatorname{SO}_\mathrm{diag}(3)$, so the most general interaction is
\begin{equation}
    S_\mathrm{int}[A,X]=\int dt \left( \mu\tilde{A}_{I}\mathcal{O}^{I}(X) + \tilde{E}_{I}\mathcal{O}_{E}^{I}(X)+\tilde{B}_{I}\mathcal{O}_{B}^{I}(X)\right).
\end{equation}
Because $\mathcal{O}^I$ is not a conserved current, the $\mu$ factor in the first term was added to ensure that the theory
has a smooth $\mu\rightarrow0$ limit~\cite{Endlich:2016jgc}. The
exact details of the composite operators $\mathcal{O}(X)$ depend
on the full theory, so they are in effect the generalized Wilson coefficients
in this EFT framework to be determined from matching with UV calculations. The matching calculation will identify the leading-order coefficients, which are sensitive only to the scale $r_\mathrm{s}$, in the series expansions of correlation functions of the composite operators. We will call these the Wilson coefficients below.\footnote{\label{fn:conservative-bilinear}The interaction above collects the leading \emph{dissipative} couplings---linear in the field, and responsible for the instability rates matched in \S\ref{sec: results}. It does not exhaust the worldline action: conservative static contact terms \emph{quadratic} in the field, of order $r_\mathrm{s}$ and living in the parity-even $j=1$ channel, are likewise allowed, and play an essential role in the degenerate sector treated in \S\ref{sec:Perturbative-corrections}.}

The probability of any transition is given by the square of the S-matrix element
\begin{equation} \label{eq: lowest order rate}
P_{i \to f} = \sum_{X_f}\vert\langle \beta_f;X_f\vert S\vert \beta_i;X_{i}\rangle\vert^{2},
\end{equation}
where the S-matrix is the time-ordered exponential of the interaction Hamiltonian
\begin{equation} \label{eq: s-matrix}
    S = \mathcal{T} \exp\biggl( -i \int dt\,{H}_\mathrm{int}(t)\biggr).
\end{equation}
The probability $P_{i\to f}$ represents an inclusive scattering process where a BH in the initial state $\lvert X_{i}\rangle$ absorbs an initial field state $\lvert\beta_{i}\rangle$ and emits a final state $\lvert\beta_f\rangle$ while transitioning into any possible final state $\lvert X_f\rangle$; the collective indices $\beta_{i,f}$ of \eqref{eq:field-expansion} may stand for the quantum numbers of bound states or of asymptotic scattering states, so that \eqref{eq: lowest order rate} describes elastic and inelastic scattering as well as absorption and emission. For asymptotic states, the divergence contained in \eqref{eq: lowest order rate} is spurious: it cancels against the (equally divergent) norm $\langle\beta_i\vert\beta_i\rangle=2\pi\delta(0)$ of the continuum initial state, and the normalized probability---for instance, the probability that a single incident quantum be absorbed---is finite, because a wave packet interacts with the BH only for a finite time. It is this normalized probability that we will match against full-theory results in \S\ref{sec: results}. For transitions between a bound state and the vacuum, by contrast, the initial state is normalizable but overlaps with the BH at all times: the squared S-matrix element grows linearly with the total observation time $T=2\pi\delta(0)$, and this infrared divergence merely signals that the finite observable is the transition \emph{rate},
\begin{equation}
    \Gamma_{i \to f} = \frac{P_{i \to f}}{T} \, .
\end{equation}
The physical growth rate of a bound state is then the difference between the emission rate and the absorption rate, but the latter vanishes in the superradiant regime $m \Omega > \omega$ at leading order.
For absorption (emission), we set the final (initial) field state to vacuum $\vert 0 \rangle$, so that $\beta$ will hereafter denote the quantum numbers of the state being absorbed (emitted).

\subsection{Transition rate from the EFT}
\label{sec: EFT rates}

Since the composite operators and field operators act on the black
hole and field Hilbert space separately, all the terms factorize. This means, in particular, that the inclusive rate obtained by expanding $S$ in \eqref{eq: s-matrix} to linear order in the interaction is proportional to the Wightman correlation functions of the composite operators:
\begin{align}
    \Gamma_{\beta} \propto \sum_{X}\langle X_{0}\vert\mathcal{O}_{a}(t)\vert X\rangle\langle X\vert\mathcal{O}_{b}(t')\vert X_{0}\rangle=\langle X_{0}\vert\mathcal{O}_{a}(t)\mathcal{O}_{b}(t')\vert X_{0}\rangle\equiv\langle\mathcal{O}_{a}\mathcal{O}_{b}'\rangle,
\end{align}
where $a,b$ run through the three operators $\left(\mathcal{O}_{E},\mathcal{O}_{B},\mathcal{O}\right)$, and in suppressing the dependence on $t$, an operator $\mathcal{O}$ evaluated at $t'$ is denoted $\mathcal{O}'$.
In the first equality, we have assumed that the Hilbert space of the
EFT is unitary, so the summation over final states is the resolution
of identity. Furthermore, the observed superradiant effects occur for classically large systems, so the BH is considered to have zero absolute temperature, and the correlation function above is effectively in its ground state.

For parity invariant systems, the even-odd parity operators must have vanishing correlations,~\textit{i.e.}\
\[
\langle\mathcal{O}_{E}\mathcal{O}_{B}'\rangle=\langle\mathcal{O}\mathcal{O}_{B}'\rangle=0.
\]
The correlation function $\langle\mathcal{O}_{E}\mathcal{O}'\rangle$, by contrast, involves two parity-even operators and is not required to vanish. Time reversal, however, constrains it to be an even function of the frequency (see Appendix~\ref{sec:Worldline-Wightman-functions}), which is incompatible with the matching in \S\ref{sec: results}, so it must be negligible in our small-spin regime. 

Since the mixed correlation functions all vanish, the absorption rate of a state $\lvert\beta\rangle$ simply decomposes as $\Gamma_{\mathrm{abs},\beta} = \Gamma^A_{\mathrm{abs},\beta} + \Gamma^B_{\mathrm{abs},\beta} + \Gamma^E_{\mathrm{abs},\beta}$, where each field $F^a_I = \mu A_I, B_I, E_I$ contributes the term
\begin{align}\label{eq: abs-rate}
    \Gamma^a_{\mathrm{abs},\beta} = \frac{1}{T} \int dt\,dt'\, \langle\mathcal{O}^L_a(t') \mathcal{O}^J_a(t) \rangle \langle \beta \rvert F^a_K (t') \lvert 0\rangle \langle 0 \rvert F
    ^a_I(t)\lvert \beta\rangle {R^K}_L(t'){R^I}_J(t).
\end{align}
The emission rates are calculated similarly, resulting in \eqref{eq: abs-rate} but with the exchange $t \leftrightarrow t'$, which matters only for the correlation function. The net production rate
is given by the difference $\Gamma=\Gamma_\mathrm{em}-\Gamma_\mathrm{abs}$. 
The expansion \eqref{eq:field-expansion} implies for the field matrix elements that
\begin{subequations}
\begin{align}
    \langle 0 \vert \mu A^I(t,0) \vert \beta \rangle &= \frac{\mu}{\sqrt{2E_\beta}} f_\beta^I(0) e^{-iE_\beta t}, \\
    \langle 0 \vert B^I(t,0) \vert \beta \rangle &= \frac{1}{\sqrt{2E_\beta}} {\epsilon^{IJ}}_{K} \partial_J f_\beta^K(0) e^{-iE_\beta t}, \\
    \langle 0 \vert E^I(t,0) \vert \beta \rangle &= i \sqrt{\frac{E_\beta}{2}} \left( \delta^I_J + \frac{1}{E_\beta^2}\partial^I\partial_J \right) f_\beta^J(0) e^{-iE_\beta t}, \label{eq:bound-E-field-matrix-element}
\end{align}
\end{subequations}
where $E_{\beta}$ is the energy of the state $\beta$, and the index contractions are made with the flat metric. Therefore, the ingredients for calculating the transition rate are
the correlation functions and the mode functions $\vec{f}_\beta(\vec{x})$. The correlation functions are explored next; the mode functions depend on the states under consideration, and are computed in \S\ref{sec: results} for the bound states of interest.

\subsection{Wightman functions}

For the Wightman correlation functions, we find it convenient to work in frequency space
\begin{equation}
    \langle\mathcal{O}_{I}^{\prime \,a}{\mathcal{O}}_{J}^{b}\rangle =\int\frac{d\omega}{2\pi}e^{i\omega(t-t')}W^{a}(\omega) \delta_{IJ}\delta^{ab},\label{eq:Wightman-def}
\end{equation}
where, by results in Appendix~\ref{sec:Worldline-Wightman-functions}, the functions $W(\omega)$, $W^B(\omega)$, and $W^E(\omega)$ are real. Because the composite operators $\mathcal{O}_I^a(t)$ are defined in the rest frame of the BH, they are insensitive to its spin, and the spherical symmetry of the system constrains the overall correlation function to be proportional to $\delta_{IJ}$. 
Furthermore, because the BH can be treated classically as a zero-temperature object, $\vert X_0 \rangle$ is its ground state, and $W^a(\omega)$ vanishes for negative frequencies~\cite{Endlich:2016jgc}.

The difference $P_\mathrm{em}-P_\mathrm{abs}$ is proportional to the Pauli-Jordan correlator
\begin{equation}
    \langle[\mathcal{O}^{\prime \,a}_I,{\mathcal{O}}^{b}_J]\rangle =\delta_{IJ}\delta^{ab}\int\frac{d\omega}{2\pi}e^{i\omega (t-t')} \Delta_{-}^a(\omega).
\end{equation}
It immediately follows that $\Delta_-^a(\omega)$, the spectral function for the correlator~\cite{Arteaga:2008ux}, is real and odd, and that $W^a(\omega) = \Delta_-^a(\omega) \theta(\omega)$ (see Appendix~\ref{sec:Worldline-Wightman-functions}), which means that the instability rate and absorption probability of a field are both related to the same unknown function. Moreover, for a system in thermal equilibrium, correlations must decay quickly enough to guarantee that $\Delta^a_-(\omega)$ is analytic at $\omega = 0$~\cite{Endlich:2012vt}.
Taking the small-spin limit, we use a first-order series expansion to approximate $\Delta_-^a(\omega) = \gamma_a \omega$, with $\gamma_a$ some unknown coefficient that needs matching for each correlation function. 

Combined with the results of the previous section, this means that the unknowns of the leading-order theory reduce to three coefficients, $\gamma$, $\gamma_B$, and $\gamma_E$, defined via
\begin{equation}
    \langle\mathcal{O}_{I}^{\prime\,a}{\mathcal{O}}_{J}^{a}\rangle =\delta_{IJ}\int\frac{d\omega}{2\pi}e^{i\omega(t-t')}W^a(\omega) \simeq\delta_{IJ}\int\frac{d\omega}{2\pi}e^{i\omega(t-t')}\gamma_{a}\omega\theta(\omega).\label{eq:ga}
\end{equation}

\section{Instability Rates from Leading-Order Mode Functions} \label{sec: results}

We now apply the general formalism to the bound states of \S\ref{sec: bound states}, retracing the calculation first performed in~\cite{Endlich:2016jgc}. The strategy is the following: we first evaluate the bound-state mode functions on the worldline and express the instability rates in terms of the unknown coefficients $\gamma_{a}$ of \eqref{eq:ga}; we then fix these coefficients by matching the EFT to the absorption probabilities of massless asymptotic states quoted in \S\ref{sec:asymptotics}. The outcome will agree with the full-theory rates \eqref{eq: instability rates for j=1 bound states} for $\ell=0,1$, but not for $\ell=2$---the mismatch that the remainder of the paper diagnoses and resolves. Since the bound states are non-relativistic, we set $E_\beta \simeq \mu$ in the matrix elements of \S\ref{sec: EFT rates} throughout this section.

\subsection{Bound state mode functions}

The leading order mode functions are the standard eigenbasis of the
vector Schr\"odinger equation given in terms
of the vector spherical harmonics $\vec{Y}_{\ell,jm}$ and radial wavefunction
\begin{equation}
    \vec{f}_{njm\ell}(r,\theta,\varphi)=R_{n\ell}(r)\vec{Y}_{\ell,jm}(\theta,\varphi),\label{eq:ansatz}
\end{equation}
where we have suppressed the label for leading order results. The quantities $\ell$
and $j$ are the orbital and total angular quantum numbers respectively. For spin-1 fields, the relationship between these numbers is $j=\ell,\ell\pm1$.
The dominant modes have $j=1$, so their orbital eigenvalues are restricted
to $\ell=0,1,2$. 

The derivatives of the mode functions are computed in Appendix~\ref{sec:Mode-functions}. We also use the $r\to0$ limit of the radial wavefunctions \eqref{eq:radial-mode}
\begin{equation} \label{eq:radial-mode-0}
    R_{n\ell}(r) = \frac{2^{\ell+1}(\alpha \mu)^{\ell + 3/2}}{(2\ell+1)!} \left(\frac{(n+\ell)!}{n^{2\ell+4}(n-\ell-1)!}\right)^{1/2} r^\ell \bigl(1 + \mathcal{O}(\alpha \mu r)\bigr).
\end{equation}
The radial wavefunction scales like $r^{\ell}$
as $r\rightarrow0$, so the relevant mode function must come with $\ell$ derivatives
in order to maintain a finite contribution on the worldline located
at $r=0$. 

Putting it all together, the perturbatively most significant contributions for each value of $\ell$ are
\begin{subequations}\label{eq:modes-at-0}
\begin{align}
    \vec{f}_{n,1,m,0}(0) &= \frac{1}{\sqrt{\pi}}(\alpha\mu)^{3/2} \frac{1}{n^{3/2}} \vec{V}_m,\\
    \vec{\nabla}\times\vec{f}_{n,1,m,1}(0) &= i\sqrt{\frac{2}{3\pi}}(\alpha\mu)^{5/2}\left(\frac{n^2-1}{n^5}\right)^{1/2}\vec{V}_{m} ,\\
    \vec{\nabla}\Bigl(\vec{\nabla}\cdot\vec{f}_{n,1,m,2}\Bigr)(0) & =-\frac{1}{3\sqrt{2\pi}}(\alpha\mu)^{7/2} \left(\frac{(n^2-1)(n^2-4)}{n^7}\right)^{1/2}\vec{V}_{m}, \label{eq:modes-at-0-l2}
\end{align}
\end{subequations}
where $\vec{V}_m$ is a constant eigenvector of rotations with eigenvalue $e^{im\varphi}$ defined in Appendix~\ref{sec:Vector-spherical-harmonics}.

The fact that the $\ell=2$ mode function is suppressed by two powers of $\alpha$ relative to the $\ell=0$ one has counterintuitive implications. Recall that $\vec{f}_\beta$ is the mode function obtained by keeping only the Coulomb potential in \eqref{eq: eqs for spatial components of mode functions}. The neglected part of the Hamiltonian was suppressed by $\alpha^2$, so the perturbative corrections to these leading-order mode functions built from the $\ell=0$ states can contribute at \textit{leading order} to the $\ell = 2$ mode. This idea will be explored in \S \ref{sec:diagonal-fix} and \S \ref{sec:Perturbative-corrections}.

\subsection{Instability rates}

Now we take the difference between the emission and absorption rates
to obtain the overall production rate
\begin{equation}
\Gamma\equiv\Gamma_\mathrm{em}-\Gamma_\mathrm{abs}.
\end{equation}
These rates represent superradiant instabilities when $\omega - m\Omega < 0$.
The difference between absorption rates and emission rates is that the vacuum field is paired with the final and initial BH state, respectively. This means the correlation function is conjugated relative to the field matrix element, which results in negating the argument in the Wightman function. Thus, the difference in the rates combines the Wightman correlators into the Pauli-Jordan
correlator. The end result is that the instability rate
simply depends on the spectral density functions $\Delta_{-}(\omega)$. As we discuss in Appendix~\ref{sec:Worldline-Wightman-functions}, the ground-state Wightman function coincides with the Pauli-Jordan function for positive frequencies, so we can use absorption probabilities to match unknown constants in these instability rates.

For the $j=1$ rates, we consider each field's contribution in turn. The instability rate due to the interaction with $\mathcal{O}_I$ is given by
\begin{subequations}\label{eq:eft-rates}
\begin{equation}
    \Gamma^{A}_{n,1,m,\ell} \simeq\frac{\mu}{4\pi\delta(0)}\int \frac{d \omega'}{2\pi}dtdt'\,e^{i\left(-\omega'-\mu+m\Omega\right)(t-t')}\vert\vec{f}_{n,1,m,\ell}(0)\vert^{2}\gamma\omega' =\delta_{\ell,0}\frac{\alpha^{3}\mu^{4}}{2\pi n^3}\gamma(m\Omega-\mu).\label{eq:sca}
\end{equation}
We can make sense of the integration kernel $e^{-i\left(\omega'-\mu+m\Omega\right)(t-t')}$
as originating from the fact that rotation operators ${R^I}_J(t)$ acting
on the spherical harmonics simply yield phase rotations $e^{im\Omega t}$~\cite{Endlich:2016jgc}. Because there are no derivatives acting on the field, there is support only for $\ell = 0$ in the $r \to 0$ limit. 

Similarly, we can compute the instability rates due to $\mathcal{O}^{E}_I$ and $\mathcal{O}^{B}_I$. The leading contributions are
\begin{align}
    \Gamma^{B}_{n,1,m,1} &= \frac{(n^2-1)\alpha^5\mu^4}{3\pi n^5}\gamma_{B}(m\Omega-\mu), \label{eq:bmode} \\
    \Gamma^{E}_{n,1,m,0} &= \frac{\alpha^{3}\mu^{4}}{2\pi n^3}\gamma_E(m\Omega-\mu), \label{eq:emode0}\\
    \Gamma^{E}_{n,1,m,2} &= \frac{(n^2-1)(n^2-4) \alpha^{7}\mu^{4}}{36\pi n^7} \gamma_E(m\Omega - \mu). \label{eq:emode2}
\end{align}
The rates vanish for any value of $\ell$ that exceeds the number of field derivatives in the $r\to0$ limit. Contributions such as the $\ell = 0$ one from the $B$-field are subleading in powers of $\alpha$ compared to the other $\ell=0$ rates shown above, so they are irrelevant for the tree-level EFT. The matching results \eqref{eq:matching} for the $\gamma$-coefficients render them all proportional to $(GM)^4$, so the leading-order rates follow the pattern $\Gamma_{njm\ell}\propto\alpha^{5+2j+2\ell}$ as observed numerically in~\cite{Rosa:2011my,Baryakhtar:2017ngi}. The leading $j=1$, $\ell = 0$ rate for vector fields is $\Gamma_{1,1,m,0} = \Gamma^A_{1,1,m,0} + \Gamma^E_{1,1,m,0}$, the leading $\ell = 1$ rate is $\Gamma_{2,1,m,1} = \Gamma^B_{2,1,m,1}$, and the leading $\ell = 2$ rate is $\Gamma_{3,1,m,2} = \Gamma^E_{3,1,m,2}$. 
\end{subequations}

The same formalism applied to the single $j=0$ channel reproduces the corresponding full-theory rate exactly; see Appendix~\ref{sec:j0-appendix}.

\subsection{Asymptotic states in the massless limit}

The rates (\ref{eq:sca}\textendash\ref{eq:emode2}) depend on the unknown Wilson coefficients $\gamma_a$, which are most easily determined by matching the EFT to analytic full-theory results for simple configurations; a more detailed discussion of this matching procedure can be found in~\cite{Endlich:2016jgc}. The best understood analytic results are the greybody factors of the
massless asymptotic states calculated in the full theory using black
hole perturbation theory (specifically, the Teukolsky formalism).\footnote{This matching is carried out with massless, ultrarelativistic states, whereas the bound states responsible for the instability are non-relativistic ($v^2\sim r_\mathrm{s}/r\ll1$). The two regimes are connected by the Sommerfeld enhancement of the long-range gravitational potential: for a massive scalar, Unruh's absorption cross section~\cite{Unruh:1976fm} factorizes into the standard $\ell=1$ Sommerfeld factor times the massless result, and the same factor arises identically in the effective theory, so that it cancels in the matching and the finite mass drops out. We take the same to hold for the massive vector.} The relevant results
are the absorption probabilities (\ref{eq:Pabs-scalar}\textendash\ref{eq:Pabs-vector}), already quoted at the end of \S\ref{sec:asymptotics}, which hold in the absorptive regime $\omega>m\Omega$; below we denote them by $P^\mathrm{scalar}_{\omega j m}$ and $P^\mathrm{vector}_{\omega j m\lambda}$ to make the quantum numbers of the spherical helicity states explicit.

While the field $\mu A^I$ does not survive the massless limit, as must be the case to preserve gauge invariance, all three degrees of freedom persist due to the Stueckelberg substitution
\begin{equation}
    A^{\nu}\mapsto A^{\nu}+\frac{\partial^{\nu}\phi}{\mu},
\end{equation}
as discussed in \S\ref{sec:asymptotics};
the massive vector decouples into a massless vector and scalar in the limit according to \eqref{eq:decoupling}. The gauge-invariant fields $E$ and
$B$ survive the massless limit unscathed, so their corresponding interaction terms carry over, and the interaction involving
the $A$ field corresponds to the scalar $\phi$, with $\mu \vec{A}\cdot\vec{\mathcal{O}} \to \vec{\nabla} \phi \cdot \vec{\mathcal{O}}$.

The asymptotic states admit the standard plane wave mode expansions
\begin{align}
    \phi(x) &=\int\frac{d^{3}{k}}{(2\pi)^{3}2\omega_{k}}\left( \hat{a}_{k}e^{ik\cdot x}+\hat{a}_{k}^{\dagger}e^{-ik\cdot x}\right) , \label{eq:scalar} \\
    A^{\nu}(x) &=\sum_{\lambda}\int\frac{d^{3}{k}}{(2\pi)^{3}2\omega_{k}}\left( \epsilon^\nu  ( k, \lambda)\hat{a}_{\lambda,k}e^{ik\cdot x}+\epsilon^{\nu*}(k, \lambda)\hat{a}_{\lambda,k}^{\dagger}e^{-ik\cdot x}\right) , \label{eq:vector-expansion} 
\end{align}
under the normalization condition
\begin{equation}
    \left[\hat{a}_{\lambda,k},\hat{a}_{\lambda',k'}^{\dagger}\right]=\left(2\pi\right)^{3}2\omega_{k}\delta^{3}(\vec{k}-\vec{k}')\delta_{\lambda \lambda'}.
\end{equation}
The single-particle spherical helicity states $\vert\omega,j,m\rangle$ have normalization~\cite{Endlich:2016jgc}
\begin{equation}\label{eq:helicity-norm}
    \langle \omega,j,m \vert \omega',j',m'\rangle = 2\pi\delta(\omega-\omega')\delta_{jj'}\delta_{mm'}
\end{equation}
and their inner product with the momentum eigenstates $\vert \vec{k},\lambda\rangle = \hat a^\dagger_{k,\lambda}\vert 0 \rangle$ is
\begin{equation}\label{eq:asymptotic-basis-change}
    \langle \vec{k},\lambda' \vert \omega,j,m,\lambda\rangle = \sqrt{\frac{2}{k}} (2\pi)^2\delta(\omega_k-\omega)\,_{-\lambda}Y^{jm}(\hat{k}) \delta_{\lambda\lambda'}.
\end{equation}
For the scalar field with no polarizations, we can set $\lambda = 0$ above to replace the spin-weighted spherical harmonic with the regular spherical harmonic. 
Together, these yield the matrix elements of spherical helicity states derived in
Appendix~\ref{sec:Asymptotic-vector-fields},
\begin{subequations}
\begin{align}
    \langle 0 \vert \partial^I\phi(t,0) \vert \omega, j, m \rangle &= i e^{-i\omega t} \sqrt{\frac{\omega^3}{6\pi}} V^I_m \delta_{j,1}, \\
    \langle 0 \vert {B}^I(t,0) \vert \omega, j, m, \lambda \rangle &= e^{-i\omega t} \sqrt{\frac{\omega^3}{6\pi}} V^I_m \delta_{j,1} \lambda, \\
    \langle 0 \vert {E}^I(t,0) \vert \omega, j, m, \lambda \rangle &= ie^{-i\omega t} \sqrt{\frac{\omega^3}{6\pi}} V^I_m \delta_{j,1}\lvert\lambda\rvert,
\end{align}
\end{subequations}
where $V^I_m$ is an eigenvector of $\operatorname{SO}(3)$ rotations about the $z$-axis, satisfying $V_{m\,I}\,{R^I}_J(\varphi) = e^{im\varphi}V_{m\,J}$, as defined in \eqref{eq:rotation-eigenvectors}. The derivation in Appendix~\ref{sec:Asymptotic-vector-fields} retains a finite mass throughout; the Lorentz factor $\omega/\mu$ that accompanies the longitudinal polarization there corrects the corresponding expressions of Ref.~\cite{Endlich:2016jgc}.
For the $\langle \mathcal{O}_I \mathcal{O}_J \rangle$ correlation function, we use $\mu \vec{A} \cdot \vec{\mathcal{O}} \to \vec{\nabla} \phi \cdot \vec{\mathcal{O}}$ to justify using the same correlation function $W(\omega)$ from \eqref{eq:ga} for the massless scalar as for the field $\mu A^I$. The calculation of the absorption probability is
\begin{align}
    P_{\omega,1,m}^\mathrm{scalar} &= \lim_{\mu\rightarrow 0}\sum_{X}\frac{\left\vert\int dt\,\langle 0;X\vert\tilde{\partial}^{I}\phi\mathcal{O}_{I}\vert \omega,1,m;X_0\rangle\right\vert^{2}}{\langle\omega,1,m\vert\omega,1,m\rangle}\nonumber \\
    &= \frac{\omega^{3}}{6\pi \bigl(2\pi\delta(0)\bigr)}\int dt\,dt'\,\frac{d\omega}{2\pi}'W(\omega')e^{i(\omega'-\omega+m\Omega)(t-t')}\nonumber \\
    &\simeq \frac{\omega^{3}}{6\pi}\gamma(\omega-m\Omega)\theta(\omega-m\Omega),
\end{align}
and matching with the scalar $\ell=1$ probability \eqref{eq:Pabs-scalar} results in
\begin{equation}
\gamma=\frac{32}{3}\pi G^{4}M^{4}.
\end{equation}
Similarly, the remaining two\footnote{Recall interference terms like $\langle \mathcal{O}_I^{\prime\,E}\mathcal
O_J^B\rangle$ vanish by parity symmetry or time-reversal symmetry, so there are only two more correlation functions to match.} coefficients, $\gamma_B$ and $\gamma_E$, are matched by the massless vector calculation,
\begin{align}
    P^\mathrm{vector}_{\omega,1,m,\lambda} & =\lim_{\mu\rightarrow0}\sum_{X}\frac{\left|\int dt\,\langle0;X\vert\bigl(\tilde{E}^{I}\mathcal{O}_{I}^{E} + \tilde{B}^I\mathcal{O}_I^B \bigr)\vert\omega,1,m,\lambda;X_{0}\rangle\right|^2}{\langle\omega,1,m,\lambda\vert\omega,1,m,\lambda\rangle}\nonumber \\
    & =\frac{\omega^{3}}{6\pi\bigl(2\pi\delta(0)\bigr)}\int dt\,dt'\,\frac{d\omega'}{2\pi}\bigl(W^{E}(\omega')+W^{B}(\omega')\bigr)e^{i(\omega'-\omega+m\Omega)(t-t')}\nonumber \\
    & \simeq\frac{\omega^3}{6\pi}(\gamma_{E}+\gamma_{B})(\omega-m\Omega)\theta(\omega-m\Omega).
\end{align}
In the Schwarzschild vacuum, Maxwell's equations are invariant under $\vec{E} \to -\vec{B}$, $\vec{B} \to \vec{E}$. For the action to respect this symmetry, the correlation functions
$\langle \mathcal{O}_I^{\prime E} \mathcal{O}_J^{E} \rangle$ and $\langle \mathcal{O}_I^{\prime B} \mathcal{O}_J^{B} \rangle$ must be equal~\cite{Goldberger:2005cd}, implying after matching with \eqref{eq:Pabs-vector} that
\begin{equation}\label{eq:matching}
    \gamma_{E}=\gamma_{B} = 2\gamma = \frac{64}{3}\pi G^{4}M^{4}.
\end{equation}
Thus, combining these results with (\ref{eq:sca}\textendash\ref{eq:emode2}), we obtain the following superradiant instability rates:
\begin{subequations}\label{eq:EFT-rates}
\begin{align}
    \Gamma_{n,1,m,0} &= \frac{16}{n^3}\alpha^7(m\Omega-\mu), \label{eq:EFT-rates0}\\
    \Gamma_{n,1,m,1} &= \frac{64(n^2-1)}{9n^5} \alpha^9(m\Omega-\mu), \label{eq:EFT-rates1}\\
    \Gamma_{n,1,m,2} &= \frac{16(n^2-1)(n^2-4)}{27n^7} \alpha^{11}(m\Omega-\mu).\label{eq:EFT-rates2}
\end{align}
\end{subequations}
As anticipated, the naive rate \eqref{eq:EFT-rates2} comes out three times the full-theory value \eqref{eq: instability rates for j=1 bound states l=2}---a mismatch we will address in the following two sections.\footnote{\label{fn:endlich-factor2}The value quoted in eq.~(95) of~\cite{Endlich:2016jgc} carries this same factor of three and is larger still by a factor of two. The latter is not a matter of convention---that reference reports the same occupation-number growth rate used here---but traces to the matched coefficients quoted in eq.~(92) there, $\gamma_E=\gamma_B=4\gamma$, twice the values in \eqref{eq:matching}: with those coefficients, the rates (\ref{eq:sca}\textendash\ref{eq:emode2}) reproduce eqs.~(93)\textendash(95) of that reference exactly. The same normalization affects the $\ell=0,1$ rates quoted there, which exceed \eqref{eq: instability rates for j=1 bound states} by factors of $5/3$ and $2$, respectively; the values obtained here supersede them.}

\section{Connection to the FKKS Ansatz} \label{sec:diagonal-fix}

The apparent discrepancy between \eqref{eq:EFT-rates2} and \eqref{eq: instability rates for j=1 bound states l=2} originates entirely in the choice of basis functions, as we now show by making contact with the separated solution of the full theory. When discussing \eqref{eq:radial-mode-0}, we pointed out that in the point-particle formalism, the radial profile of modes influences power-counting. If mode functions approach the origin as a positive power of $r$, they need to be differentiated before they can contribute to interactions with the point particle on its worldline. Since each spatial derivative is penalized in our power-counting, a ``leading-order'' mode function that is $\mathcal{O}(r^n)$ at the origin has the same significance as a parametrically smaller contribution suppressed by $\alpha^n$ that is $\mathcal{O}(1)$. The rate in question has the largest value of $\ell$ among the $j=1$ rates, and the modes are $\mathcal{O}(r^\ell)$ according to \eqref{eq:radial-mode-0}, which implies that some effect suppressed by $\alpha^2$ could misalign the state we used to derive \eqref{eq:EFT-rates2} with the state used to derive \eqref{eq: instability rates for j=1 bound states l=2}.

To explore this, we will follow the original derivation~\cite{Baumann:2019eav} of \eqref{eq:Baumann-result}.
In particular, we will focus on the functions defined by the 
angular eigenvalues $\Lambda$\footnote{The source material uses $\lambda$ for these eigenvalues, but we use $\Lambda$ instead to avoid a notational collision with the helicity index introduced in \S\ref{sec:asymptotics}.}, which arise from the separable ansatz used for the vector field. 
The authors identify $\Lambda$ with the more standard quantum numbers $j$ and $\ell$ in the far zone, where our EFT resides and where the two theories are directly comparable. Whereas $\ell$ or $j$ might be associated with a spherical harmonic, $\Lambda$ is associated with a spheroidal-type harmonic.
The identification with $\ell$ holds at leading order in $\alpha$. However, $\ell$ is only a good quantum number in the large-$r$ limit, so it is no surprise that at higher orders in $\alpha$, eigenstates of one quantity become distinct from those of the other.
We must therefore examine those corrections to determine whether any of them are significant at leading order in the point-particle EFT.

The solution to the equations of motion for the ``electric'' modes with $\ell = j \pm 1$ involves, in the differential equation
for the angular profile and in the limit of a Schwarzschild black hole, the identification
\begin{equation}
    \Lambda\left(\Lambda - \frac{\omega}{\mu}\right) \equiv j(j+1), \label{eq: definition of j}
\end{equation}
where the energy, up to next-to-leading order, is hydrogenic, 
$\omega = \mu - \alpha^2 \mu / (2n^2)$.
This results in a pair of solutions for the eigenvalue $\Lambda$ that can be 
expanded up to next-to-leading order as 
$\Lambda = \Lambda_0 + \alpha^2 \Lambda_2$, where
\begin{gather}
    \Lambda_0^- = j+1, \qquad \Lambda_0^+ = -j, \label{eq: lambda leading order}\\
    \Lambda_2^\pm = -\frac{\left\vert{\Lambda_0^\pm}\right\vert}{2n^2(2j+1)}.
\end{gather}

The identification in (\ref{eq: definition of j}) introduces the spherical harmonic $Y^{jm}$ into the FKKS ansatz~\cite{Frolov:2018ezx} for the classical vector field, $f^\nu = B^{\nu\sigma} \nabla_\sigma \big( \mathcal{R}(r) Y^{jm}(\theta,\varphi) e^{-i\omega t} \big)$\footnote{The vector mode is called $f^\nu$ to better align with our notation, though it differs from our modes \eqref{eq: expansion in pure orbital vector harmonics} in that this includes the time-dependent phase. In the source material, the mode function is $A^\nu$ instead.}, in which $\Lambda$ appears as a parameter within the polarization tensor $B^{\nu\sigma}$ given by~\cite{Baumann:2019eav} 
{\everymath={\displaystyle}
\begin{equation}
    B^{\nu\sigma} = \frac{1}{r^2 + \Lambda^2} \left(
    \begin{array}{cccc}
        - \frac{\Lambda^2}{1-r_\mathrm{s}/r} & i \Lambda r & 0 & 0 \\
        -i\Lambda r & \Lambda^2\biggl(1 - \frac{r_\mathrm{s}}{r}\biggr) & 0 & 0 \\
        0 & 0 & 0 & 0 \\
        0 & 0 & 0 & 0
    \end{array}
    \right)
    + \frac{1}{r^2} \left(
    \begin{array}{cccc}
        0 & 0 & 0 & 0 \\
        0 & 0 & 0 & 0 \\
        0 & 0 & 1 & 0 \\
        0 & 0 & 0 & \csc^2 \theta
    \end{array}\right),
\end{equation}
}
in units with $\mu = 1$ and with vanishing angular momentum.

Expanding with the characteristic scale $r\sim(\alpha\mu)^{-1}$ up to next-to-leading order in $\alpha$, the far-zone mode functions are
\begin{align}
    f^r &=  -\Lambda_0 \frac{\mathcal{R}}{r} Y^{jm}e^{-i t} + \biggl( - \Lambda_2\alpha^2 \frac{\mathcal{R}}{r} + \Lambda_0^3\frac{\mathcal{R}}{r^3} + \frac{\Lambda_0 \alpha^2}{2n^2} \frac{\mathcal{R}}{r} - i \frac{\Lambda_0 \alpha^2 t}{2n^2} \frac{\mathcal{R}}{r} + \Lambda_0^2\frac{\mathcal{R}'}{r^2}\biggr) Y^{jm}e^{-i t}, \nonumber \\
    f^\theta &=\frac{1}{r}\biggl( 1+i \frac{\alpha^2 t}{2n^2} \biggr) \mathcal{R} \frac{1}{r} \partial_\theta Y^{jm} e^{-it}, \quad \quad f^\varphi = \frac{1}{r\sin\theta}\biggl( 1+i \frac{\alpha^2 t}{2n^2} \biggr) \mathcal{R} \frac{1}{r\sin\theta} \partial_\varphi Y^{jm} e^{-it}.
\end{align}
After changing to the normalized spherical basis $(\hat{r},\hat{\theta},\hat{\varphi}) = (\vec{e}_r, r^{-1}\vec{e}_\theta, r^{-1}\csc\theta \,\vec{e}_\varphi )$, the factors of $r^{-1}$ and $r^{-1}\csc\theta$ appearing in $f^\theta$ and $f^\varphi$ are removed by the rescaling of the basis vectors and by being absorbed into the spherical gradient operator acting on $Y^{jm}$.
Then, $r\nabla Y^{jm}$ and $Y^{jm} \hat{r}$
can be expressed in the basis of ``pure-orbit'' spherical harmonics $\vec{Y}_{j\pm1,jm}$ (see Appendix~\ref{sec:Vector-spherical-harmonics}). We will normalize the mode function by dividing by the factor $\sqrt{(2j+1)|\Lambda_0|}$, obtaining
\begin{align}
    \vec{f} &= \biggl( 1+i \frac{\alpha^2 t}{2n^2} \biggr) \frac{\mathcal{R}}{r} \Biggl(\sqrt{\frac{j}{|\Lambda_0|}}\frac{j+1-\Lambda_0}{2j+1} \vec{Y}_{j-1,jm} +\sqrt{\frac{j+1}{|\Lambda_0|}}  \frac{j+\Lambda_0}{2j+1}\vec{Y}_{j+1,jm} \Biggr)e^{-i t} \nonumber\\
    &+\frac{1}{2j+1}\biggl( \frac{\Lambda_0^2}{r^2}\partial_r + \frac{\Lambda_0^3}{r^3} + \frac{\Lambda_0 |\Lambda_0|\alpha^2}{(2j+1)n^2r}\biggr) \mathcal{R} \Biggl( \sqrt{\frac{j}{|\Lambda_0|}} \vec{Y}_{j-1,jm} - \sqrt{\frac{j+1}{|\Lambda_0|}}\vec{Y}_{j+1,jm} \Biggr) e^{-it}.
\end{align}
Above, we used the identity $\lvert{\Lambda_0}\rvert+(2j+1)\Lambda_0 = 2\Lambda_0\lvert{\Lambda_0}\rvert$ to combine the two real terms in $f^r$ with the same radial profile $\mathcal{R}/r$.

The leading-order contribution comes from the first line, where the choice of $\Lambda_0$ removes one or the other vector spherical harmonic: 
\begin{equation}
    \vec{f}_{\Lambda^\pm}^{(0)}(x) = \frac{\mathcal{R}(r)}{r} \vec{Y}_{j\mp1,jm} (\theta,\varphi) e^{-i\mu t}.
\end{equation}
This must solve the far-zone equations of motion, so it must equal \eqref{eq: expansion in pure orbital vector harmonics}; therefore, $\mathcal{R}(r) = r R_{n\ell}(r)$ with $\ell = j\mp1$. Thus, the eigenvalue $\Lambda = \Lambda^+ \simeq -j$ is associated with $\ell = j-1$, and $\Lambda = \Lambda^- \simeq j+1$ is associated with $\ell = j+1$. 

With the radial functions determined and factors of $\mu$ restored, the solutions are
\begin{align}
    \vec{f}_{\Lambda^+} &= \biggl(1 - \frac{j^2\alpha^2}{(2j+1)^2 n^2} + i \frac{\alpha^2 \mu t}{2n^2}\biggr) R_{n,j-1} \vec{Y}_{j-1,jm} e^{-i\mu t} \nonumber\\
    &\qquad\qquad + \frac{\sqrt{j^3(j+1)}\alpha^2}{(2j+1)^2n^2} R_{n,j-1}\vec{Y}_{j+1,jm}e^{-i\mu t} \nonumber\\
    &\qquad\qquad+ \sqrt{\frac{j^3}{2j+1}} \biggl( \frac{R'_{n,j-1}}{\mu^2r} - (j-1) \frac{R_{n,j-1}}{\mu^2r^2}\biggr)Y^{jm}  e^{-i\mu t} \hat{r}, \label{eq:lambda-plus}
    \\
    \vec{f}_{\Lambda^-} &= \biggl(1 - \frac{(j+1)^2\alpha^2}{(2j+1)^2 n^2} + i \frac{\alpha^2 \mu t}{2n^2}\biggr) R_{n,j+1} \vec{Y}_{j+1,jm} e^{-i\mu t} \nonumber\\
    &\qquad\qquad + \frac{\sqrt{j(j+1)^3}\alpha^2}{(2j+1)^2n^2} R_{n,j+1}\vec{Y}_{j-1,jm}e^{-i\mu t} \nonumber\\
    &\qquad\qquad+ \sqrt{\frac{(j+1)^3}{2j+1}} \biggl( \frac{R'_{n,j+1}}{\mu^2r} + (j+2) \frac{R_{n,j+1}}{\mu^2r^2}\biggr)Y^{jm} e^{-i\mu t} \hat{r}.\label{eq:lambda-minus}
\end{align}
In the last line of each mode, we recombined the pure-orbit vector spherical harmonics into $Y^{jm} \hat{r}$. These mode functions $f^i_{\Lambda^\pm}$ are associated with single-particle states of the vector field. We will refer to these states as $\vert n,j,m;\pm\rangle$, where $\langle x \vert n,j,m;\pm\rangle = \vec{f}_{\Lambda^\pm}(x)$. These are similar to the single-particle states $\vert n,j,m,\ell=j\mp 1\rangle$ but, as we now show, not identical.

Now, we investigate whether any of the corrections are significant in transition amplitudes. For a correction suppressed by $\alpha^2$ to be significant, its radial function must require two fewer spatial derivatives than the leading-order part to be nonvanishing on the worldline.
First, all terms carrying an explicit factor of $\alpha^2$ have the same radial profile as the leading-order term, so they can all be neglected. That leaves the last term in each mode to consider.

With $j=1$, the mode function $R_{n,j-1}$ is a constant as $r\to0$, so the field operators in $S_\mathrm{int}$ with no derivatives can be used with it to construct nonvanishing amplitudes. The derivatives and factors of $1/r$ in the third term of \eqref{eq:lambda-plus} scale it by two factors of $\alpha$ relative to $R_{n,j-1}$. The term is singular in the $r\to0$ limit, but it is also subleading, so we expect additional loop-level terms in the EFT to be able to regularize it. Thus, to leading order, $\vert n,1,m;+\rangle = \vert n,1,m,\ell=0\rangle$.

On the other hand, $R_{n,2}$ in \eqref{eq:lambda-minus} is quadratic as $r\to0$, so it only contributes to transition amplitudes involving twice-spatially-differentiated fields in $S_\mathrm{int}$. The radial profile of the term in the third line of that equation remains finite as $r\to0$, as though the requisite derivatives had already been taken, so amplitudes constructed with the derivative-free field operators in $S_\mathrm{int}$ survive on the worldline $r=0$, and those amplitudes have the same $\alpha$-scaling as the ``leading-order'' part. This predicts a significant difference between $\vert n,1,m;-\rangle$ and $\vert n,1,m,\ell=2\rangle$.

We will calculate the instability rate of the $\vert n,1,m;-\rangle$ state, $\Gamma^-_{njm}$. Since the full theory uses this state, we should be comparing $\Gamma^-_{njm}$ to \eqref{eq: instability rates for j=1 bound states l=2}. The first part of \eqref{eq:lambda-minus} requires two derivatives to contribute on the worldline $r=0$, and the last part needs none. We use \eqref{eq:avg-radial-vsh} to evaluate the angular function $Y^{jm}\hat{r}$ on the worldline and (\ref{eq:R0},\ref{eq:R2}) for the radial components. The result is
\begin{align}
    \nabla(\nabla\cdot \vec{f}_{\Lambda^-})(0) &= - \frac{1}{6}\sqrt{\frac{2(n^2-1)(n^2-4)}{\pi n^7} \alpha^7 \mu^7} \vec{V}_m e^{-i\mu t} + \cdots\label{eq:lambda-old} \\
    \vec{f}_{\Lambda^-}(0) &= \frac{1}{9} \sqrt{\frac{2(n^2-1)(n^2-4)}{\pi n^7}\alpha^7\mu^3} \vec{V}_m e^{-i\mu t} + \cdots \label{eq:lambda-new}
\end{align}
Above, \eqref{eq:lambda-old} comes from the ``leading-order'' part of \eqref{eq:lambda-minus} and thus is identical to \eqref{eq:modes-at-0-l2} with a time-dependent phase. \eqref{eq:lambda-new} comes from the other surviving term.

The field operator $\mu\tilde{A}^i$ and the term \eqref{eq:lambda-new} produce the transition amplitude
\begin{equation}
    \langle 0 \vert \mu\tilde{A}^i(t) \vert n,1,m;-\rangle = \frac{1}{9} \sqrt{\frac{(n^2-1)(n^2-4)}{\pi n^7}} \alpha^{7/2}\mu^2 V_m^i e^{-i(\mu-m\Omega)t}, 
\end{equation}
while both \eqref{eq:lambda-old} and \eqref{eq:lambda-new} combine with $\tilde{E}^i$ in
\begin{equation}
    \langle 0 \vert \tilde{E}^i(t) \vert n,1,m;-\rangle = -i \frac{1}{18} \sqrt{\frac{(n^2-1)(n^2-4)}{\pi n^7}} \alpha^{7/2}\mu^2 V_m^i e^{-i(\mu-m\Omega)t}.
\end{equation}

The vanishing of $\langle \mathcal{O}_i \mathcal{O}_j^E\rangle$ means that there is no interference between these two amplitudes in the calculation of the instability rate $\Gamma^{-}_{n,1,m}$. Using the results from \eqref{eq:matching}, we obtain
\begin{equation}
    \Gamma^{-}_{n,1,m} = \biggl(\frac{\gamma}{81} + \frac{\gamma_E}{324}\biggr){\frac{(n^2-1)(n^2-4)}{\pi n^7}} \alpha^{7} (m\Omega-\mu) = \frac{16(n^2-1)(n^2-4)}{81 n^7} \alpha^{11} (m\Omega-\mu), \label{eq:EFT-rates-lambda}
\end{equation}
which is exactly the same as the result \eqref{eq: instability rates for j=1 bound states l=2}. 

The state $\vert n,1,m;-\rangle$ is more appropriate when considering the effects of spin-orbit coupling and frame-dragging. The full theory assigns this state the label $\ell = 2$ by looking at its behavior in the far zone. An $\ell$ eigenstate, however, cannot be carried all the way down to the near zone. What is perhaps surprising is that the distinction between $\vert n,1,m;-\rangle$ and $\vert n,1,m,2\rangle$ matters even though the EFT never leaves the far zone: the way the radial profile enters the power counting forces us to adopt the same $\Lambda$ eigenstate as the full theory. The lesson is that, when working in a point-particle EFT, some care is needed to select states that remain well-behaved in the UV, at least as far as can be diagnosed in the limit $r\to0$.

The rate $\Gamma^-_{n,1,m}$ in \eqref{eq:EFT-rates-lambda} therefore reproduces the full-theory result \eqref{eq: instability rates for j=1 bound states l=2} exactly, and the origin of the factor of three in \eqref{eq:EFT-rates2} is now clear: it came from evaluating the transition on the naive $|n,1,m,2\rangle$ state rather than on the $\Lambda^-$ eigenstate that the full theory actually uses. What we have done in this section, however, is to identify that eigenstate \emph{by hand}---importing the change of basis from the separated full-theory solution rather than deriving it. In \S\ref{sec:Perturbative-corrections} we show that this input is not needed. Indeed, the care called for above is not in tension with a first-principles treatment: being careful about the choice of states means, operationally, being careful about perturbation theory. In a degenerate subspace, the correct leading-order states are not selected by symmetry labels alone but are determined by the perturbations themselves. Working entirely within the effective theory, degenerate perturbation theory together with near-zone matching thus generates the same admixture of the degenerate partner---and hence the same rate---as an \emph{output}, with no change of basis put in by hand. The agreement between the two derivations is then a nontrivial check rather than a restatement.

\section{Degenerate Perturbation Theory and Near-Zone Matching}\label{sec:Perturbative-corrections}

In the previous section, we recovered the full-theory rate \eqref{eq: instability rates for j=1 bound states l=2} by adopting the $\Lambda^-$ eigenstate---a change of basis imported by hand from the separated solution of the full theory. In this section, we show that the effective theory arrives at the very same result on its own. The strategy is to take seriously the perturbation $V$ in \eqref{eq: eqs for spatial components of mode functions}, which our leading-order solution neglected, and to compute the corrections it induces on the mode function $\vec f_{3,1,m,2}$ using ordinary time-independent perturbation theory. No step in what follows makes use of the FKKS ansatz or of the identification of the $\Lambda^-$ eigenstate: the small $\ell=0$ admixture that \S\ref{sec:diagonal-fix} attributed to a change of basis will instead emerge as an \emph{output} of the calculation. The agreement between the two derivations is therefore a nontrivial consistency check on both.

Let us first recall why these corrections matter at leading order. The transition rates calculated in (\ref{eq:sca}\textendash\ref{eq:emode2})
carry higher powers of $\alpha$ for larger values of $\ell$ because $R_{n\ell}(r)$ is dominated by $(\alpha\mu)^{3/2}(\alpha\mu r)^\ell$ as $r\to0$: without perturbations, the $\mu A^{I}\mathcal{O}_{I}$
and $\partial_{0}A^{I}\mathcal{O}_{I}^{E}$ terms in the action contribute only
to the $\ell=0$ transition rate, since all other mode functions vanish in the $r\rightarrow0$ limit. When perturbations are included, however,
every $\ell>0$ state such that $\langle\ell=0\vert V\vert\ell>0\rangle$ is nonvanishing
picks up a correction proportional to an $\ell=0$
mode function, making the $\mu A^{I}\mathcal{O}_{I}$ and
$\partial_{0}A^{I}\mathcal{O}_{I}^{E}$ contributions non-vanishing. Because we have cast the equations of motion \eqref{eq: eqs for spatial components of mode functions} in the form of a Schr\"odinger equation for the hydrogen atom, we work with the non-relativistic energy eigenvalues $\tilde{E}_n = (E_n^2-\mu^2)/(2\mu) \approx - \alpha^2\mu/(2n^2)$. For illustration purposes, consider the contribution of an intermediate \emph{bound} state $\vert n',1,m,0\rangle$ to the correction to the state $(n,j,m,\ell) = (3,1,m,2)$:
\begin{equation}\label{eq:typical-correction}
    \vec{f}^{(1)}_{3,1,m,2} \supset \frac{\langle n',1,m,0\vert V^{(1)}\vert 3,1,m,2\rangle}{\tilde{E}_{3}-\tilde{E}_{n'}}\vec{f}^{(0)}_{n',1,m,0}\sim\frac{\alpha^4\mu}{\alpha^2\mu}\vec{f}^{(0)}_{n',1,m,0} \ ;
\end{equation}
as we will see in \S\ref{sec:pert-continuum}, continuum intermediate states contribute at the same order. The scaling of the numerator follows from the fact that the hydrogenic states are supported at the Bohr radius, where every gradient or inverse power of $r$ counts as $\alpha\mu$: the two-derivative structure of $V^{(1)}$ in \eqref{eq:pert} then scales as $\alpha^{2}\mu$, like the unperturbed kinetic and potential terms, while the accompanying factor of $r_\mathrm{s}/r\sim r_\mathrm{s}\,\alpha\mu \sim \alpha^{2}$ supplies the additional suppression. The energy denominator is instead of order $\alpha^{2}\mu$, since it involves states with different values of $n$. Since $\alpha^2 \vec f^{(0)}_{n',1,m,0}$ has the same $\alpha$-scaling on the worldline as $\vec f^{(0)}_{3,1,m,2}$, these terms contribute to $\Gamma_{3,1,m,2}$ at the same order as the unperturbed mode function, and they must be treated on an equal footing. (Because the perturbative potential is parity even, the $\ell=1$ mode function receives no such corrections from the modes with $\ell = 0,2$; see \S\ref{sec:pert-DPT} for the selection rules.)

Carrying this program out consistently requires more care than one might expect, and the purpose of this section is to spell out the three points where the naive calculation falls short. First, the states $\vert n,1,m,0\rangle$ and $\vert n,1,m,2\rangle$ are degenerate at leading order, so we must use \emph{degenerate} perturbation theory; even though $V^{(1)}$ turns out to be diagonal in the degenerate subspace, the degenerate partner still enters the corrected state at second order, promoted to leading order by the small energy denominator (\S\ref{sec:pert-DPT}). Second, the sum over intermediate states is a resolution of the identity, and completeness of the hydrogenic spectrum requires the positive-energy continuum alongside the bound states (\S\S\ref{sec:pert-bound}--\ref{sec:pert-continuum}). Third, the far-zone potential $V$ does not by itself determine the corrected state: near-zone physics contributes to the degenerate pair of states at the same order at which their degeneracy is lifted. In the effective theory, this information is encoded in a worldline contact operator whose coefficient---the static monopole susceptibility of the horizon in the parity-even $j=1$ Proca sector---must be fixed by matching. This is a two-step procedure: we first compute the static response of the horizon in the full theory (\S\ref{sec:pert-response}), and then match it onto a contact term on the worldline (\S\ref{sec:pert-matching}), whose insertions enter the calculation in two distinct places, correcting both the energy splitting between the degenerate states and their mixing.

It will be convenient to summarize the outcome before diving in. The entire correction relevant on the worldline is captured by a single dimensionless normalization factor $\mathcal{N}$, defined in \eqref{eq:leading-correction-at-origin} below, to which the intermediate channels contribute additively:
\begin{equation}\label{eq:S-decomposition}
    \mathcal{N} \,=\, \underbrace{\mathcal{N}_\mathrm{b}+\mathcal{N}_\mathrm{c}}_{\textstyle -\tfrac{1}{432}} \;+ \underbrace{\mathcal{N}_\mathrm{deg}}_{\textstyle -\tfrac{1}{3888}} =\, -\frac{5}{1944} \qquad (n=3).
\end{equation}
Here, $\mathcal{N}_\mathrm{b}$ and $\mathcal{N}_\mathrm{c}$ are the contributions of the bound and continuum intermediate states, computed in \eqref{eq:S-bound} and \eqref{eq:S-perp} respectively, while $\mathcal{N}_\mathrm{deg}$ is that of the degenerate partner, obtained in \eqref{eq:S-deg} by combining far-zone perturbation theory (\S\ref{sec:pert-second-order}) with near-zone input: the static response of the horizon, computed in \S\ref{sec:pert-response}, and the worldline contact term that reproduces it in the EFT, matched in \S\ref{sec:pert-matching}. Assembled in \S\ref{sec:pert-assembly}, these ingredients reproduce the full-theory rate \eqref{eq: instability rates for j=1 bound states l=2} \emph{exactly}---not only at $n=3$ but identically in $n$---and, as a byproduct, the fine structure of the Proca spectrum in exact agreement with the published results of~\cite{Baumann:2019eav}.

\subsection{Degenerate perturbation theory}\label{sec:pert-DPT}

Since our perturbative matrix (\ref{eq:pert}) is given in the
standard $\left(\vec{e}_{r},\vec{e}_{\theta},\vec{e}_{\varphi}\right)=\left(\partial_{r},\partial_{\theta},\partial_{\varphi}\right)$
basis, it is convenient to decompose the vector spherical harmonics used in \eqref{eq:ansatz} along the same basis. This decomposition is given in \eqref{eq:vsh-coordinate-basis} of Appendix~\ref{sec:Vector-spherical-harmonics}, in terms of the scalar harmonics $Y^{jm}$ and of the combinations $Y_{\pm}^{jm}$ of spin-weighted spherical harmonics~\cite{10.1063/1.1705135} defined in \eqref{eq:Ypm-def}.

Expressing the lowest-order modes as $\vec{f}^{(0)}_{njm\ell}(\vec{x}) = R_{n\ell}(r)\vec{Y}_{\ell,jm}(\theta,\varphi)$, where the overall vector $\vec{Y}$ is independent of $r$ even if $Y^\theta$ and $Y^\varphi$ are not, we can then rewrite \eqref{eq:pert} as
\begin{equation}\label{eq:pert-reduced}
    {V^{(1)I}}_Jf^J \!= \frac{r_\mathrm{s}}{2\mu r} \left( 2 R'' + \frac{3}{r} R' - \frac{\ell^2 + \ell + 1}{r^2} R \right) Y^I - \frac{3 r_\mathrm{s}}{2\mu r^3} R \left(Y^r +r\partial_\theta Y^\theta + r \cot\theta\, Y^\theta + r \partial_\varphi Y^\varphi \right)\delta_r^I,
\end{equation}
and by expressing the angular derivatives in terms of the Newman-Penrose covariant derivatives, one may show (see Appendix~\ref{sec:Degeneracies}) that
\begin{equation}\label{eq:angular-divergence}
    Y^r_{\ell,jm} + r\partial_\theta Y^\theta_{\ell,jm} + r \cot\theta\, Y^\theta_{\ell,jm} + r \partial_\varphi Y^\varphi_{\ell,jm} =
    \begin{cases}
        -(j+1)\sqrt{\frac{j+1}{2j+1}} Y^{jm}, & \ell = j+1 \\
        0, & \ell = j \\
        -j \sqrt{\frac{j}{2j+1}} Y^{jm}, & \ell = j-1
    \end{cases}
\end{equation}
so every term is proportional to either $\vec{Y}_{\ell,jm}$ or $Y^{jm}$. This greatly simplifies the calculation of matrix elements. Using the orthogonality of those functions on the sphere, the matrix element
\begin{equation}
    \langle n',j',m' ,\ell'\vert V^{(1)}\vert n,j,m,\ell\rangle = \int d\Omega'dr'\,r'^{2}f_{nj'm'\ell'}^{J*}(r',\Omega')V_{JK}^{(1)}(r',\Omega')f_{njm\ell}^{K}(r',\Omega'), \label{eq:matrix-element}
\end{equation}
reduces to
\begin{gather}
    \frac{\alpha}{\mu^2}\delta_{jj'}\delta_{mm'} \Biggl(\delta_{\ell\ell'}\int_0^\infty dr\, R_{n'\ell} \left( 2rR_{n\ell}'' + 3 R'_{n\ell} - (\ell^2 + \ell + 1) \frac{R_{n\ell}}{r} \right) + 3K_{j\ell\ell'} \int_0^\infty dr\, \frac{R_{n'\ell'} R_{n\ell}}{r}\Biggr),\label{eq:matrix-element-reduced}\\
    K_{j\ell\ell'} = \frac{\sqrt{(j+\ell'+1)(j+\ell+1)^3}}{4(2j+1)}\bigl(\delta_{j-1,\ell'}-\delta_{j+1,\ell'}\bigr)\bigl(\delta_{j-1,\ell}+\delta_{j+1,\ell}\bigr).\nonumber
\end{gather}
Because the only scale appearing in the radial integrals is the Bohr radius $(\alpha\mu)^{-1}$, the matrix element \eqref{eq:matrix-element} indeed scales as $\alpha^4\mu$ by dimensional analysis, as anticipated in \eqref{eq:typical-correction}.

The structure of \eqref{eq:pert-reduced} implies a simple selection rule in $\ell$. The first term is proportional to $\vec{Y}_{\ell,jm}$ itself, and is therefore diagonal in $\ell$. The second term sources only the radial component and, by \eqref{eq:angular-divergence}, is nonvanishing only when acting on the harmonics with $\ell=j\pm1$---precisely the ones with a nonzero radial component, $Y^{r}_{j\pm1,jm}\propto Y^{jm}$. As a result, the purely tangential harmonic $\vec{Y}_{j,jm}$ does not mix with $\vec{Y}_{j\pm1,jm}$, as also required by parity: $\vec{Y}_{j,jm}$ has parity $(-1)^{j+1}$, the harmonics $\vec{Y}_{j\pm1,jm}$ have parity $(-1)^{j}$, and the potential is parity even. The second term does, however, connect $\vec{Y}_{j-1,jm}$ and $\vec{Y}_{j+1,jm}$ to each other---this is the mixing encoded in the coefficient $K_{j\ell\ell'}$ of \eqref{eq:matrix-element-reduced}. Within a sector of fixed $j$ and parity $(-1)^{j}$, the perturbation can therefore convert an $\ell=j+1$ state into an $\ell=j-1$ state, and vice versa.

Since the leading-order eigenvalues depend only on the principal quantum
number $n$, states with different values of $\ell$ are degenerate, and we must employ degenerate perturbation theory~\cite{Sakurai1993Modern}. Given the selection rules derived above, the degenerate subspace relevant to our problem is $D=\{\vert n,1,m,0\rangle, \vert n,1,m,2\rangle\}$ at fixed $n$. The first step of degenerate perturbation theory is to diagonalize $V^{(1)}$ within $D$. The last integral in \eqref{eq:matrix-element-reduced} turns out to vanish when $n'=n$ and $\ell' \ne \ell$ (see Appendix~\ref{sec:Degeneracies}), so that $\langle n,j',m' ,\ell'\vert V^{(1)}\vert n,j,m,\ell\rangle \propto \delta_{jj'}\delta_{mm'}\delta_{\ell\ell'}$: the perturbation is diagonal in the degenerate subspace, the ``good'' zeroth-order states are the $\ell$ eigenstates themselves, and no rotation is needed at first order.

It is tempting to conclude that the degeneracy is then harmless, and that the intermediate-state sum may simply skip the $n'=n$ term. This is the first of the three pitfalls anticipated at the beginning of this section. The diagonal elements of $V^{(1)}$ lift the degeneracy at $\mathcal{O}(\alpha^4\mu)$: evaluating \eqref{eq:matrix-element-reduced} at $n'=n$ for the two members of $D$, one finds\footnote{For $\ell=0$, the two $R^2/r$ terms in \eqref{eq:matrix-element-reduced} enter with coefficients $-(\ell^2+\ell+1)=-1$ and $3K_{1,0,0}=+1$ and cancel exactly, removing the logarithmic short-distance divergence that either term carries on its own. For $\ell=2$, the combined coefficient is $-7+3K_{1,2,2}=-11$ and every piece is finite.}
\begin{equation}\label{eq:diagonal-shifts}
    \tilde{E}^{(1)}_{n,0} = -\frac{2(3n-1)}{n^4}\,\alpha^4\mu\,, \qquad\qquad 
    \tilde{E}^{(1)}_{n,2} = -\frac{23n-30}{15\,n^4}\,\alpha^4\mu\,,
\end{equation}
so that the splitting between the two levels is
\begin{equation}\label{eq:far-zone-splitting}
    \tilde{E}^{(1)}_{n,2}-\tilde{E}^{(1)}_{n,0} = \frac{67}{15\,n^3}\,\alpha^4\mu \, .
\end{equation}
The more penetrating $\ell=0$ state, with $R_{n,0}(0)\neq0$, is pushed down the most, as one would expect. We should warn the reader right away that \eqref{eq:far-zone-splitting} is \emph{not} yet the physical fine-structure splitting:\footnote{We follow the gravitational-atom literature~\cite{Baumann:2019eav} in referring to all $\mathcal{O}(\alpha^{4}\mu)$ corrections to the spectrum as ``fine structure''. Note, however, a difference with the hydrogen atom: there, the $\mathcal{O}(\alpha^{4})$ energies depend only on $n$ and $j$, so that states with the same $j$ but different $\ell$---such as $2s_{1/2}$ and $2p_{1/2}$---remain degenerate at this order, an accidental degeneracy of the Dirac--Coulomb problem. No analogous degeneracy protects the gravitational problem: the shifts \eqref{eq:diagonal-shifts} depend on both $\ell$ and $j$, and in particular split the two members of $D$.} as we will see in \S\ref{sec:pert-matching}, the near zone shifts the $\ell=0$ level at this very same order, and the corrected splitting is $\tfrac{9}{5n^3}\alpha^4\mu$, in agreement with the published Proca spectrum~\cite{Baumann:2019eav}. For the time being, we simply record the far-zone value: only its scaling with $\alpha$ is relevant for the argument of this subsection, not the precise coefficient.

Because the two members of $D$ are split only at $\mathcal{O}(\alpha^4\mu)$---parametrically below the $\mathcal{O}(\alpha^2\mu)$ energy denominators that separate states with different values of $n$---the component of the corrected state $\vert n,1,m,2\rangle^{(1)}$ \emph{along its degenerate partner} is not determined by the first-order equations. It is instead fixed at second order (see, e.g.,~\cite{Sakurai1993Modern}). To write it down, recall from \eqref{eq: eqs for spatial components of mode functions} that $V^{(1)}$ is only the leading term of the full perturbation $V$, which admits a far-zone expansion in powers of $r_\mathrm{s}/r$, $V = V^{(1)}+V^{(2)}+\cdots$. In the case at hand, this expansion actually terminates after two terms: multiplying (\ref{eq:vecr}\textendash\ref{eq:vecph}) by $(1-r_\mathrm{s}/r)/(2\mu)$ renders the equations linear in $r_\mathrm{s}$, so that $V = V^{(1)}+V^{(2)}$ with
\begin{equation}\label{eq:pert2}
    {V^{(2)I}}_J = -\frac{r_\mathrm{s}}{r}\left({V^{(1)I}}_J - \frac{r_\mathrm{s}}{2\mu r}\,\delta^I_J\,\nabla^2\right)
\end{equation}
scaling as $\alpha^6\mu$ in the far zone. In terms of these operators, the corrected state contains the term $c_{n,0}\vert n,1,m,0\rangle$ with\footnote{Neither $V^{(1)}$ nor $V^{(2)}$ is self-adjoint---for instance, $3K_{1,2,0}=2\sqrt{2}$ while $3K_{1,0,2}=-\sqrt{2}$---so the order of the matrix elements in \eqref{eq:cn0-formula} matters and must be kept as written. The formula remains valid term by term because the unperturbed hydrogenic states form an orthonormal basis.}
\begin{equation}\label{eq:cn0-formula}
    c_{n,0}=\frac{1}{\tilde{E}_{n,2}-\tilde{E}_{n,0}}\left[\langle n,1,m,0\vert V^{(2)}\vert n,1,m,2\rangle + \sum_{\psi\notin D}\frac{\langle n,1,m,0\vert V^{(1)}\vert \psi\rangle\,\langle \psi\vert V^{(1)}\vert n,1,m,2\rangle}{\tilde{E}_{n}-\tilde{E}_{\psi}}+\cdots\right],
\end{equation}
where $\tilde{E}_{n,2}-\tilde{E}_{n,0}$ is the energy splitting within $D$, the sum over $\psi$ runs over the complete set of intermediate states $\vert\psi\rangle$ outside $D$ (an abstract label, as these will include both bound and continuum states), and the ellipsis stands for higher-order corrections---including the near-zone contact contributions derived in \S\ref{sec:pert-matching}, which will turn out to enter at the same order as the two terms displayed. Note that the first-order piece of the numerator starts with $V^{(2)}$: the corresponding matrix element of $V^{(1)}$ vanishes, as shown above.

The power counting of \eqref{eq:cn0-formula} is the crux of the matter. The angular structure of $V^{(2)}$ parallels that of $V^{(1)}$, and steps analogous to those leading to \eqref{eq:matrix-element-reduced} yield its off-diagonal element in the degenerate subspace,
\begin{equation}\label{eq:V2-element}
    \langle n,1,m,0\vert V^{(2)}\vert n,1,m,2\rangle = -\frac{4\sqrt{2}}{15}\,\frac{\sqrt{(n^{2}-1)(n^{2}-4)}}{n^{5}}\,\alpha^{6}\mu \, ,
\end{equation}
which is finite and of order $\alpha^6\mu$; divided by the $\mathcal{O}(\alpha^4\mu)$ splitting, it contributes to $c_{n,0}$ at order $\alpha^2$. The second-order sum enters at exactly the same order: with matrix elements $\langle V^{(1)}\rangle\sim\alpha^4\mu$ and denominators $\tilde{E}_n-\tilde{E}_\psi\sim\alpha^2\mu$,
\begin{equation}
    c_{n,0}\;\sim\;\frac{1}{\alpha^{4}\mu}\left[\,\alpha^{6}\mu+\frac{(\alpha^{4}\mu)^{2}}{\alpha^{2}\mu}\,\right]\;\sim\;\alpha^{2} \, .
\end{equation}
Although the sum is second order in $V^{(1)}$, the small splitting supplies a compensating enhancement of $\alpha^{-2}$ that promotes it to the same order as the single insertion of $V^{(2)}$, so neither term may be dropped. Note also that $c_{n,0}\sim\alpha^{2}$ is precisely the size of the admixtures estimated in \eqref{eq:typical-correction}; since $R_{n,0}(r)$ approaches a constant as $r\to0$, the degenerate partner contributes on the worldline at the same, leading order as the intermediate states considered there. By contrast, second-order admixtures along states \emph{outside} $D$ involve two powers of the denominators $\tilde{E}_{n}-\tilde{E}_{\psi}\sim\alpha^{2}\mu$ and no small splitting: their coefficients are of order $(\alpha^{4}\mu)^{2}/(\alpha^{2}\mu)^{2}=\alpha^{4}$, suppressed by $\alpha^{2}$ relative to \eqref{eq:typical-correction}, and they can be safely neglected. The degenerate partner is thus the unique second-order term promoted to leading order. This is the subtlety that the vanishing of the off-diagonal element $\langle n,0\vert V^{(1)} \vert n,2 \rangle$ conceals: first-order diagonality does not end the story.

Before evaluating these contributions, one more far-zone matrix element deserves comment: the \emph{diagonal} element $\langle n,0\vert V^{(2)}\vert n,0\rangle$. At first sight it appears irrelevant to the correction we are after: a diagonal $\mathcal{O}(\alpha^{6}\mu)$ energy shift affects the splitting in the denominator of \eqref{eq:cn0-formula} only at relative order $\alpha^{2}$. Its significance lies elsewhere: unlike \eqref{eq:V2-element}, this matrix element is UV divergent. The radial integrand behaves as $dr/r$ near the origin so that, cutting the integral off at a radius $r_\mathrm{min}$,
\begin{equation}\label{eq:V2-diagonal-log}
    \langle n,0\vert V^{(2)}\vert n,0\rangle=\frac{8}{n^{3}}\,\alpha^{6}\mu\,\ln\!\left(\frac{1}{\alpha\mu\, r_\mathrm{min}}\right)+\cdots=\frac{\mu\, r_\mathrm{s}^{3}}{4}\,R_{n,0}(0)^{2}\,\ln\!\left(\frac{1}{\alpha\mu\, r_\mathrm{min}}\right)+\cdots \, ,
\end{equation}
where the dots stand for terms that remain finite as $r_\mathrm{min}\to0$; rescaling the (otherwise arbitrary) argument of the logarithm only shifts these finite terms. The coefficient of the logarithm is proportional to $R_{n,0}(0)^{2}$---the structure of a $\delta$-function potential localized at the origin. This divergence is the effective theory telling us, in the standard way, that the far-zone potential alone does not define the physics of the degenerate block: worldline operators, with coefficients fixed by a near-zone matching, are part of the effective description. Crucially, nothing guarantees that these operators first enter at the order of the divergence. The matching of \S\ref{sec:pert-matching} will show that the leading contact term carries a single power of $r_\mathrm{s}$: it shifts the $\ell=0$ level---and hence the denominator of \eqref{eq:cn0-formula}---already at $\mathcal{O}(\alpha^{4}\mu)$, and feeds into the numerator at $\mathcal{O}(\alpha^{6}\mu)$ through a cross term with $V^{(1)}$. We will supply this near-zone input in \S\ref{sec:pert-matching}.

For concreteness, we will carry out the calculation at $n=3$, the lowest overtone that supports an $\ell=2$ mode; we will restore general $n$ in \S\ref{sec:pert-assembly}. Collecting the conclusions of this subsection, the complete first-order correction to the mode function reads
\begin{align}\label{eq:corrected-state}
    \vec{f}^{(1)}_{3,1,m,2} &= \sum_{n'\neq3}\frac{\langle n',1,m,0\vert V^{(1)}\vert 3,1,m,2\rangle}{\tilde{E}_{3}-\tilde{E}_{n'}}\,\vec{f}^{(0)}_{n',1,m,0} \nonumber\\
    &\quad + \int_0^\infty \frac{d\tilde{E}'}{2\pi}\,\frac{\langle \tilde{E}',1,m,0\vert V^{(1)}\vert 3,1,m,2\rangle}{\tilde{E}_{3}-\tilde{E}'}\,\vec{f}^{(0)}_{\tilde{E}',1,m,0} \,+\, c_{3,0}\,\vec{f}^{(0)}_{3,1,m,0} \,+\,\cdots ,
\end{align}
where the second term runs over the positive-energy ($\tilde{E}'>0$) Coulomb scattering states demanded by completeness of the hydrogenic spectrum, the third is the degenerate partner discussed above, and the dots stand for admixtures of other states, whose radial profiles vanish at the origin and which therefore drop out on the worldline. Each of the three terms displayed is proportional to an $\ell=0$ profile, which is $\mathcal{O}(1)$ at the origin, and they map one-to-one onto the three contributions anticipated in \eqref{eq:S-decomposition}: the sum over bound intermediate states yields $\mathcal{N}_\mathrm{b}$ and is evaluated in \S\ref{sec:pert-bound}; the continuum integral yields $\mathcal{N}_\mathrm{c}$ and is evaluated in \S\ref{sec:pert-continuum}; the degenerate partner yields $\mathcal{N}_\mathrm{deg}$. This last contribution is the most laborious of the three, and we will devote two subsections to it. According to \eqref{eq:cn0-formula}, the coefficient $c_{3,0}$ collects three inputs: the far-zone matrix element of $V^{(2)}$, already computed in \eqref{eq:V2-element}; the second-order sum over intermediate states, which we evaluate in \S\ref{sec:pert-second-order}; and the near-zone contact contributions to both its numerator and its denominator, which we fix by matching in \S\ref{sec:pert-matching}. These inputs are assembled into $c_{3,0}$---and hence into $\mathcal{N}_\mathrm{deg}$---in \S\ref{sec:pert-assembly}, where we also combine the three contributions to $\mathcal{N}$, obtain the resulting instability rate, and extend every step to general $n$.

\subsection{Bound-state intermediate states}\label{sec:pert-bound}

Let us start with the sum over bound states---the first term in \eqref{eq:corrected-state}. Using \eqref{eq:matrix-element-reduced}, the matrix element is simply
\begin{equation}\label{eq:bound-matrix-element}
    \langle n',1,m,0\vert V^{(1)}\vert 3,1,m,2\rangle = \frac{2\sqrt{2}\alpha}{\mu^2}\int_0^\infty dr\, \frac{R_{n',0}(r)R_{3,2}(r)}{r}.
\end{equation}
Using \eqref{eq:radial-mode}, the explicit forms of the radial wavefunctions are
\begin{subequations}\label{eq:radial-profiles}
\begin{align}
    R_{n',0}(r) & =\frac{2(\alpha\mu)^{3/2}}{n^{\prime 5/2}}\exp\left(-\frac{\alpha\mu r}{n'}\right)\sum_{i=0}^{n'-1}\frac{(-1)^{i}}{i!}\left(\begin{array}{c}
        n'\\
        n'-1-i
    \end{array}\right)\left(\frac{2\alpha\mu r}{n'}\right)^{i},\label{eq:radial-profile-n0}\\
    R_{3,2}(r) & =\frac{4r^{2}(\alpha\mu)^{7/2}}{81\sqrt{30}}\exp\left(-\frac{\alpha\mu r}{3}\right).\label{eq:radial-profile-32}
\end{align}
\end{subequations}
Hence, the radial integral evaluates to
\begin{align}
    \int_0^\infty dr\,\frac{R_{n',0}(r)R_{3,2}(r)}{r} &= \frac{8(\alpha\mu)^3}{81\sqrt{30n'^{5}}}\sum_{i=0}^{n'-1}\frac{(-1)^{i}}{i!}\left(\begin{array}{c}
    n'\\
    i+1
    \end{array}\right)\left(\frac{2}{n'}\right)^{i}\int_0^\infty dx\,x^{i+1}\exp\left(-\left(\frac{1}{n'}+\frac{1}{3}\right)x\right)\nonumber \\
    &= \frac{4}{9}\sqrt{\frac{2n'}{15}}\frac{\left(n'-3\right)^{n'-1}}{\left(n'+3\right)^{n'+1}}(\alpha\mu)^3.
\end{align}
The bound-state part of the $\ell = 0$ (order $\alpha^{7/2}$) correction to $\vec{f}_{3,1,m,2}(r,\theta,\varphi)$ is then
\begin{equation}\label{eq:leading-correction}
    \vec{f}^{(1)}_{3,1,m,2}(r,\theta,\varphi) \supset - \frac{32\alpha^2}{\sqrt{15}} \vec{Y}_{0;1m}\sum_{n'\neq3} \frac{n'^{5/2}(n'-3)^{n'-2}}{(n'+3)^{n'+2}}R_{n',0}(r) \, ,
\end{equation}
where the $n'=3$ term is excluded: within the degenerate subspace, its role is played by the coefficient $c_{3,0}$ of \eqref{eq:cn0-formula}.

Every term in \eqref{eq:corrected-state} that survives on the worldline---whether it involves bound intermediate states, continuum ones, or the degenerate partner---carries an $\ell=0$ radial profile that approaches a constant at the origin. It is therefore convenient to parametrize the complete correction at the origin by a single dimensionless normalization factor $\mathcal{N}$,
\begin{equation}\label{eq:leading-correction-at-origin}
    \vec{f}^{(1)}_{3,1,m,2}(0) = - \frac{32\mathcal{N}}{\sqrt{15\pi}} \alpha^{7/2}\mu^{3/2}\vec{V}_{m} \, ,
\end{equation}
where we used $\vec{Y}_{0;1m} = \vec{V}_m/\sqrt{4\pi}$ from \eqref{eq:vector-spherical-harmonic-clebsch} in Appendix~\ref{sec:Vector-spherical-harmonics}.\footnote{When explicit numerical values of the quantum numbers are displayed, we separate the orbital number by a semicolon: \textit{e.g.}, $\vec{Y}_{0;1m}$ denotes $\vec{Y}_{\ell,jm}$ with $\ell=0$ and $j=1$.} The prefactor in \eqref{eq:leading-correction-at-origin} is chosen so that the bound-state contribution below, \eqref{eq:S-bound}, takes the form of a sum with unit coefficient: this choice absorbs the factor $1/\sqrt{4\pi}$ from the vector spherical harmonic $\vec{Y}_{0;1m}$ and the factor $2(\alpha\mu)^{3/2}$ common to all the $R_{n',0}(0)$, so that the rationality---or otherwise---of each contribution can be read off directly from $\mathcal{N}$, a property we will exploit shortly. The channels of \eqref{eq:corrected-state} contribute additively to $\mathcal{N}$, as anticipated in \eqref{eq:S-decomposition}: setting \eqref{eq:leading-correction-at-origin} equal to $\vec{f}^{(1)}(0)=R^{(1)}(0)\,\vec{Y}_{0;1m} = R^{(1)}(0) \vec{V}_m/\sqrt{4\pi}$ and solving for $\mathcal{N}$, we find that an admixture whose $\ell=0$ radial profile approaches the constant value $R^{(1)}(0)$ at the origin contributes $-\tfrac{\sqrt{15}}{64}\,R^{(1)}(0)/(\alpha^{7/2}\mu^{3/2})$ to $\mathcal{N}$. In particular, evaluating \eqref{eq:leading-correction} at the origin with $R_{n',0}(0)=2(\alpha\mu)^{3/2}/n'^{3/2}$ gives the bound-state contribution
\begin{equation}\label{eq:S-bound}
    \mathcal{N}_\mathrm{b}=\sum_{n'\neq3}\frac{n'\left(n'-3\right)^{n'-2}}{\left(n'+3\right)^{n'+2}} \approx -0.0045191 \, .
\end{equation}
Had we stopped here, we would be in trouble: superradiant instability rates are expected to be rational numbers times $\alpha^{\#}(m\Omega-\mu)$, and the sum \eqref{eq:S-bound} shows no sign of being rational. Far from being a nuisance, this observation is a smoking gun---an irrational $\mathcal{N}$ signals an incomplete intermediate-state sum.

\subsection{Continuum intermediate states}\label{sec:pert-continuum}

The missing states are the positive-energy Coulomb scattering states, which the resolution of the identity in the $\ell=0$ sector of the hydrogenic problem,
\begin{equation}\label{eq:completeness}
    \mathbbm{1}_{\ell=0}=\sum_{n'}\vert n',0\rangle\langle n',0\vert+\int_{0}^{\infty}\frac{d\tilde{E}'}{2\pi}\,\vert \tilde{E}',0\rangle\langle \tilde{E}',0\vert \, ,
\end{equation}
forces upon the sum over intermediate states in \eqref{eq:corrected-state}. Here the continuum states are normalized as $\langle\tilde{E},0\vert\tilde{E}',0\rangle=2\pi\delta(\tilde{E}-\tilde{E}')$, in the same convention as \eqref{eq:helicity-norm}, and their radial mode functions are given explicitly by
\begin{equation}\label{eq:coulomb-wave}
    R_{\tilde{E}',0}(r)=\frac{2\mu\sqrt{2\pi\alpha}}{\sqrt{1-e^{-2\pi\alpha\mu/k}}}\,e^{ikr}\,{}_{1}F_{1}\!\left(1-\frac{i\alpha\mu}{k};\,2;\,-2ikr\right) \, ,\qquad \tilde{E}'=\frac{k^{2}}{2\mu} \, ,
\end{equation}
a real function of $r$ despite appearances. One might hope that these states decouple. They don't. Because the matrix elements $\langle \tilde{E}',1,m,0\vert V^{(1)}\vert 3,1,m,2\rangle$ are supported at the Bohr radius, the integral in \eqref{eq:corrected-state} is dominated by energies $\tilde{E}'\sim\vert\tilde{E}_{3}\vert\sim\alpha^{2}\mu$---equivalently, momenta $k\sim\alpha\mu$---for which the energy denominators are just as small as those of the bound-state sum. For all such momenta, the value of \eqref{eq:coulomb-wave} at the origin, where ${}_{1}F_{1}=1$, is
\begin{equation}\label{eq:coulomb-origin}
    R_{\tilde{E}',0}(0)=\frac{2\mu\sqrt{2\pi\alpha}}{\sqrt{1-e^{-2\pi\alpha\mu/k}}}=2\mu\sqrt{2\pi\alpha}\,\Bigl[1+\mathcal{O}\bigl(e^{-2\pi\alpha\mu/k}\bigr)\Bigr] \, ,
\end{equation}
so a bin of continuum states of width $\Delta\tilde{E}'\sim\alpha^{2}\mu$ carries an effective contribution of the order $\sqrt{\Delta\tilde{E}'/2\pi}\,R_{\tilde{E}',0}(0)\sim\alpha^{3/2}\mu^{3/2}$ near the origin, which is the same size as its bound-state counterparts $R_{n,0}(0)=2(\alpha\mu)^{3/2}/n^{3/2}$. The continuum therefore contributes to $\mathcal{N}$ at the same order as the bound states, and it must be included. 

Note that it is the attractive Coulomb interaction that makes this conclusion uniform in $k$: the ratio of \eqref{eq:coulomb-origin} to the corresponding free-particle value, $R_{\tilde{E}',0}(0)^{2}/(4\mu k)=2\pi\eta/\bigl(1-e^{-2\pi\eta}\bigr)$ with $\eta=\alpha\mu/k$, is the standard Sommerfeld enhancement factor~\cite{Iengo:2009ni}, which approaches unity for $k\gg\alpha\mu$ but grows like $2\pi\eta$ near threshold, compensating the $\sqrt{k}$ suppression of the free s-wave and keeping \eqref{eq:coulomb-origin} constant all the way down to $k=0$.

Fortunately, there is no need to integrate over the continuum term by term. The key observation, which goes back to Dalgarno and Lewis~\cite{Dalgarno:1955}, is that the \emph{summed} correction can be computed directly, without ever constructing its individual terms. The first two terms in \eqref{eq:corrected-state} are both proportional to $\vec{Y}_{0;1m}$, with a radial profile given by
\begin{equation}\label{eq:R1-def}
    R^{(1)}(r)\equiv\sum_{n'\neq3}\frac{\langle n',1,m,0\vert V^{(1)}\vert 3,1,m,2\rangle}{\tilde{E}_{3}-\tilde{E}_{n'}}\,R_{n',0}(r)+\int_{0}^{\infty}\frac{d\tilde{E}'}{2\pi}\,\frac{\langle \tilde{E}',1,m,0\vert V^{(1)}\vert 3,1,m,2\rangle}{\tilde{E}_{3}-\tilde{E}'}\,R_{\tilde{E}',0}(r) \, .
\end{equation}
Every radial function on the right-hand side is an eigenfunction of $-\frac{1}{2\mu}\left(\partial_{r}^{2}+\frac{2}{r}\partial_{r}\right)-\frac{\alpha}{r}$, the $\ell=0$ radial part of the unperturbed Schr\"odinger operator in \eqref{eq: eqs for spatial components of mode functions}. Acting on \eqref{eq:R1-def} with this operator minus $\tilde{E}_{3}$ therefore cancels the energy denominators, and the completeness relation \eqref{eq:completeness} reassembles the remaining matrix elements into a local source (the excluded $n'=3$ term would contribute a piece proportional to $\langle 3,1,m,0\vert V^{(1)}\vert 3,1,m,2\rangle$, which vanishes):
\begin{equation}\label{eq:sternheimer}
    \left[-\frac{1}{2\mu}\left(\partial_{r}^{2}+\frac{2}{r}\partial_{r}\right)-\frac{\alpha}{r}-\tilde{E}_{3}\right]R^{(1)}(r)=-\frac{2\sqrt{2}\,\alpha}{\mu^{2}}\,\frac{R_{3,2}(r)}{r^{3}} \, ,
\end{equation}
where the right-hand side is the $\ell=0$ projection of $-V^{(1)}\vec{f}^{(0)}_{3,1,m,2}$, with the same radial structure as the integrand of \eqref{eq:bound-matrix-element}. Solving this ordinary differential equation is thus equivalent to performing the bound-state sum \emph{and} the continuum integral in one fell swoop. The boundary conditions make the solution unique: regularity at the origin and decay at infinity select the physical profile, while orthogonality to $R_{3,0}$ fixes the one remaining ambiguity---the freedom to add a multiple of the homogeneous solution $R_{3,0}$, which belongs to the degenerate subspace excluded from the sum and is accounted for separately by $c_{3,0}$. It is easy to check that the solution is remarkably simple,
\begin{equation}\label{eq:Rperp}
    R^{(1)}(r)=\frac{4\sqrt{15}}{405}\,\alpha^{7/2}\mu^{3/2}\,e^{-\alpha\mu r/3} \, ,
\end{equation}
a pure exponential. Its value at the origin immediately yields, via \eqref{eq:leading-correction-at-origin},
\begin{equation}\label{eq:S-perp}
    \mathcal{N}_\mathrm{b}+\mathcal{N}_\mathrm{c}=-\frac{\sqrt{15}}{64}\cdot\frac{4\sqrt{15}}{405}=-\frac{1}{432} \, .
\end{equation}
The continuum contribution, $\mathcal{N}_\mathrm{c}=-\tfrac{1}{432}-\mathcal{N}_\mathrm{b}\approx+0.0022043$, is comparable in magnitude to the bound-state sum---roughly half---and of opposite sign. More importantly, the complete sum is \emph{rational}: the rationality puzzle raised below \eqref{eq:S-bound} is resolved by completeness itself.

\subsection{Second-order mixing with the degenerate partner}\label{sec:pert-second-order}

We now turn to the last term in \eqref{eq:corrected-state}: the admixture of the degenerate partner, with coefficient $c_{3,0}$ given by \eqref{eq:cn0-formula}. Of its far-zone ingredients, the matrix element of $V^{(2)}$ is already in hand, \eqref{eq:V2-element}; what remains is the second-order sum over intermediate states, which the technique of the previous subsection closes as well. Denoting this sum by
\begin{equation}\label{eq:N-def}
    \Sigma \equiv \sum_{\psi\notin D}\frac{\langle 3,1,m,0\vert V^{(1)}\vert \psi\rangle\,\langle \psi\vert V^{(1)}\vert 3,1,m,2\rangle}{\tilde{E}_{3}-\tilde{E}_{\psi}}=\Sigma^{(0)}+\Sigma^{(2)} \, ,
\end{equation}
the intermediate states split into an $\ell_\psi=0$ and an $\ell_\psi=2$ channel---the only orbital quantum numbers that $V^{(1)}$ connects to both members of $D$---and each channel again contains a complete set of bound and continuum states.

The $\ell_\psi=0$ channel requires no new work. To see why, read \eqref{eq:N-def} from the right: the sum over this channel builds up the state $\sum_{\psi}\vert\psi\rangle\langle\psi\vert V^{(1)}\vert 3,1,m,2\rangle/(\tilde{E}_{3}-\tilde{E}_{\psi})$, with $\psi$ restricted to the $\ell_\psi=0$ states outside $D$. These are the same matrix elements, the same energy denominators, and the same set of intermediate states---bound and continuum alike---that define $R^{(1)}(r)$ in \eqref{eq:R1-def}: indeed, excluding $\psi\in D$ removes from this channel only the degenerate partner $\vert 3,1,m,0\rangle$, which is precisely the state excluded there. The radial profile of the summed state is therefore the function \eqref{eq:Rperp} that we have already determined, and $\Sigma^{(0)}$ collapses to the single matrix element of $V^{(1)}$ between the profiles $R_{3,0}$ and $R^{(1)}$, both in the $\ell=0$ channel. Evaluating it with \eqref{eq:matrix-element-reduced} at $\ell'=\ell=0$, we find
\begin{equation}\label{eq:N0}
    \Sigma^{(0)}=\frac{\alpha}{\mu^{2}}\int_{0}^{\infty}dr\,R_{3,0}\left(2rR^{(1)\prime\prime}+3R^{(1)\prime}\right)=-\frac{4\sqrt{5}}{1215}\,\alpha^{6}\mu \, ,
\end{equation}
where the terms proportional to $\int_{0}^{\infty}dr\,R_{3,0}R^{(1)}/r$ are absent: at $\ell'=\ell=0$, this integral enters \eqref{eq:matrix-element-reduced} twice---once through the first integral, with coefficient $-(\ell^{2}+\ell+1)=-1$, and once through the mixing term, with coefficient $3K_{1,0,0}=+1$---and the two contributions cancel exactly, just as they did in the computation of the diagonal shifts $\tilde{E}^{(1)}_{n,0}$ in \eqref{eq:diagonal-shifts}.

The $\ell_\psi=2$ channel is handled by the same strategy, with one extra step. Let $S^{(1)}(r)$ denote the radial profile of the summed correction in this channel---the analogue of \eqref{eq:R1-def}, with the intermediate states now restricted to $\ell_\psi=2$---so that $\Sigma^{(2)}$ reduces to a single matrix element of $V^{(1)}$ between the profiles $R_{3,0}$ and $S^{(1)}$, evaluated in \eqref{eq:N2} below. To derive the equation obeyed by $S^{(1)}$, we retrace the steps that led to \eqref{eq:sternheimer}. Acting with the $\ell=2$ radial operator minus $\tilde{E}_{3}$ cancels the energy denominators and leaves $-\sum_{\psi}R_{\psi}(r)\,\langle\psi\vert V^{(1)}\vert 3,1,m,2\rangle$, with $\psi$ running over the $\ell_\psi=2$ states outside $D$. If this sum ran over \emph{all} the $\ell=2$ states, completeness would convert it into (minus) the full projection of $V^{(1)}\vert 3,1,m,2\rangle$ onto the $\ell=2$ channel, whose radial profile is
\begin{equation}\label{eq:ell2-source}
    s_{2}(r)=\frac{\alpha}{\mu^{2}r^{2}}\left(2rR_{3,2}''+3R_{3,2}'-\frac{11R_{3,2}}{r}\right) .
\end{equation}
This is where the extra step comes in: the state $\vert 3,1,m,2\rangle$ itself carries $\ell=2$ but belongs to $D$, and is therefore excluded from the sum. Its contribution must be subtracted by hand from the completeness result, which adds $+R_{3,2}(r)\,\langle 3,1,m,2\vert V^{(1)}\vert 3,1,m,2\rangle=+\tilde{E}^{(1)}_{3,2}\,R_{3,2}(r)$ to the right-hand side, where we used $\langle 3,1,m,2\vert V^{(1)}\vert 3,1,m,2\rangle=\int_{0}^{\infty}dr\,r^{2}R_{3,2}\,s_{2}=\tilde{E}^{(1)}_{3,2}$. No such term arose in the $\ell=0$ channel, where the analogous subtraction was proportional to the vanishing off-diagonal element $\langle 3,1,m,0\vert V^{(1)}\vert 3,1,m,2\rangle$. The Dalgarno--Lewis equation for this channel therefore reads
\begin{equation}\label{eq:sternheimer2}
    \left[-\frac{1}{2\mu}\left(\partial_{r}^{2}+\frac{2}{r}\partial_{r}-\frac{6}{r^{2}}\right)-\frac{\alpha}{r}-\tilde{E}_{3}\right]S^{(1)}(r)=-s_{2}(r)+\tilde{E}^{(1)}_{3,2}\,R_{3,2}(r) \, ,
\end{equation}
subject to regularity at the origin, decay at infinity, and orthogonality to $R_{3,2}$. The projection is not just bookkeeping: since the operator on the left-hand side annihilates $R_{3,2}$, a solution exists only if the right-hand side is orthogonal to $R_{3,2}$, and the projected source satisfies this condition automatically. The same observation supplies the solution method. Because $R_{3,2}$ solves the associated homogeneous equation, substituting $S^{(1)}(r)=R_{3,2}(r)\,u(r)$ into \eqref{eq:sternheimer2} collapses the left-hand side to $-\bigl(r^{2}R_{3,2}^{2}\,u'\bigr)'/\bigl(2\mu\,r^{2}R_{3,2}\bigr)$, reducing the problem to two quadratures: a first integration determines $u'$, with the integration constant fixed by regularity at the origin, and a second determines $u$. All the integrals involved are elementary---powers of $r$ times an exponential, plus a single $\int dx/x$ responsible for the logarithm below---and yield the closed-form solution
\begin{equation}\label{eq:S1-profile}
    S^{(1)}(r)=\frac{\alpha^{7/2}\mu^{3/2}}{54675\sqrt{30}}\,x\left(1350+873x-260x^{2}-4140\,x\ln x\right)e^{-x/3}+c\,R_{3,2}(r) \, ,\qquad x\equiv\alpha\mu r \, ,
\end{equation}
as can be verified by direct substitution. The logarithm---rendered dimensionless by the Bohr scale---is characteristic of the Coulomb Green's function evaluated at a bound-state energy, and the coefficient $c$ of the homogeneous solution is fixed by the orthogonality condition. Its precise value, which absorbs the $\gamma_{E}$ and $\ln\tfrac{3}{2}$ constants generated by the logarithmic integrals (and shifts under a rescaling of the argument of the logarithm), turns out to be irrelevant: $\int_{0}^{\infty}dr\,R_{3,0}R_{3,2}/r=0$,\footnote{This is not the standard orthonormality of the radial wavefunctions, which involves the measure $r^{2}dr$ and pairs of states with the same $\ell$ (indeed, $\int_{0}^{\infty}dr\,r^{2}R_{3,0}R_{3,2}\neq0$). It is instead the vanishing of the last integral in \eqref{eq:matrix-element-reduced} at $n'=n$ and $\ell'\neq\ell$, established in Appendix~\ref{sec:Degeneracies}---the same property that made $V^{(1)}$ diagonal in the degenerate subspace in \S\ref{sec:pert-DPT}.} so the $R_{3,2}$ direction drops out of the off-diagonal matrix element
\begin{equation}\label{eq:N2}
    \Sigma^{(2)}=\frac{2\sqrt{2}\,\alpha}{\mu^{2}}\int_{0}^{\infty}dr\,\frac{R_{3,0}\,S^{(1)}}{r}=+\frac{22\sqrt{5}}{6075}\,\alpha^{6}\mu
    \qquad\Longrightarrow\qquad
    \Sigma=\frac{2\sqrt{5}}{6075}\,\alpha^{6}\mu \, .
\end{equation}
The remaining logarithms cancel in the final integral---as they must, since $\mathcal{N}$ is algebraic.

At this point, every far-zone quantity is in hand, and it is worth pausing to take stock. Inserting \eqref{eq:V2-element} and \eqref{eq:N2} into \eqref{eq:cn0-formula}, the far-zone ingredients assemble into
\begin{equation}\label{eq:c30-farzone}
    c_{3,0}=\frac{\langle 3,1,m,0\vert V^{(2)}\vert 3,1,m,2\rangle+\Sigma}{\tilde{E}_{3,2}-\tilde{E}_{3,0}}+\cdots=-\frac{74\sqrt{5}}{18225}\,\frac{\alpha^{6}\mu}{\tilde{E}_{3,2}-\tilde{E}_{3,0}}+\cdots \, ,
\end{equation}
where the dots stand for the near-zone contact contributions announced below \eqref{eq:V2-diagonal-log}. These are anything but an afterthought: as we show in the next subsection, they correct \emph{both} the numerator and the denominator of \eqref{eq:c30-farzone} at leading order---indeed, the log-divergent diagonal element \eqref{eq:V2-diagonal-log} was the warning that the degenerate block is UV-sensitive at exactly this order. We therefore refrain from evaluating \eqref{eq:c30-farzone} until the near-zone input is in hand.
\subsection{Static response of the horizon in the massless limit}\label{sec:pert-response}

The logarithmic divergence \eqref{eq:V2-diagonal-log} has the standard effective-theory interpretation: contact operators localized on the worldline, with coefficients that encode near-zone physics, contribute to the degenerate block at the same order as the far-zone potential. To fix those coefficients we must, for once, leave the far zone. The task is simpler than it might appear, because the matching can be carried out in the \emph{static and massless} limit. Within the near zone, the frequency and the mass enter the mode equations only through the dimensionless combinations $(Er_\mathrm{s})^{2}$ and $(\mu r_\mathrm{s})^{2}$; for our bound states, $E\simeq\mu$ and $\mu r_\mathrm{s}=2\alpha$, so both are of order $\alpha^{2}$ and negligible at $r\sim r_\mathrm{s}$. We may therefore set $E=\mu=0$ when solving the equations in the near zone.\footnote{This is a formal device rather than a statement about the physical spectrum: it in no way conflicts with the fact that bound states owe their existence to the mass term, which remains fully active at the Bohr scale where the states are supported. It merely expresses that the near-zone field configuration---and hence the Wilson coefficients extracted from it---is insensitive to $E$ and $\mu$, up to corrections analytic in $(Er_\mathrm{s})^{2}$ and $(\mu r_\mathrm{s})^{2}$ that enter beyond the orders we need.} In particular, setting $\mu=0$ is a \emph{decoupling} limit, and one might worry that it throws away the longitudinal polarization and, with it, part of the response we are trying to compute. This would indeed be a delicate step had we taken the limit at the level of the action, which is the setting where the St\"uckelberg construction of \S\ref{sec:asymptotics} becomes necessary. Here, instead, we take the limit directly in \eqref{eq:vec_eoms}: this is a closed system for the three spatial mode functions $f^{I}$, in which no polarization is ever projected out, and it depends on $E$ and $\mu$ analytically, so its solution space retains all three physical modes as $E,\mu\to0$ and no order-of-limits ambiguity can arise. In particular, the longitudinal mode survives the limit as a pure-gradient configuration, in agreement with the St\"uckelberg expectation \eqref{eq:decoupling} that the massless limit of a Proca field contains the gradient of a massless scalar; we will encounter this configuration shortly, as one of the two basis solutions of the static problem. All we need, then, is the following: place a static, long-wavelength vector field in the parity-even $j=1$ sector around a Schwarzschild BH, impose regularity at the horizon, and read off the induced tails.

Eliminating $f^{t}$ through the exact constraint \eqref{eq:constr}, expanding the spatial components as $\vec{f}=u_{0}(r)\,\vec{Y}_{0;1m}+u_{2}(r)\,\vec{Y}_{2;1m}$---the subscript on each channel amplitude $u_{\ell}(r)$ being its orbital number, and $\vec{Y}_{0;1m}=\vec{V}_{m}/\sqrt{4\pi}$ a constant vector (see Appendix~\ref{sec:Mode-functions})---and setting $E=\mu=0$ as argued above, the Proca equation reduces to the pair of coupled equations
\begin{subequations}\label{eq:near-zone-system}
\begin{align}
    x^{2}(x-1)\,u_{0}''+x(2x-1)\,u_{0}'-2\sqrt{2}\,u_{2}&=0 \, ,\label{eq:near-zone-system-0}\\
    x^{2}(x-1)\,u_{2}''+x(2x-1)\,u_{2}'+(5-6x)\,u_{2}+\sqrt{2}\,u_{0}&=0 \, ,\label{eq:near-zone-system-2}
\end{align}
\end{subequations}
where $x=r/r_\mathrm{s}$ and the equations are \emph{exact} in $r_\mathrm{s}/r$.\footnote{The qualifier ``near zone'' should accordingly be taken with a grain of salt. Had we kept $E$ and $\mu$ finite, the terms we dropped would become important at $r\sim1/E$ and $r\sim1/\mu$, marking the outer edge of the near zone; once $E=\mu=0$, no such scale survives, and \eqref{eq:near-zone-system} holds at every $r>r_\mathrm{s}$. We are thus solving the static problem \emph{exactly}, over all of space. What is genuinely ``near-zone'' about this calculation is the information it supplies to the EFT: the boundary condition at the horizon, which far-zone perturbation theory cannot access, and which fixes the otherwise free tail coefficients in \eqref{eq:response-def} below.} As a consistency check, we point out that the off-diagonal couplings in \eqref{eq:near-zone-system} are the mixing coefficients $3K_{1,2,0}=2\sqrt{2}$ and $3K_{1,0,2}=-\sqrt{2}$ of \eqref{eq:matrix-element-reduced}; the constant $5$ in \eqref{eq:near-zone-system-2} encodes the $\ell=2$ diagonal coefficient $-7+3K_{1,2,2}=-11$; and the absence of any $\mathcal{O}(r_\mathrm{s}/r)$ term proportional to $u_{0}$ (\textit{i.e.}, without derivatives) in \eqref{eq:near-zone-system-0} is the $-1+1$ cancellation of the $\ell=0$ diagonal noted in \S\ref{sec:pert-DPT}.

Let us now discuss the boundary conditions at the horizon. Since the system \eqref{eq:near-zone-system} is fourth order, regularity at the horizon selects two solutions up to overall normalization. Concretely, evaluating \eqref{eq:near-zone-system} at $x=1$ yields $u_{0}'(1)=2\sqrt{2}\,u_{2}(1)$ and $u_{2}'(1)=u_{2}(1)-\sqrt{2}\,u_{0}(1)$, after which all higher Taylor coefficients follow recursively. Horizon-regular solutions therefore form a \emph{two}-parameter family, labeled by the free horizon values $u_{0}(1)$ and $u_{2}(1)$. A moment of reflection shows that this is exactly as it should be: the static problem admits two independent source multipoles---the growing data introduced in \eqref{eq:response-def} below---and the response matrix relating tails to sources can only be extracted if each source can be turned on independently while preserving regularity.

We now construct a convenient basis for this two-parameter family. The first basis solution is a pure-gradient (longitudinal) configuration: in the static and massless limit, $f_{\nu}=\partial_{\nu}\bigl(\chi\,Y^{1m}\bigr)$ solves \eqref{eq:near-zone-system} whenever the St\"uckelberg profile $\chi(r)$ obeys the static massless-scalar equation in the $\ell=1$ sector, $\bigl(x(x-1)\,\chi'\bigr)'=2\chi$---a Legendre equation, whose horizon-regular solution is simply $\chi\propto x-\tfrac{1}{2}$.\footnote{The label $\ell=1$ refers to the \emph{scalar} harmonic: a scalar with total angular momentum $j=1$ necessarily has orbital number $\ell=1$. The corresponding \emph{vector} configuration $\partial_{\nu}\bigl(\chi\,Y^{1m}\bigr)$ instead spreads over the orbital channels $\ell=0$ and $\ell=2$, as the channel amplitudes below make explicit.} The corresponding channel amplitudes are
\begin{equation}\label{eq:sol-gradient}
    u_{0}=\sqrt{3}-\frac{2}{\sqrt{3}\,x} \, , \qquad u_{2}=\frac{1}{\sqrt{6}\,x} \, ,
\end{equation}
as one can verify by direct substitution into \eqref{eq:near-zone-system}. The overall normalization is of course arbitrary---the system is linear and homogeneous---and the same is true of the second basis solution, which is independent of the first (its $u_{2}$ grows as $x^{2}$ at large distances, whereas that of \eqref{eq:sol-gradient} decays) and is also elementary:
\begin{subequations}\label{eq:sol-quadrupole}
\begin{align}
    u_{0}&=\frac{x}{444}+\frac{\ln x}{296}-\frac{1}{666}-\left(\frac{31}{7992}+\frac{\ln x}{444}\right)\frac{1}{x}+\frac{1}{1776\,x^{2}} \, ,\\
    u_{2}&=\frac{1}{\sqrt{2}}\left[\frac{x^{2}}{444}+\frac{x}{1776}+\frac{1}{888}+\left(\frac{1}{3996}+\frac{\ln x}{888}\right)\frac{1}{x}-\frac{1}{888\,x^{2}}\right] .
\end{align}
\end{subequations}
This solution is analytic on the horizon, with $u_{0}(1)=-\tfrac{41}{15984}$ and $u_{2}(1)=\tfrac{49}{15984\sqrt{2}}$, and its large-$x$ expansion terminates as displayed. This solution is also defined only up to admixtures of \eqref{eq:sol-gradient}, since any such combination is equally regular. The response matrix introduced below is constructed so that nothing depends on either the normalizations or this residual freedom. 

At large $x$, each channel of the flat ($r_\mathrm{s}=0$) problem has one growing and one decaying solution, $x^{\ell}$ and $x^{-\ell-1}$, so a generic horizon-regular solution behaves as
\begin{equation}\label{eq:response-def}
    u_{\ell}\;\to\;a_{\ell}\left(\frac{r}{r_\mathrm{s}}\right)^{\ell}+\,b_{\ell}\left(\frac{r_\mathrm{s}}{r}\right)^{\ell+1}+\cdots \, ,\qquad \ell=0,2 \, ,
\end{equation}
where the coefficients $a_{\ell}$ parametrize the growing (source) data and the $b_{\ell}$ the induced tails. The dots stand for all the other terms visible in \eqref{eq:sol-quadrupole}---additional powers and logarithms---whose coefficients are fixed, order by order in $r_\mathrm{s}/r$, by solving \eqref{eq:near-zone-system} perturbatively around the flat solutions: they carry no independent near-zone information, and the genuine boundary data of the problem are the four coefficients $(a_{0},b_{0},a_{2},b_{2})$.\footnote{\label{fn:continuation}When $\ell$ is an integer, the series of $r_\mathrm{s}/r$ corrections descending from the growing solution can reach the decaying slot $x^{-\ell-1}$ and contaminate it. A clean separation between the two series is obtained by analytically continuing $\ell$ to generic real values, where they never overlap~\cite{Kol:2011vg}.} The near-zone information is then encoded in how the tails respond to the sources. Because the two channels are coupled---only $j$, $m$, and parity are symmetry labels of the static problem, whereas $\ell$ is not conserved since Eqs. \eqref{eq:near-zone-system} are coupled---the response is a matrix rather than a pair of numbers, $b_{\ell}=\sum_{\ell'}G_{\ell\ell'}\,a_{\ell'}$, and we now read off its entries from the two basis solutions.

The gradient solution \eqref{eq:sol-gradient} carries $(a_{0},b_{0},a_{2},b_{2})=\bigl(\sqrt{3},-\tfrac{2}{\sqrt{3}},0,0\bigr)$, since its $u_{2}$ contains neither of the fall-offs $x^{2}$ and $x^{-3}$: it is the unique horizon-regular solution, up to normalization, that is sourced purely by the monopole. It follows that $G_{00}=b_{0}/a_{0}=-\tfrac{2}{3}$ and $G_{20}=0$. The second solution \eqref{eq:sol-quadrupole} is the only one that turns on the quadrupole source, $a_{2}=\tfrac{1}{444\sqrt{2}}$, but it does so accompanied by monopole data, $a_{0}=-\tfrac{1}{666}$,\footnote{Quoting $a_{0}$ and $b_{0}$ separately requires a convention, because \eqref{eq:sol-quadrupole} contains logarithms: replacing $\ln x\to\ln(\lambda x)$ reshuffles the non-logarithmic coefficients of the $x^{0}$ and $x^{-1}$ slots of $u_{0}$, and with them the values of $a_{0}$ and $b_{0}$. We fix this ambiguity by writing all logarithms as $\ln(r/r_\mathrm{s})$. Nothing physical hinges on this choice: the shift of \eqref{eq:sol-quadrupole} induced by $\lambda\neq1$ is proportional to the gradient solution \eqref{eq:sol-gradient}, as one can readily check, and therefore drops out of the invariant combination $b_{0}-G_{00}\,a_{0}$ from which $G_{02}$ is extracted below.} and with tails $(b_{0},b_{2})=\bigl(-\tfrac{31}{7992},0\bigr)$. The vanishing of $b_{2}$ immediately gives $G_{22}=0$, while the defining relation $b_{0}=G_{00}\,a_{0}+G_{02}\,a_{2}$ yields $G_{02}=\bigl(b_{0}-G_{00}\,a_{0}\bigr)/a_{2}=-\tfrac{13\sqrt{2}}{6}$---a combination invariant under shifts of \eqref{eq:sol-quadrupole} by any multiple of \eqref{eq:sol-gradient}, so that the residual freedom noted below \eqref{eq:sol-quadrupole} drops out. In summary, the static linear response in the parity-even $j=1$ sector is captured by the following coefficients:
\begin{equation}\label{eq:response-matrix}
    G_{00}=-\frac{2}{3} \, ,\qquad G_{02}=-\frac{13\sqrt{2}}{6} \, ,\qquad G_{20}=G_{22}=0 \, .
\end{equation}

These values admit a simple physical interpretation once we recall that, in the static and massless limit in which the matching is performed, we are no longer dealing with a Proca field: by the St\"uckelberg decomposition \eqref{eq:decoupling}, the system describes a massless vector together with a massless scalar. In the sector at hand, the scalar carries orbital number $\ell=1$ and the massless vector total angular momentum $j=1$; the labels $\ell=0,2$ of the channel amplitudes refer to the orbital decomposition of the \emph{massive} theory, and should not be confused with the multipoles of the massless constituents. The susceptibilities relevant to our problem are thus the $\ell=1$ susceptibility of the scalar and the $j=1$ (dipole) polarizability of the massless vector, and both vanish for four-dimensional BHs in general relativity, just like their better-known spin-2 counterparts, the Love numbers (see~\cite{Hui:2020xxx,Rodriguez:2026iot} and references therein).\footnote{This property is specific to general relativity in four spacetime dimensions---it fails, for instance, in higher dimensions~\cite{Kol:2011vg} and in modified theories of gravity (see, \textit{e.g.},~\cite{Cardoso:2018ptl})---and its seemingly fine-tuned character~\cite{Porto:2016zng} has motivated a considerable body of work aimed at explaining it in terms of symmetries; see, \textit{e.g.},~\cite{Charalambous:2021mea,Charalambous:2021kcz,Hui:2021vcv,BenAchour:2022uqo,Hui:2022vbh,Charalambous:2022rre,Katagiri:2022vyz,Berens:2022ebl,Sharma:2024hlz,Charalambous:2024tdj,Rai:2024lho,Charalambous:2024gpf,Combaluzier-Szteinsznaider:2024sgb,Gounis:2024hcm,Lupsasca:2025pnt,Berens:2025okm,Parra-Martinez:2025bcu}.} This vanishing is the origin of the two zeros in \eqref{eq:response-matrix}: the $x^{-3}$ tail of $u_{2}$ can only descend from the decaying branches of the massless constituents---for the gradient solution, for instance, from the decaying branch $\chi\sim x^{-2}$---and the coefficients of those branches are precisely the susceptibilities that vanish, i.e. $G_{20}=G_{22}=0$.

Against this backdrop, the nonvanishing $G_{00}$ may come as a surprise. The resolution is that $G_{00}$ is \emph{not} the static susceptibility of any massless field: in this sector, neither constituent admits an $\ell=0$ multipole, so there is no vanishing theorem for $G_{00}$ to violate. The $1/x$ tail of $u_{0}$ is instead a by-product of the change of variables, generated by the \emph{growing} scalar branch because the gradient is taken on the Schwarzschild background: with $\chi\propto x-\tfrac{1}{2}$, the metric factor in $f^{r}=(1-\tfrac{1}{x})\,\chi'\,Y^{1m}$ and the $\chi/x$ terms in the angular components produce $1/x$ contributions in $u_{0}$, even though $\chi$ itself has no tail at all.\footnote{Consistently, under the analytic continuation of footnote~\ref{fn:continuation}, this term moves with the growing branch rather than with the response slot $x^{-\ell-1}$.} This does not make $G_{00}$ any less physical: the far-zone expansion \eqref{eq:response-def} treats the $1/x$ slot of $u_{0}$ as free data, so the ratio $b_{0}/a_{0}=-\tfrac{2}{3}$, fixed by horizon regularity, must be supplied by the near zone. In the next subsection, we recast this near-zone data as the Wilson coefficient of a contact operator on the worldline.

\subsection{Matching the worldline contact term}\label{sec:pert-matching}

The response coefficients \eqref{eq:response-matrix} are full-theory data, obtained by solving the static, massless limit of the field equations on the full Schwarzschild geometry with regularity imposed at the horizon, where the point-particle description does not apply. No finite order of far-zone perturbation theory can substitute for this data, since the expansion in $r_\mathrm{s}/r$ is blind to the boundary condition at the horizon---which selects the physical solution among the many admitted by the exterior equations---and commits short-distance errors wherever it is extrapolated beyond its domain of validity, as when the radial integral in \eqref{eq:V2-diagonal-log} is carried down to $r=0$. Both deficits, however, are localized in the UV at $r\lesssim r_\mathrm{s}$, and are therefore remedied by the same means as in any EFT: at any fixed order in the power counting, a finite number of local operators on the worldline supplies the missing boundary data and absorbs the spurious short-distance contributions.\footnote{These are the conservative contact terms anticipated in footnote~\ref{fn:conservative-bilinear}.}

Our remaining task is therefore to construct these operators explicitly, and doing so requires answering three questions. Which worldline operators encode the static response? How do they translate into a potential that can be fed into the quantum-mechanical perturbation theory of \S\S\ref{sec:pert-DPT}--\ref{sec:pert-second-order}? And which of their insertions contribute to $c_{n,0}$ at the order we are working? Let's address these questions in turn.

The operators in question belong to the conservative sector of the worldline theory. The interactions introduced in \S\ref{sec: eft} are linear in the field and couple it to the composite operators $\mathcal{O}^{I}(X)$: they describe the exchange of energy with the horizon, and their correlators determine the absorption rates matched in \S\ref{sec: results}. Static response is a different piece of physics---no energy is exchanged in a static configuration---and is encoded instead in worldline operators \emph{quadratic} in the field, with ordinary c-number coefficients and no reference to the microscopic degrees of freedom~\cite{Hui:2020xxx}. In the parity-even $j=1$ sector, the dictionary between such bilinears and the entries of the response matrix follows from the behavior of the orbital channels at the origin: a smooth profile has $u_{0}(0)\neq0$ while $u_{2}$ vanishes as $r^{2}$, so the field $\tilde{A}_{I}$ evaluated on the worldline probes the $\ell=0$ channel, whereas the $\ell=2$ channel is reached only by acting with two (symmetrized, trace-free) spatial derivatives. The leading operator is therefore the monopole bilinear
\begin{equation}\label{eq:monopole-op}
    \Delta S_\mathrm{pp}=\frac{\kappa}{2}\int d\tau\,\tilde{A}_{I}\tilde{A}^{I}+\cdots=\frac{\kappa}{2}\int d\tau\,A_{I}A^{I}+\cdots \, ,
\end{equation}
where the second equality follows from the definition $\tilde{A}_{I}\equiv{R_{I}}^{J}A_{J}$ of \S\ref{sec: eft}, the rotation matrices canceling out in the bilinear. The dots stand for bilinears involving derivatives of the field, which encode the remaining entries of the response matrix; we will verify below, on dimensional grounds, that they enter well beyond the order of interest. The coefficient $\kappa$ admits a simple physical interpretation as a susceptibility: expanding \eqref{eq:monopole-op} around a slowly varying background value, $A_{I}=\bar{A}_{I}+\delta A_{I}$, produces a term linear in the fluctuation,
\begin{equation}\label{eq:induced-monopole}
    \Delta S_\mathrm{pp}\supset\kappa\int d\tau\,\bar{A}_{I}\,\delta A^{I} \, ,
\end{equation}
\textit{i.e.}, a worldline source for $\delta A_{I}$ of strength $\kappa\bar{A}_{I}$: an applied field induces a monopole moment in the parity-even $j=1$ channel, proportional to the field itself, with $\kappa$ the corresponding susceptibility.

Let us then translate \eqref{eq:monopole-op} into the language of \S\ref{sec:pert-DPT}. Inserted at tree level into the quadratic action of the field, a worldline bilinear is a contact modification of the one-particle problem: \eqref{eq:monopole-op} shifts the mass term of the mode equations to $\mu^{2}-\kappa\,\delta^{3}(\vec{x})$ and, dividing by $2\mu$ to reach the Schr\"odinger form \eqref{eq: eqs for spatial components of mode functions}, it is equivalent to the following additional contribution to the potential
\begin{equation}\label{eq:deltaVd-def}
    \delta V=-\frac{\kappa}{2\mu}\,\delta^{3}(\vec{x}) \, .
\end{equation}
In the Cartesian components of \eqref{eq: eqs for spatial components of mode functions}, the $\delta$-function evaluates the field on the worldline, where a regular profile scales as $r^{\ell}$, so only the $\ell=0$ channel (which belongs exclusively to the parity-even $j=1$ sector) is affected.

To fix $\kappa$, we repeat the computation of \S\ref{sec:pert-response} on the EFT side, where the near zone has been excised and replaced by \eqref{eq:deltaVd-def}, which we treat perturbatively. Neglecting $\delta V$ at lowest order, the static $\ell=0$ far-zone problem is \emph{exactly} free: there is no potential term for $u_{0}$ because of the $-1+1$ cancellation noted below \eqref{eq:near-zone-system}, and the mixing with the $\ell=2$ channel is $\mathcal{O}(r_\mathrm{s}/r)$ suppressed compared to the derivatives of $u_0$. Setting  $\vec{f}=u_0 \vec{Y}_{0;1m}$ where $\vec{Y}_{0;1m}$ are constant vectors (see discussion in App. \ref{sec:Mode-functions}), the unperturbed $\ell=0$ solutions are spanned by $u_{0}=\mathrm{const}$ and $u_{0}\propto1/r$. Perturbing around the constant solution, $\vec{f}_{(0)} =a_0 \vec{Y}_{0;1m}$, the first order correction is determined by the  equation $-\tfrac{1}{2\mu}\nabla^{2}\vec{f}_{(1)} = - \delta V\vec{f}_{(0)}$, and the delta function on the right hand side sources a $b_0 r_s/r$ profile with
\begin{equation}\label{eq:EFT-tail}
    b_{0}\,r_\mathrm{s}=\frac{\kappa}{4\pi}\,a_{0} \, .
\end{equation}
In the full theory, horizon regularity fixed the same tail-to-source ratio to be $b_{0}/a_{0}=G_{00}= -2/3$; solving \eqref{eq:EFT-tail} for $\kappa$ and matching to the full-theory result yields
\begin{equation}\label{eq:kappa-matched}
    \kappa= 4\pi r_s \frac{b_0}{a_0} = 4\pi \,r_\mathrm{s} G_{00}=-\frac{8\pi}{3}\,r_\mathrm{s} \, .
\end{equation}

We now put the contact term to work. Our aim is the energy splitting it induces within the degenerate $n$~block---and in particular its effect on the $\ell=0$--$\ell=2$ separation---so, by first-order perturbation theory, what we need are its matrix elements between the unperturbed bound states $\lvert n,j,m,\ell\rangle$. Because $\delta V$ in \eqref{eq:deltaVd-def}, as noted below \eqref{eq:deltaVd-def}, reaches only the $\ell=0$ channel, the relevant matrix elements are those between two $\ell=0$ states with mode functions $\vec{f}_{n,1,m,0}=R_{n,0}(r)\,\vec{Y}_{0;1m}$. Writing the matrix element as an integral, inserting \eqref{eq:deltaVd-def}, and recalling that $\vec{Y}_{0;1m}$ is constant with $\lvert\vec{Y}_{0;1m}\rvert^{2}=1/4\pi$, we obtain
\begin{align}\label{eq:deltaVd}
    \langle n',1,m,0\vert \delta V\vert n,1,m,0\rangle
    &=\int d^{3}x\,\vec{f}^{\,*}_{n',1,m,0}\cdot\,\vec{f}_{n,1,m,0} \delta V \nonumber\\
    &=-\frac{\kappa}{2\mu}\int d^{3}x\,\delta^{3}(\vec{x})\,\lvert\vec{Y}_{0;1m}\rvert^{2}\,R_{n',0}(r)R_{n,0}(r) \nonumber\\
    &=-\frac{\kappa}{8\pi\mu}\,R_{n',0}(0)R_{n,0}(0)
    =+\frac{r_\mathrm{s}}{3\mu}\,R_{n',0}(0)R_{n,0}(0) \, ,
\end{align}
where in the last step we used the matched value \eqref{eq:kappa-matched}.
It is worth commenting on the sign of this correction before moving on: the negative (screening) monopole response $G_{00}=-\tfrac{2}{3}$ maps onto a \emph{repulsive} contact term. By \eqref{eq:deltaVd}, this contact term shifts the $\ell=0$ level diagonally by
\begin{equation}\label{eq:deltaE}
    \delta\tilde{E}_{n,0}=+\frac{r_\mathrm{s}}{3\mu}R_{n,0}(0)^{2}=+\frac{8}{3n^{3}}\,\alpha^{4}\mu \, ,\qquad \delta\tilde{E}_{n,2}=0 \, ,
\end{equation}
so that the physical splitting within the degenerate block becomes
\begin{equation}\label{eq:corrected-splitting}
    \bigl(\tilde{E}_{n,2}-\tilde{E}_{n,0}\bigr)_\mathrm{tot}=\left(\frac{67}{15n^{3}}-\frac{8}{3n^{3}}\right)\alpha^{4}\mu=\frac{9}{5n^{3}}\,\alpha^{4}\mu \, .
\end{equation}
Thus, the repulsive contact term pushes the $\ell=0$ level up and \emph{reduces} the far-zone splitting \eqref{eq:far-zone-splitting}, to the value anticipated in \S\ref{sec:pert-DPT}. Because \eqref{eq:corrected-splitting} is no longer a far-zone statement, it admits a sharp external check: it reproduces exactly the fine structure of the Proca spectrum computed in the full theory, eq.~(2.31) of~\cite{Baumann:2019eav}. We defer the channel-by-channel comparison to \S\ref{sec:pert-assembly} (see Table~\ref{tab:fine-structure}), where it will join the instability rate as the second full-theory check on the near-zone input.

Before putting $\delta V$ to work, let us settle a debt: the logarithmic divergence \eqref{eq:V2-diagonal-log}, which has been hanging over the calculation since \S\ref{sec:pert-DPT}. Its coefficient is proportional to $R_{n,0}(0)^{2}$ with an $n$-independent constant of proportionality---by \eqref{eq:deltaVd}, precisely the diagonal matrix-element structure of the contact term, for every $n$ at once. The divergence is therefore reabsorbed into $\kappa$, at the cost of letting the coefficient depend on the cutoff: demanding that the diagonal element $\langle n,0\vert V^{(2)}+\delta V\vert n,0\rangle$ be independent of $r_\mathrm{min}$ requires
\begin{equation}\label{eq:kappa-running}
    r_\mathrm{min}\frac{d\kappa}{dr_\mathrm{min}}=-2\pi\,\mu^{2}r_\mathrm{s}^{3} \, ,
\end{equation}
and the $n$-independence guarantees that this single, state-independent counterterm does the job for the entire tower of levels. Relative to the matched value \eqref{eq:kappa-matched}, the running \eqref{eq:kappa-running} is a fractional correction of order $(\mu r_\mathrm{s})^{2}\ln\alpha\sim\alpha^{2}\ln\alpha$, and it feeds into the spectrum only at $\mathcal{O}(\alpha^{6}\mu\ln\alpha)$---beyond our working order, so \eqref{eq:kappa-matched} is all we will ever need.
Note also that the beta function \eqref{eq:kappa-running} is proportional to $\mu^{2}$ and thus shuts off in the massless limit, consistent with the fact that the static Love numbers of four-dimensional BHs, which vanish, also do not run---logarithmic running being a hallmark of their higher-dimensional counterparts~\cite{Kol:2011vg}. It is the \emph{running} of $\kappa$, not its \emph{value}, that must disappear as $\mu\to0$: the matched coefficient $\kappa=-\tfrac{8\pi}{3}r_\mathrm{s}$ is $\mathcal{O}(r_\mathrm{s})$ and $\mu$-independent, and so it should be, since $\kappa$ is not the susceptibility of any massless field but the response of the horizon in the parity-even $j=1$ monopole channel of the \emph{massive} vector (cf.\ the discussion below \eqref{eq:response-matrix}). One might still be puzzled that, after the St\"uckelberg substitution $A_{\nu}\to A_{\nu}+\partial_{\nu}\phi / \mu$ of \eqref{eq:decoupling}, the bilinear \eqref{eq:monopole-op} generates a worldline operator $\tfrac{\kappa}{2\mu^{2}}\,\partial_{I}\phi\,\partial^{I}\phi$ whose coefficient blows up as $\mu\to0$. This is not a pathology but the familiar strong coupling problem of the longitudinal polarization in the massless limit: reinstated as the St\"uckelberg field, it couples to the worldline with strength $\mu^{-1}$, and because the horizon bilinear it couples to is not a conserved source, this coupling does not switch off but grows as $\mu\to0$~\cite{Hinterbichler:2011tt}. The $\mu^{-2}$ is compensated by the $\mu$-dependence of the St\"uckelberg map between the two sets of variables, so that no physical quantity---in particular no level shift---diverges as $\mu\to0$, as guaranteed by the analyticity of the near-zone problem in $(\mu r_\mathrm{s})^{2}$.

What about the operators hidden in the dots of \eqref{eq:monopole-op}? Because the static near-zone problem involves no scale other than $r_\mathrm{s}=2GM$---the only one in \eqref{eq:near-zone-system}, since $E$ and $\mu$ have dropped out---dimensional analysis fixes each Wilson coefficient to a pure number times a \emph{non-negative integer} power of $r_\mathrm{s}$. Measuring all radii in units of $r_\mathrm{s}$ in \eqref{eq:response-def}, a source of strength $s_{\ell'}$ that grows as $s_{\ell'}\,r^{\ell'}$ induces the tail
\begin{equation}\label{eq:tail-scaling}
    u_{\ell}\;\supset\;G_{\ell\ell'}\,r_\mathrm{s}^{\,\ell+\ell'+1}\,\frac{s_{\ell'}}{r^{\ell+1}} \, ,
\end{equation}
so the operator encoding $G_{\ell\ell'}$ carries $r_\mathrm{s}^{\,\ell+\ell'+1}$: one power for the monopole (as \eqref{eq:kappa-matched} confirms), three for the $G_{02}$ mixing, five for the quadrupole. In matrix elements between Bohr-scale states each extra power of $r_\mathrm{s}$ is accompanied by a derivative of the mode functions at the origin, costing a factor $\alpha\mu$, so every unit beyond the first suppresses the result by $r_\mathrm{s}\,\alpha\mu\sim\alpha^{2}$. The monopole thus enters at fine-structure order, $r_\mathrm{s}R_{n,0}(0)^{2}/\mu\sim\alpha^{4}\mu$, the mixing bilinears are pushed to $\sim\alpha^{8}\mu$, and $G_{20}=G_{22}=0$ require no operators at all. In particular, no worldline operator can generate an $\mathcal{O}(\alpha^{6}\mu)$ \emph{off-diagonal} element in the degenerate block, and the near zone enters our calculation only through the single number $G_{00}$.

We can now answer the third question. The perturbation acting on the degenerate block is the sum
\begin{equation}\label{eq:W-def}
    W=V^{(1)}+V^{(2)}+\delta V+\cdots \, ,
\end{equation}
where the dots stand for the tower of worldline operators hidden in \eqref{eq:monopole-op}, which we have just argued cannot contribute at the order we are working. Eq.~\eqref{eq:cn0-formula} holds with $V^{(1)}+V^{(2)}$ replaced by $W$, so that the insertions of $\delta V$ enter alongside those of $V^{(1)}$ and $V^{(2)}$---this is precisely the content of the ellipsis there.

Most of these insertions vanish, and it is worth seeing why in some detail. Being a contact term, $\delta V$ probes a state only through the value of its mode function on the worldline: by \eqref{eq:deltaVd-def},
\begin{equation}\label{eq:deltaV-annihilates}
    \delta V\,\vec{f}_{n,1,m,\ell}(\vec{x})=-\frac{\kappa}{2\mu}\,\delta^{3}(\vec{x})\,\vec{f}_{n,1,m,\ell}(0) \, ,
\end{equation}
and a regular profile behaves as $R_{n,\ell}(r)\propto r^{\ell}$ near the origin, so that $\vec{f}_{n,1,m,\ell}(0)\propto R_{n,\ell}(0)$ vanishes for every $\ell\neq0$. Thus $\delta V\vert n,1,m,2\rangle=0$ identically---not merely in matrix elements against the states of $D$---which kills at once the off-diagonal element of $\delta V$ within $D$, its diagonal element on the $\ell=2$ state, and every second-order term in which $\delta V$ acts directly on $\vert n,1,m,2\rangle$, including the double insertion of $\delta V$ itself. Only two insertions survive, and they enter \eqref{eq:cn0-formula} in two distinct places.

The first is the diagonal element on the $\ell=0$ state, already computed in \eqref{eq:deltaE}: it corrects the splitting in the \emph{denominator}, which henceforth reads \eqref{eq:corrected-splitting}. The second lives in the second-order \emph{numerator}: it is the analogue of the sum $\Sigma$ defined in \eqref{eq:N-def}, with one of the two insertions of $V^{(1)}$ traded for $\delta V$,
\begin{equation}\label{eq:Sigma-dV-def}
    \Sigma_{\delta V}\equiv\sum_{\psi\notin D}\frac{\langle 3,1,m,0\vert \delta V\vert \psi\rangle\,\langle \psi\vert V^{(1)}\vert 3,1,m,2\rangle}{\tilde{E}_{3}-\tilde{E}_{\psi}} \, ,
\end{equation}
the mirror term---with $V^{(1)}$ on the left and $\delta V$ on the right---vanishing by \eqref{eq:deltaV-annihilates}. Because $\delta V$ selects $\ell_{\psi}=0$, the sum over $\psi$ in \eqref{eq:Sigma-dV-def} is precisely the one that assembles the profile $R^{(1)}$ of \eqref{eq:R1-def}, and with the matrix element of $\delta V$ given by \eqref{eq:deltaVd} we find
\begin{equation}\label{eq:Sigma-dV}
    \Sigma_{\delta V}=\frac{r_\mathrm{s}}{3\mu}\,R_{3,0}(0)\,R^{(1)}(0)
    =+\frac{16\sqrt{5}}{3645}\,\alpha^{6}\mu\;=\;-\,\langle 3,1,m,0\vert V^{(2)}\vert 3,1,m,2\rangle \, :
\end{equation}
at $n=3$, the near-zone cross term cancels the far-zone $V^{(2)}$ element \emph{exactly}---an accident of the lowest overtone, as we will see in the next subsection, but a welcome one. The numerator of \eqref{eq:c30-farzone} thus collapses to the second-order sum alone, and dividing by the corrected splitting \eqref{eq:corrected-splitting} we obtain
\begin{equation}\label{eq:c30}
    c_{3,0}=\frac{\Sigma}{\tfrac{9}{5\cdot 27}\,\alpha^{4}\mu}=\frac{2\sqrt{5}}{6075}\cdot 15\;\alpha^{2}=\frac{2\sqrt{5}}{405}\,\alpha^{2} \, .
\end{equation}
This is the last ingredient of the corrected mode function \eqref{eq:corrected-state}: all that remains is to assemble the pieces, which we do next.

\subsection{The assembled rate}\label{sec:pert-assembly}

We now have every ingredient in hand. With $c_{3,0}$ given by \eqref{eq:c30} and $R_{3,0}(0)=2(\alpha\mu)^{3/2}/3^{3/2}$, the degenerate partner contributes
\begin{equation}\label{eq:S-deg}
    \mathcal{N}_\mathrm{deg}=-\frac{\sqrt{15}}{64}\cdot\frac{2\sqrt{5}}{405}\cdot\frac{2}{3^{3/2}}=-\frac{1}{3888} \, ,
\end{equation}
and, adding the contributions in \eqref{eq:S-perp}, the total normalization factor is
\begin{equation}\label{eq:S-total}
    \mathcal{N}=-\frac{1}{432}-\frac{1}{3888}=-\frac{5}{1944} \, ,
\end{equation}
as advertised in \eqref{eq:S-decomposition}---rational, as it had to be.

The transition amplitudes follow as before. The electric field also involves the gradient of the divergence of $\vec{f}$, shown in \eqref{eq:bound-E-field-matrix-element}; for the $\ell=0$ admixture this is singular as $r\to0$, but it is suppressed by $\alpha^{2}$ compared to \eqref{eq:leading-correction-at-origin} and is captured by subleading worldline operators, exactly as the near-zone contact terms of \S\ref{sec:pert-matching} capture the leading ones; see the discussion in Appendix~\ref{sec:Mode-functions} around \eqref{eq:2-derivatives-singular}. The perturbative contributions to the matrix elements
of single particle absorption are
\begin{align}
    \langle0\vert A^{I}(t,0)\vert3,1,m,2\rangle & \simeq -\frac{32\mathcal{N}}{\sqrt{30\pi}}\alpha^{7/2}\mu V_{m}^{I}e^{-i\mu t},\label{eq:pert-potential}\\
    \langle0\vert E^{I}(t,0)\vert3,1,m,2\rangle & \simeq-i\frac{1}{\sqrt{30\pi}}\left(\frac{10}{81}+32\mathcal{N}\right){\alpha^{7/2}}{\mu}^{2}V_{m}^{I}e^{-i\mu t},\label{eq:pert-electric}
\end{align}
and the rate is given by
\begin{align}\label{eq:pert-rate}
    \Gamma_{3,1,m,2} & =\frac{1}{T}\sum_{X}\bigg\vert\langle\beta;X\vert\int dt\left( \mu\tilde{A}^{I}\mathcal{O}_{I}+\tilde{E}^{I}\mathcal{O}_{I}^{E}\right) \vert0;X_{0}\rangle\bigg\vert^{2}\nonumber\\
    &= \left(\frac{10}{19683\pi}\gamma_E + \frac{64}{243\pi}\gamma_E \mathcal{N} + \frac{512}{15\pi}(\gamma + \gamma_E) \mathcal{N}^2\right) \alpha^{7}\mu^4(m\Omega-\mu) \nonumber\\
    &\simeq \left( \frac{640}{59049} + \frac{4096}{729}\mathcal{N} + \frac{16384}{15}\mathcal{N}^2 \right)\alpha^{11}(m\Omega-\mu) \, .
\end{align}
Substituting $\mathcal{N}=-5/1944$, the rate collapses onto
\begin{equation}\label{eq:pert-rate-final}
    \Gamma_{3,1,m,2}=\frac{640}{177147}\,\alpha^{11}(m\Omega-\mu) \, ,
\end{equation}
in \emph{exact} agreement with the full-theory result \eqref{eq: instability rates for j=1 bound states l=2} and with the $\Lambda^-$ rate \eqref{eq:EFT-rates-lambda}. The agreement has a neat structural counterpart: completing the square in \eqref{eq:pert-rate} gives
\begin{equation}\label{eq:double-root}
    \Gamma_{3,1,m,2}=\left[\frac{640}{177147}+\frac{16384}{15}\left(\mathcal{N}+\frac{5}{1944}\right)^{2}\right]\alpha^{11}(m\Omega-\mu) \, ,
\end{equation}
so the full-theory value is the \emph{minimum} of the rate polynomial, attained precisely at our value of $\mathcal{N}$. Any incomplete evaluation of $\mathcal{N}$ necessarily overestimates the rate.

Nothing in this derivation is special to $n=3$, with one caveat. Repeating each step at general $n$ (all the ingredients remain in closed form), the combination entering the numerator of \eqref{eq:cn0-formula} becomes
\begin{equation}\label{eq:V2-plus-Sigma-dV}
    \langle n,1,m,0\vert V^{(2)}\vert n,1,m,2\rangle+\Sigma_{\delta V}
    =\frac{4\sqrt{2}}{135}\,\frac{(n-3)(n+6)\sqrt{(n^{2}-1)(n^{2}-4)}}{(n+1)(n+2)\,n^{5}}\,\alpha^{6}\mu \, ,
\end{equation}
which vanishes at $n=3$ but not beyond: the exact cancellation between near and far zone found in \eqref{eq:Sigma-dV} is an accident of the lowest overtone, and away from $n=3$ this combination must be retained in the numerator. Carrying it along, one finds
\begin{equation}\label{eq:general-n-split}
    \mathcal{N}_\mathrm{b}+\mathcal{N}_\mathrm{c}=\frac{n(n+3)}{(n+1)(n+2)}\,\mathcal{N}(n) \, ,\qquad
    \mathcal{N}_\mathrm{deg}=\frac{2}{(n+1)(n+2)}\,\mathcal{N}(n) \, ,
\end{equation}
where each piece is separately $n$-dependent, but the prefactors sum to unity by the elementary identity $n(n+3)+2=(n+1)(n+2)$, consistently with $\mathcal{N}=\mathcal{N}_\mathrm{b}+\mathcal{N}_\mathrm{c}+\mathcal{N}_\mathrm{deg}$. The total takes the closed form $\mathcal{N}(n)=-\tfrac{\sqrt{15}}{288}\sqrt{2(n^{2}-1)(n^{2}-4)/n^{7}}$, which is precisely the normalization factor that reproduces, through \eqref{eq:leading-correction-at-origin}, the worldline value \eqref{eq:lambda-new} of the $\Lambda^{-}$ eigenstate, and therefore
\begin{equation}\label{eq:pert-rate-general-n}
    \Gamma_{n,1,m,2}=\frac{16(n^{2}-1)(n^{2}-4)}{81\,n^{7}}\,\alpha^{11}(m\Omega-\mu)=\Gamma^{-}_{n,1,m}
\end{equation}
for every $n$. The perturbatively dressed $\ell=2$ state thus reproduces the worldline amplitude of the $\Lambda^{-}$ eigenstate for all $n$: the change of basis that \S\ref{sec:diagonal-fix} imported from the FKKS ansatz has now been \emph{derived}, entirely within the effective theory.\footnote{The agreement is between physical worldline amplitudes, not term by term between mode functions: in \eqref{eq:lambda-minus}, the same-$n$, $\ell=0$ admixture carries the radial profile $R_{n,2}\sim r^{2}$ and vanishes at the origin, the nonzero value \eqref{eq:lambda-new} arising instead from the longitudinal $\hat{r}$ term. The two descriptions assign different internal bookkeeping to the same state.}

The near-zone input passes a second, independent test, this one spectroscopic rather than dynamical. The corrected splitting \eqref{eq:corrected-splitting} can be confronted with the fine structure of the Proca spectrum computed in the full theory, eq.~(2.31) of~\cite{Baumann:2019eav}. The comparison, channel by channel, is shown in Table~\ref{tab:fine-structure}: the far-zone diagonal shifts alone disagree with the published spectrum in the $(\ell,j)=(0,1)$ channel, and adding the near-zone shift \eqref{eq:deltaE} restores exact agreement---while the two channels for which the response matrix predicts \emph{no} near-zone shift agree without it. The near-zone monopole term is not optional.

\begin{table}[t]
    \centering
    \renewcommand{\arraystretch}{1.5}
    \begin{tabular}{cccc}
    \toprule
    $(\ell,j)$ & far zone $\tilde{E}^{(1)}_{n,\ell}$ & near zone $\delta\tilde{E}_{n,\ell}$ & total \\
    \midrule
    $(0,1)$ & $-\tfrac{6}{n^{3}}+\tfrac{2}{n^{4}}$ & $+\tfrac{8}{3n^{3}}$ & $-\tfrac{10}{3n^{3}}+\tfrac{2}{n^{4}}$ \\
    $(1,1)$ & $-\tfrac{7}{3n^{3}}+\tfrac{2}{n^{4}}$ & $0$ & $-\tfrac{7}{3n^{3}}+\tfrac{2}{n^{4}}$ \\
    $(2,1)$ & $-\tfrac{23}{15n^{3}}+\tfrac{2}{n^{4}}$ & $0$ & $-\tfrac{23}{15n^{3}}+\tfrac{2}{n^{4}}$ \\
    \bottomrule
    \end{tabular}
    \caption{Fine-structure shifts of the $j=1$ Proca levels in units of $\alpha^{4}\mu$, in the non-relativistic variable $\tilde{E}$ (in which the universal relativistic-kinematics term $-\alpha^{4}\mu/8n^{4}$ drops out). In each channel, the total coincides exactly with the full-theory result $f_{n\ell j}\,\alpha^4 \mu/n^{3}$, with $f_{n\ell j}$ given in eq.~(2.31) of~\cite{Baumann:2019eav}: the near-zone monopole shift \eqref{eq:deltaE} is required for agreement in the $(0,1)$ channel, and the predicted null shifts in the other two channels are confirmed.}
    \label{tab:fine-structure}
\end{table}

Let us summarize. Ordinary degenerate perturbation theory, carried out with a complete set of intermediate states and supplemented by a single piece of near-zone data---the static monopole response $G_{00}$ of the horizon, fixed by matching and independently confirmed by the fine structure of the Proca spectrum---reproduces the full-theory instability rate exactly, for all $n$, with no change of basis performed by hand and no appeal to the FKKS ansatz. The expectation expressed above \eqref{eq:pert-potential}---that additional operators in the effective action are needed to fully capture the corrections---is thereby vindicated in a sharpened form: the required operator is the conservative $\mathcal{O}(r_\mathrm{s})$ monopole contact term \eqref{eq:deltaVd}, and its coefficient is not a free parameter but an output of the matching. We will come back to the broader lessons of this exercise in the next section.

\section{Discussion} \label{sec: discussion}

The point-particle EFT provides a streamlined and easy way to calculate superradiant instability rates, and we have seen that it reproduces the bound-state rates of the full theory---including the $\alpha^{11}$ rate of the $j=1$, $\ell=2$ mode, which the full theory extracts from the separable FKKS ansatz---exactly, and for every overtone. The way we arrived at this result, however, deserves a comment. Our implementation of the effective theory integrated out the graviton completely at the outset: we started from the exact equation for the mode functions in the metric resummed by the worldline mass, expanded it in powers of $\alpha$ only afterwards, and organized the resulting corrections by old-fashioned degenerate perturbation theory. This strategy is economical, but the analysis of \S\ref{sec:Perturbative-corrections} makes it clear that it largely obscures the power counting. Because the background resums all orders in $r_\mathrm{s}/r$, contributions that a manifest power counting would cleanly separate---the second-order far-zone potentials, the continuum intermediate states, and the $\mathcal{O}(r_\mathrm{s})$ near-zone contact term---enter the leading-order rate on an equal footing through the degenerate mixing, and nothing in the calculation flags ahead of time which effects can safely be neglected. Indeed, the truncations that omit the near-zone input yield answers that look deceptively reasonable.

It would therefore be very interesting to organize this calculation in a way that keeps the power counting explicit at every step: integrating out the graviton by means of the method of regions~\cite{Beneke:1997zp}, and arranging the perturbative expansion in terms of Feynman diagrams with internal graviton lines, in the spirit of non-relativistic EFTs of gravity~\cite{Goldberger:2004jt}. One would expect, for instance, the hydrogenic bound states to arise from a resummed ladder of zero-frequency graviton exchanges,\footnote{For a static worldline in the test-field limit, potential gravitons suffice: the coupling to radiation modes is controlled by $G\mu^{2}=\alpha\,\mu/M$, which vanishes in the limit considered here.} and the near-zone coefficient $G_{00}$ to remain the Wilson coefficient of the worldline bilinear \eqref{eq:monopole-op}, but we have not worked out the details of this correspondence. Every contribution to the rate would then be represented by a Feynman diagram that would carry a definite power of $\alpha$ from the start. We plan to return to this question in the near future.

This exercise has also taught us a couple of broader lessons about point-particle EFTs. First, the shape of the radial profile of bound state mode functions has an impact on the power-counting rules we use to obtain leading-order results. This seems like the sort of effect that could occur in several different point-particle EFTs, and it is a particularly difficult issue to identify. Second, static near-zone responses can enter bound-state observables at fine-structure order even when the Love numbers of the neighboring channels vanish: here, a single matched coefficient---the monopole response $G_{00}$---both corrected the fine structure and closed the instability rate.

There are a few obvious next steps for the EFT method. First, relaxing the small-angular velocity approximation of the BH spin would offer a wider range of interesting results. Second, there are frontiers accessible by the EFT method in which non-effective approaches have struggled to gain an analytic foothold.

Concerning the first direction, one would ultimately like to obtain results for BHs with
finite spins and even extremal BHs, since the instability rates
are enhanced by the larger spins and are thus phenomenologically more relevant.
The outstanding difficulty with tackling such problems is that the
differential equations do not admit a simple separable ansatz. The
most recent approach~\cite{Baumann:2019eav} used the so-called FKKS
ansatz~\cite{Frolov:2018ezx}. This separation of variables captures all three polarizations of the Proca field~\cite{Dolan:2018dqv}, but it hinges on the hidden symmetries of the Kerr--NUT--(A)dS spacetimes and thus has no obvious analogue for more general backgrounds or other rotating bodies. Since our framework never relies on separability, it is a natural candidate for making analytic progress in this regime.

Indeed, in our framework, the simple ansatz (\ref{eq:ansatz}) only
yields decoupled equations for the leading-order $1/r$ potential.
The rest of the terms are coupled among different components, but
they are treated as perturbations and need not be solved exactly.
When finite spin is introduced, we get a new length scale, the Kerr spin parameter
$a=({r_\mathrm{s}}/{2})\chi$, where $0\leqslant\chi<1$ is order $1$ (the
upper bound is the extremal case). The spin dependent terms in the
inverse metric components and the connection coefficients have either
${a^{2}}/{r^{2}}$ or ${r_\mathrm{s}a}/{r^{2}}$, so both of them
scale as $\alpha^{4}$ and can be treated as perturbations for finite
$\chi$.

However, the Wilson coefficients cannot be expanded in the same way,
since the expansion in $\alpha$ is only valid for fields at long distances
while Wilson coefficients encode short distance dynamics.
Fortunately, the matching can be performed at finite---and even extremal---spin using the formalism of~\cite{Goldberger:2020fot}, in which the composite operators are defined in a manner that makes them sensitive to the BH spin, in contrast to our method that includes spin by boosting the fields to the comoving frame.

Concerning the second direction, recall that, as mentioned in \S\ref{sec:Introduction}, a key feature of this framework is the factorization of the problem of superradiance into its essential components. Superradiance is a broad phenomenon
in which dissipation acts as a mechanism to convert rotational energy and angular momentum into radiation, and the effective theory isolates precisely these ingredients, collecting all the system-specific details into a handful of Wilson coefficients. The same effective description therefore applies to any rotating body endowed with a suitable hierarchy of scales, with only the values of the coefficients left to change. Under the right conditions, even those values can be recycled: for a Schwarzschild--de Sitter BH with sufficiently small cosmological curvature, for instance, the Wilson coefficients are unaffected at leading order, so our flat-space matching carries over unchanged and promises analytic access to a regime that has so far been amenable only to numerical treatments. We leave all of this for future work.

\section{Acknowledgments}

The authors would like to thank Horng-Sheng Chia, Kurt Hinterbichler, Ira Rothstein, and Grigory Tarnopolsky for helpful discussions. KC would also like to thank his family for their support during the development of this paper. The results discussed in \S 6 were derived with the help of Claude Opus 4.8 and Claude Fable 5. This work was supported in part by the Department of Energy Award DE-SC0010118 (KC and RP) and DE- FG02-04ER41338 and FG02- 06ER41449 (JL)

\begin{appendix}

\section{Worldline Wightman Functions\label{sec:Worldline-Wightman-functions}}

In this appendix, we derive the general properties of the worldline Wightman functions defined in \eqref{eq:Wightman-def}, which are used throughout \S\ref{sec: eft}. Given two Hermitian operators, their correlation function at different times is given by
\begin{equation}
    \langle\mathcal{O}(t)\tilde{\mathcal{O}}(0)\rangle=\int\frac{d\omega}{2\pi}W(\omega)e^{-i\omega t}.
\end{equation}
Because the operators are Hermitian, they satisfy
\begin{equation}
    \label{eq: conjugate Wightman}
    \langle\tilde{\mathcal{O}}(0)\mathcal{O}(t)\rangle=\int\frac{d\omega}{2\pi}W^{*}(-\omega)e^{-i\omega t}.
\end{equation}
For our problem, the fact that black holes are classically at zero absolute temperature means that the expectation value is in the ground state. This, with unitary time evolution of the internal degrees of freedom, implies that
\begin{equation}
    \langle \mathcal{O}(t)\tilde{\mathcal{O}}(0)\rangle = \sum_{n} e^{-i(E_n-E_0)t}\langle0\vert\mathcal{O}(0)\vert n\rangle\langle n\vert\tilde{\mathcal{O}}(0)\vert0\rangle,
\end{equation}
so the Fourier transform only has support for positive frequencies, or
\begin{equation}
    \label{eq: Wightman support}
    W(\omega)=\theta(\omega)F(\omega)
\end{equation}
for some unknown complex function $F$.

The Pauli-Jordan function or spectral density~\cite{Arteaga:2008ux} is defined by
\begin{equation}
    \langle [\mathcal{O}(t), \tilde{\mathcal{O}}(0)] \rangle = \int\frac{d\omega}{2\pi}\Delta_-(\omega)e^{-i\omega t}.
\end{equation}
It possesses an obvious relationship with the Wightman function,
\begin{equation}
    \label{eq: Pauli-Jordan form}
    \Delta_-(\omega) = W(\omega) - W^*(-\omega),
\end{equation}
which can be verified by expanding the commutator and applying the identity \eqref{eq: conjugate Wightman}.

The composite operators describe small-distance, internal degrees of freedom. If the object is thermalized, correlation functions of the composite operators are believed to decay at large temporal separations faster than any power as the thermal average erases information about the internal state. Consequently, the Fourier transform describing the spectral density
\begin{equation}
    \Delta_-(\omega) = \int dt\, e^{i\omega t} \langle [\mathcal{O}(t),\tilde{\mathcal{O}}(0)]\rangle
\end{equation}
converges at $\omega = 0$. Moreover, the integral still converges after taking any number of derivatives with respect to $\omega$; the spectral density function is analytic at $\omega = 0$~\cite{Endlich:2012vt}.

\subsection{Autocorrelations}

We now specialize to correlations of a single operator, $\tilde{\mathcal{O}}=\mathcal{O}$. Even when the expectation value is not taken in the ground state, the corresponding Wightman function $W(\omega)$ must be real and non-negative. To see this, consider the Fourier transform of the Hermitian operator $\mathcal{O}(t)$:
\begin{equation}
    \mathcal{O}(\omega) = \int dt \, e^{i\omega t} \mathcal{O}(t).
\end{equation}
From the definition, it is clear that $\mathcal{O}^\dag(\omega) = \mathcal{O}(-\omega)$. Furthermore,
\begin{align}
    \langle \mathcal{O}(\omega) \mathcal{O}(-\omega) \rangle 
    &= \int dt' dt \, e^{i \omega(t - t')} \langle \mathcal{O}(t) \mathcal{O}(t') \rangle \nonumber\\
    &= \int \frac{d\omega'}{2\pi} W(\omega') \big(2 \pi \delta(\omega' - \omega)\big)^2 = 2\pi\delta(0) W(\omega).
\end{align}
But that means
\begin{equation}
    W(\omega) = \frac{\langle \mathcal{O}(\omega) \mathcal{O}(-\omega) \rangle}{2\pi \delta(0)} = \frac{\langle \mathcal{O}(\omega) \mathcal{O}^\dag(\omega) \rangle}{2\pi \delta(0)} \geqslant 0,
\end{equation}
where the last step follows because the numerator is the squared norm of the state $\mathcal{O}^\dagger(\omega)\vert X_0\rangle$, while $2\pi\delta(0)=T$ is the (positive) total observation time. Since $W(\omega)$ is real, the relation \eqref{eq: Pauli-Jordan form} reduces to
\begin{equation}
    \Delta_-(\omega) = W(\omega)-W(-\omega),
\end{equation}
which implies with \eqref{eq: Wightman support} that $W(\omega) = \theta(\omega) \Delta_-(\omega)$, so the Wightman function and the Pauli-Jordan function are essentially the same. We make use of this fact by using quantities obtained by matching absorption probabilities, proportional to $W(\omega - m \Omega)$, in the calculation of emission rates, proportional to $\Delta_-(m\Omega - \omega)$.

\subsection{Opposite spatial inversion parity}

If the ground state is invariant under parity, the Wightman functions
of opposite parity operators vanish straightforwardly as
\begin{equation}
\langle\mathcal{O}(t)\tilde{\mathcal{O}}(0)\rangle=\langle\hat{P}\mathcal{O}(t)\hat{P}^{-1}\hat{P}\tilde{\mathcal{O}}(0)\hat{P}^{-1}\rangle=-\langle\mathcal{O}(t)\tilde{\mathcal{O}}(0)\rangle=0.
\end{equation}
This relies on the fact that the worldline can be chosen to be located
at the spatial origin, which makes it insensitive to any spatial
parity transformations.

\subsection{Opposite time reversal parity}

However, this argument does not apply to time reversal since the worldline
operators also pick up a negative sign in their temporal arguments
under time reversal. Nevertheless, we can still assume that the expectation
values are invariant under time reversal $\hat{T}$, which is true
if the ground state is unique, since
\begin{equation}
    \hat{T}\hat{H}\vert X_{0}\rangle=\hat{T}\hat{H}\hat{T}^{-1}\hat{T}\vert X_{0}\rangle=\hat{H}\hat{T}\vert X_{0}\rangle=0.
\end{equation}
Uniqueness means that $\hat{T}\vert X_{0}\rangle=e^{i\theta}\vert X_{0}\rangle$
can differ by a phase at most, and it would not affect any correlation
functions. To make explicit the antilinearity of $\hat{T}$, we write
the correlation function as
\begin{align}
    \langle X_{0}\vert\mathcal{O}(t)\tilde{\mathcal{O}}(0)X_{0}\rangle & =\langle\hat{T}X_{0}\vert\mathcal{O}(t)\tilde{\mathcal{O}}(0)\hat{T}X_{0}\rangle\nonumber\\
    & =\langle X_{0}\vert\hat{T}^{\dagger}\mathcal{O}(t)\tilde{\mathcal{O}}(0)\hat{T}X_{0}\rangle^{*}\nonumber\\
    & =\langle X_{0}\vert\hat{T}^{-1}\mathcal{O}(t)\tilde{\mathcal{O}}(0)\hat{T}X_{0}\rangle^{*}.
\end{align}
By inserting $\hat{T}^{-1}\hat{T}$ between the operators, we find that
\begin{equation}
    \langle\mathcal{O}(t)\tilde{\mathcal{O}}(0)\rangle=\langle \hat{T}^{-1}\hat{T}\mathcal{O}(t)\hat{T}^{-1}\hat{T}\tilde{\mathcal{O}}(0)\hat{T}^{-1}\hat{T}\rangle^{*}=-\langle\mathcal{O}(-t)\tilde{\mathcal{O}}(0)\rangle^{*}.
\end{equation}
This means that the spectral function satisfies
\begin{equation}
    W(\omega)=-W^{*}(\omega),
\end{equation}
which makes $W$ purely imaginary. Combined with the support property \eqref{eq: Wightman support}, it takes the form
\begin{equation}
W(\omega)=i\theta(\omega)\tilde{F}(\omega),
\end{equation}
for some real function $\tilde{F}(\omega)$.
This implies with \eqref{eq: Pauli-Jordan form} that the Pauli-Jordan function is even,
taking the form
\begin{equation}
    \Delta_{-}(\omega) =i\bigl(\theta(\omega)\tilde{F}(\omega)+\theta(-\omega)\tilde{F}(-\omega)\bigr)=i \tilde{F}(|\omega|).
\end{equation}
The Pauli-Jordan function of two operators with opposite time-reversal parity is therefore an even function of $\omega$. In particular, if it is analytic at $\omega=0$, it contains no linear term, so such a correlator cannot contribute at the linear order in $\omega$ used in the matching of \S\ref{sec: results}. This is the constraint on $\langle\mathcal{O}_{E}\mathcal{O}'\rangle$ invoked in \S\ref{sec: EFT rates}.

As a simple illustration, consider the conjugate variables $\hat{x}$ and $\hat{p}$ of a harmonic oscillator, which carry opposite time-reversal parity and whose ground state is invariant under $\hat{T}$. A direct computation gives $\langle0\vert\hat{x}(t)\hat{p}(0)\vert0\rangle=\tfrac{i}{2}\,e^{-i\omega_{0}t}$, \textit{i.e.}, $W(\omega)=i\pi\delta(\omega-\omega_{0})$: purely imaginary and supported at positive frequencies, in accordance with the general arguments above.

\section{Bases of Vector Spherical Harmonics\label{sec:Vector-spherical-harmonics}}

There are many commonly encountered bases for the vector spherical harmonics~\cite{Thorne:1980ru}. 
First, we examine the ``pure-spin'' basis:
\begin{align}
\vec{Y}^{E,jm} &=\frac{r\vec{\nabla}Y^{jm}}{\sqrt{j(j+1)}},\\
\vec{Y}^{B,jm} &=\frac{\vec{r}\times\vec{\nabla}Y^{jm}}{\sqrt{j(j+1)}},\\
\vec{Y}^{R,jm} &=Y^{jm}\hat{r}.
\end{align}
This is a convenient basis for outgoing waves, and the easiest one to differentiate. However, the pure-spin harmonics are not eigenfunctions of the Laplacian, which is why the bound states use the ``pure-orbital'' harmonics $\vec{Y}_{\ell,jm}$ instead, satisfying $-r^2\nabla^2\vec{Y}_{\ell,jm}=\ell(\ell+1)\vec{Y}_{\ell,jm}$. These harmonics can be decomposed into scalar spherical harmonics by
\begin{equation}\label{eq:vector-spherical-harmonic-clebsch}
    \vec{Y}_{\ell, j m} = \sum_{m_s=-1}^1 \sum_{m_\ell=-\ell}^\ell C_{1, \ell}(j, m; m_s, m_\ell) \vec{V}_{m_s} Y^{\ell m_\ell}
\end{equation}
with $C_{s\ell}(j,m; m_s, m_\ell)$ a Clebsch-Gordan coefficient (here with $s=1$). The constant rotational eigenvector $\vec{V}_m$ above satisfies $R_z(\varphi)^\intercal\vec{V}_m = e^{im\varphi}\vec{V}_m$ (where $R_z(\varphi)$ is an $\operatorname{SO}(3)$ rotation about the $z$-axis by the angle $\varphi$) and is given by 
\begin{equation}\label{eq:rotation-eigenvectors}
    \vec{V}_0 = \left(
    \begin{array}{c}
        0 \\
        0 \\
        1
    \end{array}
    \right),
    \qquad
    \vec{V}_{\pm1} = \frac{1}{\sqrt{2}}\left(
    \begin{array}{c}
        \mp1 \\
        -i \\
        0
    \end{array}
    \right)
\end{equation}
in the standard Cartesian basis. 
The two bases for vector spherical harmonics are related by
\begin{equation}
    \left(\begin{array}{c}
    \vec{Y}_{j+1,jm}\\
    \vec{Y}_{j,jm}\\
    \vec{Y}_{j-1,jm}
    \end{array}\right)=\frac{1}{\sqrt{2j+1}}\left(\begin{array}{ccc}
    -\sqrt{j+1} & 0 & \sqrt{j}\\
    0 & -i\sqrt{2j+1} & 0\\
    \sqrt{j} & 0 & \sqrt{j+1}
    \end{array}\right)\left(\begin{array}{c}
    \vec{Y}^{R,jm}\\
    \vec{Y}^{B,jm}\\
    \vec{Y}^{E,jm}
    \end{array}\right).
\end{equation}
Furthermore, these are related to the Newman-Penrose basis (for
3-D space) by
\begin{equation}
\left(\begin{array}{c}
\vec{Y}^{R,jm}\\
\vec{Y}^{B,jm}\\
\vec{Y}^{E,jm}
\end{array}\right)=\frac{1}{\sqrt{2}}\left(\begin{array}{ccc}
\sqrt{2}Y^{jm} & 0 & 0\\
0 & -i_{-1}Y^{jm} & -i_{1}Y^{jm}\\
0 & _{-1}Y^{jm} & -_{1}Y^{jm}
\end{array}\right)\left(\begin{array}{c}
\hat{r}\\
\hat{m}\\
\hat{m}^{*}
\end{array}\right),
\end{equation}
where the Newman-Penrose basis is defined by
\begin{equation}
\left(\begin{array}{c}
\hat{r}\\
\hat{m}\\
\hat{m}^{*}
\end{array}\right)=\frac{1}{\sqrt{2}}\left(\begin{array}{ccc}
\sqrt{2} & 0 & 0\\
0 & 1 & i\\
0 & 1 & -i
\end{array}\right)\left(\begin{array}{c}
\hat{e}'_{r}\\
\hat{e}'_{\theta}\\
\hat{e}'_{\varphi}
\end{array}\right),
\end{equation}
with the normalized spherical polar basis vectors $\left(\hat{e}'_{r},\hat{e}'_{\theta},\hat{e}'_{\varphi}\right)=\left(\partial_{r},\frac{1}{r}\partial_{\theta},\frac{1}{r\sin\theta}\partial_{\varphi}\right)$. The Newman-Penrose basis will be useful in Appendix~\ref{sec:Degeneracies}.

Combining the results, the transformation rules are given by
\begin{align}
\left(\begin{array}{c}
\vec{Y}_{j+1,jm}\\
\vec{Y}_{j,jm}\\
\vec{Y}_{j-1,jm}
\end{array}\right) 
= \frac{1}{\sqrt{2j+1}}\left(\begin{array}{ccc}
-\sqrt{j+1}Y^{jm} & -\sqrt{j}Y_{-}^{jm} & i\sqrt{j}Y_{+}^{jm}\\
0 & -\sqrt{2j+1}Y_{+}^{jm} & i\sqrt{2j+1}Y_{-}^{jm}\\
\sqrt{j}Y^{jm} & -\sqrt{j+1}Y_{-}^{jm} & i\sqrt{j+1}Y_{+}^{jm}
\end{array}\right)\left(\begin{array}{c}
\hat{e}'_{r}\\
\hat{e}'_{\theta}\\
\hat{e}'_{\varphi}
\end{array}\right),
\end{align}
where the new basis of spherical harmonics is defined as
\begin{equation}\label{eq:Ypm-def}
Y_{+}^{jm}=\frac{1}{2}\left(_{1}Y^{jm}+{}_{-1}Y^{jm}\right),\ Y_{-}^{jm}=\frac{1}{2}\left(_{1}Y^{jm}-{}_{-1}Y^{jm}\right).
\end{equation}
The Schwarzschild connection coefficients and hence the differential
equations that we use are in the standard $\left(\vec{e}_{r},\vec{e}_{\theta},\vec{e}_{\varphi}\right)=\left(\partial_{r},\partial_{\theta},\partial_{\varphi}\right)$
basis. The final step is rescaling the angular basis vectors
to find
\begin{align}\label{eq:vsh-coordinate-basis}
\left(\begin{array}{c}
\vec{Y}_{j+1,jm}\\
\vec{Y}_{j,jm}\\
\vec{Y}_{j-1,jm}
\end{array}\right) & =\frac{1}{\sqrt{2j+1}}\left(\begin{array}{ccc}
-\sqrt{j+1}Y^{jm} & -\frac{\sqrt{j}}{r}Y_{-}^{jm} & \frac{i\sqrt{j}}{r\sin\theta}Y_{+}^{jm}\\
0 & -\frac{\sqrt{2j+1}}{r}Y_{+}^{jm} & \frac{i\sqrt{2j+1}}{r\sin\theta}Y_{-}^{jm}\\
\sqrt{j}Y^{jm} & -\frac{\sqrt{j+1}}{r}Y_{-}^{jm} & \frac{i\sqrt{j+1}}{r\sin\theta}Y_{+}^{jm}
\end{array}\right)\left(\begin{array}{c}
\vec{e}_{r}\\
\vec{e}_{\theta}\\
\vec{e}_{\varphi}
\end{array}\right).
\end{align}

\section{Bound State Mode Functions\label{sec:Mode-functions}}

For vector fields with spin 1, we expect the angular quantum
numbers to be related by $\ell=j,j\pm1$. Using Appendix~\ref{sec:Vector-spherical-harmonics} to change the spherical harmonic basis~\cite{Thorne:1980ru},
we may express the mode functions $\vec{f}_{njm\ell} = R_{n\ell} \vec{Y}_{\ell,jm}$ in terms of the scalar spherical harmonic $Y^{jm}$ as 
\begin{align} 
    \vec{f}_{njm(j-1)} &= \frac{1}{\sqrt{j(2j+1)}}R_{n(j-1)}\left(r\vec{\nabla}Y^{jm}+jY^{jm}\hat{r}\right),\\
    \vec{f}_{njmj} &= -i \frac{1}{\sqrt{j(j+1)}}R_{nj}\,\vec{r}\times\vec{\nabla}Y^{jm}, \\
    \vec{f}_{njm(j+1)} &= \frac{1}{\sqrt{(j+1)(2j+1)}}R_{n(j+1)}\left(r\vec{\nabla}Y^{jm}-(j+1)Y^{jm}\hat{r}\right).
\end{align}
Then we can easily take the divergence and curl, yielding
\begin{align}
    \vec{\nabla}\cdot\vec{f}_{njm(j-1)} &= \sqrt{\frac{j}{2j+1}}\left(\frac{dR_{n(j-1)}}{dr} - (j-1)\frac{R_{n(j-1)}}{r}\right)Y^{jm}, \\
    \vec{\nabla}\cdot\vec{f}_{njmj} &= 0, \\
    \vec{\nabla}\cdot\vec{f}_{njm(j+1)} &= -\sqrt{\frac{j+1}{2j+1}}\left(\frac{dR_{n(j+1)}}{dr} + (j+2)\frac{R_{n(j+1)}}{r}\right)Y^{jm}, \\
    \vec{\nabla}\times\vec{f}_{njm(j-1)} &= \frac{1}{\sqrt{j\left(2j+1\right)}}\left( \frac{dR_{n(j-1)}}{dr}-(j-1)\frac{R_{n(j-1)}}{r}\right) \,\vec{r}\times\vec{\nabla}Y^{jm}, \\
    \vec{\nabla}\times\vec{f}_{njmj} &= \frac{i}{\sqrt{j(j+1)}}\Biggl( \left(\frac{dR_{nj}}{dr} + \frac{R_{nj}}{r}\right)\,r\vec{\nabla}Y^{jm} + j(j+1)\frac{R_{nj}}{r}\,Y^{jm}\hat{r}\Biggr), \\
    \vec{\nabla}\times\vec{f}_{njm(j+1)} &= \frac{1}{\sqrt{\left(j+1\right)\left(2j+1\right)}}\left(\frac{dR_{n(j+1)}}{dr} + (j+2)\frac{R_{n(j+1)}}{r}\right) \,\vec{r}\times\vec{\nabla}Y^{jm}.
\end{align}
Finally, the gradients of the nonvanishing divergences are
\begin{align}
    \vec{\nabla}\left(\vec{\nabla}\cdot\vec{f}_{njm(j-1)}\right) &= \sqrt{\frac{j}{2j+1}}\left( \frac{1}{r} \frac{dR_{n(j-1)}}{dr} - (j-1)\frac{R_{n(j-1)}}{r^2} \right) r\vec{\nabla}Y^{jm} \\
    & \qquad {} + \sqrt{\frac{j}{2j+1}} \left( \frac{d^2R_{n(j-1)}}{dr^2} - (j-1) \frac{1}{r} \frac{dR_{n(j-1)}}{dr} + (j-1) \frac{R_{n(j-1)}}{r^2}\right) Y^{jm} \hat{r} \nonumber,\\
    \vec{\nabla}\left(\vec{\nabla}\cdot\vec{f}_{njm(j+1)}\right) &= -\sqrt{\frac{j+1}{2j+1}}\left( \frac{1}{r} \frac{dR_{n(j+1)}}{dr} + (j+2)\frac{R_{n(j+1)}}{r^2} \right) r\vec{\nabla}Y^{jm} \\
    & \qquad -\sqrt{\frac{j+1}{2j+1}} \left( \frac{d^2R_{n(j+1)}}{dr^2}+ (j+2) \frac{1}{r} \frac{dR_{n(j+1)}}{dr} - (j+2) \frac{R_{n(j+1)}}{r^2}\right) Y^{jm} \hat{r}. \nonumber
\end{align}
All possible vector quantities that can be constructed linearly from $\vec{f}_{njm\ell}$ and its Laplacian-free derivatives are represented above.

\subsection{Point particle limit}

Because we will ultimately be taking the $r\to0$ limit where $R_{n\ell} \sim r^\ell$, we will consider $\vec{f}_{njm\ell}$ for $\ell=0$, $\vec{\nabla}\times\vec{f}_{njm\ell}$ for $\ell=0,1$, and $\vec{\nabla}\bigl(\vec{\nabla}\cdot\vec{f}_{njm\ell}\bigr)$ for $\ell=0,1,2$ (the $\ell=1$ case vanishes identically, since the $\ell=j$ mode is divergence-free); any larger values of $\ell$ must leave factors of $r$ behind that cause the expression to vanish in the limit. Next, we will focus on the dominant $j=1$ results and consider the constant vector
\begin{equation}
    r\vec{\nabla}Y^{1,m} + Y^{1,m}\hat{r} = \sqrt{3}\,\vec{Y}_{0;1m} = \sqrt{\frac{3}{4\pi}}\, \vec{V}_m.
\end{equation}
The second equation can be obtained from \eqref{eq:vector-spherical-harmonic-clebsch}. This allows us to express the mode functions and their derivatives in terms of the vectors $\vec{V}_m$, $\hat{r}$, and $\hat{r} \times \vec{V}_m$. 

The zero-derivative case is quite simple,
\begin{equation}\label{eq:0-derivatives}
    \vec{f}_{n,1,m,0} = \frac{1}{\sqrt{4\pi}} R_{n,0} \vec{V}_m.
\end{equation}
With one derivative, we have
\begin{align}
    \vec{\nabla}\times\vec{f}_{n,1,m,0} &= \frac{1}{\sqrt{4\pi}} \frac{dR_{n,0}}{dr} \,\hat{r} \times \vec{V}_m, \label{eq:1-derivative-subleading}\\
    \vec{\nabla}\times\vec{f}_{n,1,m,1} &= i \sqrt{\frac{3}{8\pi}}\left( \frac{dR_{n,1}}{dr} + \frac{R_{n,1}}{r} \right) \vec{V}_m - \frac{i}{\sqrt{2}} \left( \frac{dR_{n,1}}{dr} - \frac{R_{n,1}}{r} \right) Y^{1,m} \hat{r}.\label{eq:1-derivative}
\end{align}
Finally, with two derivatives, 
\begin{align}
    \vec{\nabla}\left(\vec{\nabla}\cdot\vec{f}_{n,1,m,0}\right) &= \frac{1}{\sqrt{4\pi}} \frac{1}{r}\frac{dR_{n,0}}{dr} \vec{V}_m + \frac{1}{\sqrt{3}}\left( \frac{d^2R_{n,0}}{dr^2} - \frac{1}{r}\frac{dR_{n,0}}{dr}\right) Y^{1,m}\hat{r}, \label{eq:2-derivatives-singular}\\
    \vec{\nabla}\left(\vec{\nabla}\cdot\vec{f}_{n,1,m,2}\right) &= -\frac{1}{\sqrt{2\pi}} \left(\frac{1}{r}\frac{dR_{n,2}}{dr} + 3\frac{R_{n,2}}{r^2}\right) \vec{V}_m - \sqrt{\frac{2}{3}}\left( \frac{d^2R_{n,2}}{dr^2} + 2 \frac{1}{r}\frac{dR_{n,2}}{dr} - 6 \frac{R_{n,2}}{r^2}\right) Y^{1,m}\hat{r}. \label{eq:2-derivatives}
\end{align}

We now face two problems. If the terms $Y^{1,m}\hat{r}$ and $\hat{r}\times\vec{V}_m$ with angular dependence persist as $r\to0$, the limit does not exist, since approaching the origin from different directions gives the function a different value. Worse, we find that \eqref{eq:2-derivatives-singular} scales asymptotically like $r^{-1}$. This latter issue is solved by perturbation theory. The radial Coulomb function defined in \eqref{eq:radial-mode} is a power series in $\alpha \mu r$, so each spatial derivative scales like $\alpha \ll 1$. This means that the leading-order result for the state $\lvert n,1,m,0\rangle$ uses the function $\vec{f}_{n,1,m,0} \sim \alpha^{3/2}$ with no derivatives, and the other two terms $\nabla \times \vec{f}_{n,1,m,0}\sim\alpha^{5/2}$ and $\nabla(\nabla \cdot \vec{f}_{n,1,m,0})\sim\alpha^{7/2}$ contribute towards loop corrections to the transition amplitude. New loop-level terms in the EFT should take care of the divergences. In our tree-level EFT, we can safely ignore the subleading terms.

The issue of angular dependence can be resolved by being a bit more savvy about taking the $r\to0$ limit. In the point particle action, fields are localized to the particle position $\vec{y}$ by employing a Dirac delta function,
\begin{equation}
    F\bigl(\vec{y}\bigr) = \int d^3{x}\,\delta^3\bigl(\vec{x}-\vec{y}\bigr) F(\vec{x}).
\end{equation}
The delta function is isotropic; consider it as a limit of sharply peaked Gaussian distributions, which are functions of $\lvert \vec{x}\rvert^2 = r^2$ alone. Consequently, when moving to spherical coordinates, we can make the substitution
\begin{equation}
    \delta^{3}(\vec{x}) = \frac{\delta(r)}{4\pi r^2}.
\end{equation}
In the case $F(\vec{x}) = R(r)$, this preserves the expected result
\begin{equation}
    \int d^3x\, F(\vec{x})\delta^3(\vec{x}) = \int d\Omega \int dr\, r^2 R(r) \frac{\delta(r)}{4\pi r^2} = R(0).
\end{equation}
Now, if the function has nontrivial angular dependence at the origin, we know exactly what to do. Indeed, in the simple case of a separable function $F(\vec{x}) = R(r) Y(\theta,\varphi)$, we have
\begin{equation}
    \int d^3x\, F(\vec{x}) \delta^3(\vec{x}) = \frac{R(0)}{4\pi} \int d\Omega\, Y(\theta,\varphi),
\end{equation}
so a field localized to the worldline at $r=0$ is simply averaged over the sphere to remove any angular dependence.

Using the facts that the pure-orbit vector spherical harmonics are orthonormal with respect to integrals over the sphere and that $\vec{V}_m = \sqrt{4\pi} \vec{Y}_{0;1m}$ for $m=-1,0,1$ is a basis of constant unit vectors, we can expand $\vec{Y}_{\ell,jm}$ in that basis to take its average as
\begin{equation}
    \int d{\Omega}\, \vec{Y}_{\ell,jm} = \sum_{m'}\vec{V}_{m'} \int d\Omega\,\vec{Y}_{\ell,jm} \cdot \vec{V}^{*}_{m'} = \sqrt{4\pi} \vec{V}_m \delta_{\ell,0}\delta_{j,1}. \label{eq:int-vsh}
\end{equation}
Thus, only vector spherical harmonics with nonvanishing projection onto $\vec{Y}_{0;1m}$ contribute on the worldline. Consider vector fields of the type $\vec{f}(\vec{x}) = R(r) Y^{jm}\hat{r}$. Only when $j=1$, we have $Y^{1,m}\hat{r} = \vec{Y}_{0;1m}/\sqrt3 - \vec{Y}_{2;1m}\sqrt{2/3}$, so
\begin{equation}
    \int d^3{x}\,\delta^3(\vec{x}) R(r) Y^{jm}\hat{r} = \frac{1}{2\sqrt{3\pi}} R(0) \vec{V}_m \delta_{j,1} \label{eq:avg-radial-vsh}
\end{equation}
on the worldline. This vanishes for modes with $j\ne1$. 
On the other hand, $\vec{r}\times\nabla Y^{jm} = i\sqrt{j(j+1)} \vec{Y}_{j,jm}$ is orthogonal to $\vec{Y}_{0;1m}$ for all values of $j$. This means the fields proportional to $\hat{r}\times\vec{V}_m$ vanish on the worldline; equivalently, this happens because the average value of $\hat{r}$ on the sphere is 0. 

For the equations \eqref{eq:1-derivative} and \eqref{eq:2-derivatives}, this analysis turns out to be moot. Writing $R_{n,1}(r) \sim R'_{n,1}(0) r$ and $R_{n,2}(r) \sim \frac{1}{2}R''_{n,2}(0) r^2$, we find that the terms proportional to $Y^{1,m} \hat{r}$ vanish anyway. The field \eqref{eq:1-derivative-subleading} vanishes as well, and \eqref{eq:2-derivatives-singular} is irrelevant at tree level, so all that remains are three fields proportional to $\vec{V}_m$. However, we do use this result in \S\ref{sec:diagonal-fix} to work with the correction to the mode function discussed there.

For completeness, we use \eqref{eq:radial-mode-0} for
\begin{align}
    R_{n,0}(0) &= \sqrt{\frac{4(\alpha\mu)^{3}}{n^3}}, \label{eq:R0} \\
    R_{n,1}'(0) &= \sqrt{\frac{4(n^2-1)(\alpha\mu)^5}{9n^5}}, \label{eq:R1} \\
    \frac{1}{2} R_{n,2}''(0) &= \sqrt{\frac{(n^2-1)(n^2-4)(\alpha\mu)^7}{225 n^7}}, \label{eq:R2}
\end{align}
from which the results \eqref{eq:modes-at-0} follow.

\section{Diagonality of the Perturbation in the Degenerate Subspace\label{sec:Degeneracies}}

The goal of this appendix is twofold. First, we derive the identity \eqref{eq:angular-divergence}, used in \S\ref{sec:pert-DPT} to reduce the matrix elements of the perturbation to radial integrals. Second, we show that the leading far-zone perturbation $V^{(1)}$ of \eqref{eq:pert} is diagonal within each degenerate multiplet: for a given $n$,
\begin{equation}
\int d\Omega\, dr\,r^{2}\,f_{n,j',m',\ell'}^{*I}\,V^{(1)}_{IJ}\,f_{n,j,m,\ell}^{J}\propto\delta_{jj'}\delta_{\ell\ell'}\delta_{mm'},
\end{equation}
as claimed in \S\ref{sec:pert-DPT}.
Using \eqref{eq:pert-reduced}, the integrand may be decomposed into scalars on the 2D sphere as
\begin{align}
 r^2 f_{n,j',m',\ell'}^{*I}V^{(1)}_{IJ}f_{n,j,m,\ell}^{J} &= r^2R_{n\ell'}\hat{d}(r,\partial_{r})R_{n\ell}\vec{Y}_{\ell,jm}\cdot\vec{Y}^{*}_{\ell',j'm'} \nonumber\\
 &\qquad -\frac{3\alpha R_{n\ell}R_{n\ell'}}{\mu^2 r}Y_{\ell',j'm'}^{*r}\left( Y_{\ell,jm}^{r}+r\partial_{\theta}Y_{\ell,jm}^{\theta}+r\cot\theta\, Y_{\ell,jm}^{\theta}+r\partial_{\varphi}Y_{\ell,jm}^{\varphi}\right),\label{eq: matel}
\end{align}
where $\hat{d}(r,\partial_{r})=\frac{r_\mathrm{s}}{2\mu r}\left(2\partial_{r}^{2}+\frac{3}{r}\partial_{r}-\frac{\ell^{2}+\ell+1}{r^{2}}\right)$ collects the diagonal elements of the potential matrix, and the radial functions $R_{n\ell}$ are real, so the conjugation acts only on the primed harmonic. Since $\int d\Omega\vec{Y}^*_{\ell,jm}\cdot\vec{Y}_{\ell',j'm'}\propto\delta_{\ell\ell'}\delta_{jj'}\delta_{mm'}$,
the first term on the right-hand side needs no further analysis. The second is related to the divergence of a vector field on the sphere, so we can use the language of Newman-Penrose derivatives~\cite{10.1063/1.1931221} to rewrite it in terms of scalar spherical harmonics $Y^{jm}$. Below, we find the expression to be proportional to $(Y^{j'm'})^{*}Y^{jm}$, so its spherical integral is proportional to $\delta_{jj'}\delta_{mm'}$, and it turns out that the radial integral provides a factor of $\delta_{\ell\ell'}$. First, the Newman-Penrose derivatives 
are given by
\begin{align}
\text{ð}\eta & =-\left(\partial_{\theta}-s\cot\theta+\frac{i}{\sin\theta}\partial_{\varphi}\right)\eta\\
\bar{\text{ð}}\eta & =-\left(\partial_{\theta}+s\cot\theta-\frac{i}{\sin\theta}\partial_{\varphi}\right)\eta
\end{align}
for a function of spin weight $s$. We may re-express the angular derivatives as
\begin{align}
\partial_{\theta}\eta & =-\frac{1}{2}\left(\text{ð}+\bar{\text{ð}}\right)\eta,\\
\frac{i}{\sin\theta}\partial_{\varphi}\eta & =\frac{1}{2}\left(-\text{ð}+\bar{\text{ð}}\right)\eta+s\cot\theta\eta.
\end{align}
Therefore, using the results from Appendix~\ref{sec:Vector-spherical-harmonics}, we can evaluate part of the sum in the parentheses of (\ref{eq: matel}) as
\begin{equation}
r\partial_{\theta}Y_{\ell,jm}^{\theta}+r\cot\theta \,Y_{\ell,jm}^{\theta}+r\partial_{\varphi}Y_{\ell,jm}^{\varphi} = \begin{cases}
\frac{\sqrt{j}}{2\sqrt{2j+1}}\left(\bar{\text{ð}}\,{}_{1}Y^{jm}-\text{ð}\,{}_{-1}Y^{jm}\right) & \ell=j+1,\\
\frac{1}{2}\left(\bar{\text{ð}}\,{}_{1}Y^{jm}+\text{ð}\,{}_{-1}Y^{jm}\right) & \ell=j,\\
\frac{\sqrt{j+1}}{2\sqrt{2j+1}}\left(\bar{\text{ð}}\,{}_{1}Y^{jm}-\text{ð}\,{}_{-1}Y^{jm}\right) & \ell=j-1,
\end{cases}
\end{equation}
and using the relations~\cite{10.1063/1.1705135} between spin-weighted spherical harmonics
and $\left(\text{ð},\bar{\text{ð}}\right)$ derivatives
\begin{align}
    \text{ð}\,{}_{s}Y^{jm} & =\sqrt{(j-s)(j+s+1)}\,{}_{s+1}Y^{jm},\\
    \bar{\text{ð}}\,{}_{s}Y^{jm} & =-\sqrt{(j+s)(j-s+1)}\,{}_{s-1}Y^{jm},
\end{align}
we can simplify the full expression including $Y^r_{\ell,jm}$ to
\begin{equation}
Y^r_{\ell,jm} + r\partial_{\theta}Y_{\ell,jm}^{\theta}+r\cot\theta\, Y_{\ell,jm}^{\theta}+r\partial_{\varphi}Y_{\ell,jm}^{\varphi} =
\begin{cases}
    -\frac{(j+1)\sqrt{j+1}}{\sqrt{2j+1}}Y^{jm} & \ell=j+1,\\
    0 & \ell=j,\\
    -\frac{j\sqrt{j}}{\sqrt{2j+1}}Y^{jm} & \ell=j-1.
\end{cases}
\end{equation}
Overall, the form is a scalar spherical
harmonic $Y^{jm}$ multiplied by functions of the $j$ quantum number.
Since $Y_{\ell',j'm'}^{*r}$ is proportional to $Y^{*j'm'}$, which is orthogonal to $Y^{jm}$, we conclude that the spherical integral is indeed proportional to $\delta_{jj'}\delta_{mm'}$.

Finally, to make sure the overall result is proportional to $\delta_{\ell\ell'}$,
we only need to verify that the overlap between the $\ell=j\pm1$ modes vanishes,
since the whole expression vanishes when $\ell = j$ or $\ell'=j'$. The radial integral is proportional to an integral
over generalized Laguerre polynomials~\cite{abramowitz+stegun}  $L_{n-\ell-1}^{2\ell+1}(\rho)$, where $\rho = 2\alpha\mu r/n$, so we can use their orthogonality relation to write
\begin{align}
    \int_0^\infty dr\,\frac{R_{n(j-1)}R_{n(j+1)}}{r}& \propto \int_0^\infty d\rho\, e^{-\rho}\rho^{2j-1}L_{n-j-2}^{2j+3}(\rho)L_{n-j}^{2j-1}(\rho)\nonumber\\
    &= \sum_{\nu=0}^{n-j-2}\left(\begin{array}{c}
        n-j-\nu+1\\
        3
    \end{array}\right)\int_0^\infty d\rho\, e^{-\rho}\rho^{2j-1}L_{\nu}^{2j-1}(\rho)L_{n-j}^{2j-1}(\rho)\nonumber\\
    &= \frac{(n+j-1)!}{(n-j)!}\sum_{\nu=0}^{n-j-2}\left(\begin{array}{c}
        n-j-\nu+1\\
        3
    \end{array}\right)\delta_{\nu,n-j}\,=\,0,
\end{align}
where we have utilized the relation 
\begin{equation}\label{eq:Laguerre-recurrence}
    L_{n}^{k+p}(\rho)=\sum_{\nu=0}^{n}\left(\begin{array}{c}
        n-\nu+p-1\\
        p-1
    \end{array}\right)L_{\nu}^{k}(\rho)
\end{equation}
to change the upper index of the generalized Laguerre polynomial. 
The zero in the last step arises because the Kronecker delta enforces $\nu=n-j$, a value that lies outside the summation range $\nu\leqslant n-j-2$: every term of the sum vanishes identically. This shows that the basis is diagonalized with respect to the perturbation.

A word of caution: the diagonality established here is a first-order statement, and it does not make the degeneracy harmless. As discussed in \S\ref{sec:pert-DPT}, the degenerate partner still enters the corrected state---at second order in $V^{(1)}$, and at first order in both $V^{(2)}$ and the near-zone contact term $\delta V$---promoted to leading order by the small splitting; the resulting admixture is computed in \S\S\ref{sec:pert-second-order}--\ref{sec:pert-matching}, and the near-zone shift it involves is independently confirmed by the fine structure of the Proca spectrum (Table~\ref{tab:fine-structure}).

It remains to prove the identity \eqref{eq:Laguerre-recurrence}, which we do inductively for arbitrary integers $p\geqslant1$ from the recurrence relation~\cite{Szego:1975:OP}
\begin{equation}
L_{n}^{k+1}(\rho)=\sum_{\nu=0}^{n}L_{\nu}^{k}(\rho),
\end{equation}
which serves as the base case with $p = 1$.
Assuming that the relation holds for $p$, we observe that it does
for $p+1$ as well by applying the hockey-stick identity from combinatorics in the final step:
\begin{align}
L_{n}^{k+p+1}(\rho) &= \sum_{\nu=0}^{n}L_{\nu}^{k+p}(\rho)= \sum_{\nu=0}^{n}\sum_{\nu'=0}^{\nu}\left(\begin{array}{c}
\nu-\nu'+p-1\\
p-1
\end{array}\right)L_{\nu'}^{k}(\rho)\nonumber\\
 &= \sum_{\nu'=0}^{n} \sum_{\nu=\nu'}^{n}\left(\begin{array}{c}
\nu-\nu'+p-1\\
p-1
\end{array}\right) L_{\nu'}^{k}(\rho)= \sum_{\nu'=0}^{n}\left(\begin{array}{c}
n-\nu'+p\\
p
\end{array}\right)L_{\nu'}^{k}(\rho).
\end{align}

\section{Asymptotic Vector Fields}\label{sec:Asymptotic-vector-fields}

In this appendix, we construct the polarization vectors $\epsilon^\nu_\lambda(\vec k)$ and derive the matrix elements of spherical helicity states quoted in \S\ref{sec: results}. Our starting points are the plane-wave expansion \eqref{eq:vector-expansion} and the overlap \eqref{eq:asymptotic-basis-change} between the spherical helicity states $\vert \omega, j, m, \lambda \rangle$ and the momentum eigenstates~\cite{Endlich:2016jgc}.

We begin with the polarization vectors. Using the orthonormality condition $\epsilon_\lambda^{*\nu}(\vec k) \epsilon^{\lambda'}_\nu(\vec k) = \delta_{\lambda\lambda'}$ and the Lorenz condition $k_\nu \epsilon^\nu_\lambda(\vec k) = 0$ in a Cartesian basis aligned such that $k^\nu = (\omega,0,0,k)$, we can identify the longitudinal polarization vector $\epsilon^\mu_0(\vec k)$ via $\omega\epsilon^0_0(\vec k) = k\epsilon_0^3(\vec k)$ and $1 + (\epsilon_0^0(\vec k))^2 = (\epsilon_0^3(\vec k))^2$, using the Minkowski metric to lower indices because these states are defined at infinite separation from the BH. The transverse polarizations have 3-vectors orthogonal to $(0,0,k)$, and choosing them to be eigenvectors of the helicity operator $\vec{J}\cdot\vec{P}/|\vec{P}|$ with eigenvalues $\lambda$ results in
\begin{equation}\label{eq:polarization-vectors}
\epsilon_0^\nu(\vec k) = \frac{1}{\sqrt{1-v^2}}
\left(\begin{array}{c}
v \\
0 \\
0 \\
1
\end{array}\right) = \frac{1}{\sqrt{1-v^2}} \left(\begin{array}{c}
v\\
V_0^I
\end{array}\right),
\qquad
\epsilon_{\pm1}^\nu(\vec k) = \frac{1}{\sqrt{2}}\left(\begin{array}{c}
0\\
\pm1\\
i\\
0
\end{array}\right) = -\left( \begin{array}{c}
0\\
V^I_{\pm1}
\end{array}\right),
\end{equation}
with $v = k/\omega$.
Notice that the longitudinal mode is rescaled by the Lorentz factor $1/\sqrt{1-v^2}=\omega/\mu$. The eigenvectors $V^I_m$ for rotations about the $z$-axis, satisfying $V_{m\,I}\,{R^I}_J(\varphi) = e^{i m \varphi} V_{m\,J}$, are given by \eqref{eq:rotation-eigenvectors} in Appendix~\ref{sec:Vector-spherical-harmonics}. The polarization vectors rotate with $\hat k$. In the development of the absorption probability, these polarizations must be integrated against a spin-weighted spherical harmonic on the sphere, with result
\begin{equation} \label{eq:polarization-integral}
    \int d{\Omega}\, {}_{-\lambda}Y^{jm}(\theta,\varphi) \epsilon_\lambda^I(\vec k) =  \delta_{j,1} \left(\frac{\omega}{\mu} \delta_{\lambda,0} + (1-\delta_{\lambda,0})\right)\sqrt{\frac{4\pi}{3}} V_m^I.
\end{equation}
The choice of transverse polarization vectors in \eqref{eq:polarization-vectors} ensures that \eqref{eq:polarization-integral} holds. We note that the corresponding integral in Ref.~\cite{Endlich:2016jgc} omits the Lorentz factor $\omega/\mu$ for the longitudinal mode; the matrix elements derived below include it.

The corresponding polarizations for the magnetic field $B^I = \frac{1}{2} \epsilon^{IJK}F_{JK} = \epsilon^{IJK} \partial_J A_K$ are $i\hat k \times \vec \epsilon_\lambda(\vec k)$. The longitudinal mode vanishes, and the transverse modes are phase-shifted. Compactly, we can write
\begin{equation}
    i\epsilon^{IJK} k_J \epsilon_K^\lambda(\vec k) = \lambda k \epsilon^I_\lambda(\vec k).
\end{equation}
Similarly, for the electric field $E^I = - \nabla^I A^0 - \frac{\partial}{\partial t} {A^I}$, we substitute $\epsilon^0 = \epsilon^J k_J/\omega$ to obtain
\begin{equation} \label{eq:electric-polarization}
    i\omega_k \left( \delta^I_J - \frac{k^I k_J}{\omega_k^2} \right) \epsilon^J_\lambda(\vec k) = i\omega_k \left(\delta^I_J - v^2 \hat k^I \hat k_J \right) \epsilon_\lambda^J(\vec k) = i\omega_k\left(1 - \delta_{\lambda,0} v^2 \right) \epsilon^I_\lambda(\vec k).
\end{equation}
This scales the longitudinal mode by $1-v^2 = \mu^2/\omega^2$, two powers of the inverse Lorentz factor.

For the field $A^\mu(t,x)$, we are interested in obtaining the matrix element $\langle{0}\vert \tilde{A}^I(t,0) \vert{\omega,j,m,\lambda}\rangle$, where $\tilde{A}_I(t,0) = {R_I}^{\,J}(\Omega t) A_J(t,0)$, as defined in \S\ref{sec: eft}. Since the states have definite azimuthal quantum numbers, we can reason that 
\begin{equation}
    {R^I}_J(\varphi) \langle 0 \vert A^J \vert m \rangle = \langle 0 \vert U^\dagger(\varphi) A^I U(\varphi) \vert m \rangle = e^{i m \varphi} \langle 0 \vert A^I \vert m \rangle,
\end{equation}
so we expect the matrix elements to be proportional to $V_m^I$ and can implement the rotations by introducing a factor $e^{im\Omega t}$.

Now, putting all this together:
\begin{align}\label{eq:vector-potential-scattering}
    \langle 0 \vert \mu {A}^I(t,0) \vert \omega, j, m, \lambda \rangle &= \mu \sum_{\lambda'} \int \frac{k^2 d{\omega_k} d{\Omega}}{(2\pi)^32\omega_k v} \epsilon_\lambda^I(\vec k) e^{-i\omega_k t} \langle 0 \vert a_{\lambda,k} \vert \omega,j,m,\lambda \rangle \nonumber \\
    &= e^{-i \omega t} \mu \sqrt{\frac{k}{8\pi^2}} \int d{\Omega}\, {}_{-\lambda}Y^{jm}(\theta,\varphi) \epsilon_\lambda^I(\vec k) \nonumber \\
    &= e^{-i\omega t} \sqrt{\frac{k}{6\pi}} V^I_m \delta_{j,1} \left( \omega \delta_{\lambda,0} + \mu (1 - \delta_{\lambda,0}) \right).
\end{align}
Above, the Lorentz factor replaces $\mu$ with $\omega$ in the longitudinal mode. A remnant of the longitudinal mode thus survives the $\mu\to0$ limit, decoupling as a massless scalar. Indeed, since $\epsilon_0^I(k) \propto k^I$, we find for a massive scalar with $\ell = 1$ (recall that for a scalar $\ell = j$) that
\begin{equation}
    \langle 0 \vert \partial^I \phi(t,0) \vert \omega,\ell,m \rangle = i e^{-i\omega t}\sqrt{\frac{k}{6\pi}} V_m^I \delta_{\ell,1} k,
\end{equation}
and in the massless limit this coincides with the $\mu A^I$ contribution up to a phase.

As seen above, the magnetic and electric fields differ from \eqref{eq:vector-potential-scattering} by factors with no angular dependence, so those factors simply scale the result. Explicitly,
\begin{gather}
    \langle 0 \vert {B}^I(t,0) \vert \omega, j, m, \lambda \rangle = e^{-i\omega t} \sqrt{\frac{k}{6\pi}} V^I_m \delta_{j,1} \lambda k, \\
    \langle 0 \vert {E}^I(t,0) \vert \omega, j, m, \lambda \rangle = ie^{-i\omega t} \sqrt{\frac{k}{6\pi}} V^I_m \delta_{j,1} \left( \mu\delta_{\lambda,0} + \omega (1 - \delta_{\lambda,0}) \right).
\end{gather}
We see that the longitudinal mode for the electric field is scaled by a relative factor of $\mu/\omega = \sqrt{1 - v^2}$ because it is suppressed by two of these factors in the electric field as shown in \eqref{eq:electric-polarization} and then boosted by one as shown in \eqref{eq:polarization-integral}.

\section{The $j=0$ Instability Rate\label{sec:j0-appendix}}

Though not the goal of this paper, we have all the tools to find $j=0$ vector results, and it might be illustrative to see the method applied to a different example. In this case, the EFT prediction is in full agreement with the literature~\cite{Baumann:2019eav,Baryakhtar:2017ngi}. 

If $j=0$, the composite operator coupled to the field is an $\operatorname{SO}(3)$ scalar, which must be combined with a scalar field operator in the Lagrangian. The only scalar field linear in $A^I$ is $\partial_I A^I = i E A^0$ by the constraint \eqref{eq:constr}, so we use the expressions for the divergence of the bound-state mode function $\vec{f}_\beta$ in Appendix~\ref{sec:Mode-functions} to obtain the transition amplitude
\begin{equation}
    \langle 0 \vert \partial_I A^I(t,0) \vert n,0,0,\ell\rangle = -\delta_{\ell,1}\sqrt{\frac{n^2-1}{2\pi n^5} \alpha^5\mu^4} e^{-i\mu t}.
\end{equation}
Above, $|m| \leqslant j$ sets $m=0$; moreover, a vector mode with $\ell=0$ necessarily carries $j=1$, so $j=0$ requires $\ell=1$: there is only one vector channel with $j=0$. This is what the vector contributes to the calculation regardless of the UV completion of the theory.

We can write the interaction term involving this field as $\mu \bigl(\vec{\nabla}\cdot\vec{A}\bigr)\mathcal{O}_{\mathrm{s}}$, introducing a new composite operator. To match it with the absorption of massless particles, we use the Stueckelberg substitution $\mu A^\nu \to \mu A^\nu + \partial^\nu \phi$ before letting $\mu \to 0$, resulting in $\nabla^2 \phi \mathcal{O}_{\mathrm{s}} = \partial_t^2\phi\mathcal{O}_{\mathrm{s}} = \phi \partial_t^2\mathcal{O}_{\mathrm{s}}$ (the first equality uses the massless equation of motion, the second an integration by parts under the time integral). This means that $\mathcal{O}_0 = \partial_t^2\mathcal{O}_{\mathrm{s}}$ is the composite operator that is paired with a scalar $\phi$. If the Wightman function for this operator is $W_0(\omega)$, then the Wightman function for $\mathcal{O}_{\mathrm{s}}$ is $W_0(\omega)/\omega^4$.

We obtain the absorption probability exactly as before. The EFT method predicts that the term $\phi \mathcal{O}_0$ contributes
\begin{equation}
    P_\mathrm{abs} = \frac{1}{\langle\omega,\ell,m\vert\omega,\ell,m\rangle} \int \frac{d\omega'}{2\pi} W_0(\omega') \left| \int dt\, e^{i\omega' t} \langle 0 \vert \phi(t,0) \vert \omega,\ell,m\rangle \right|^2.
\end{equation}
to the absorption probability of spherical helicity modes $\vert \omega,\ell,m\rangle$. Using \eqref{eq:scalar} and \eqref{eq:asymptotic-basis-change}, we can show that for a massless scalar field in flat spacetime,
\begin{equation}
    \langle 0 \vert\phi(t,0)\vert\omega,\ell,m\rangle = e^{-i\omega t} \sqrt{\frac{\omega}{2\pi}} \delta_{\ell,0}\delta_{m,0}. 
\end{equation}
Each derivative on the scalar field operator adds to its mass dimension and therefore introduces an additional factor of $r_\mathrm{s}\omega \ll 1$, so we expect the contribution due to $\phi$ above to be the leading order part of the absorption process. Thus, the absorption probability for a massless scalar mode with $\ell = 0$ is
\begin{equation}
    P_{\omega,0,0}^\mathrm{scalar} = \frac{\omega}{2\pi} W_0(\omega) = 16G^{2}M^{2}\omega^{2}\theta(\omega)\label{eq:prob-scalar0}
\end{equation}
with the second equation the known scalar absorption rate for a non-rotating BH~\cite{Page:1976df}. Thus, $W_0(\omega) = 32\pi G^2M^2\omega \theta(\omega)$. This is proportional to $\omega \theta(\omega)$ as expected for the small-frequency Wightman function.

The instability rate for a vector bound state with $j=0$ is 
\begin{equation}
    \Gamma_{n,0,0,1} = \frac{1}{T} \int \frac{d\omega'}{2\pi} \frac{W_0(\omega') - W_0(-\omega')}{\omega'^4} \left| \int dt\, e^{-i\omega' t} \langle 0 \vert \mu \partial_I A^I(t,0) \vert n,0,0,1\rangle \right|^2.
\end{equation}
Hence, the EFT method predicts
\begin{equation}
    \Gamma_{n,0,0,1} = \frac{16(n^2-1)}{n^5} \alpha^7 (-\mu),
\end{equation}
which is in full agreement with \eqref{eq:Baumann-result}. There is no emission for modes with zero angular momentum, so the rate is negative.

\end{appendix}

\bibliographystyle{JHEP}
\bibliography{biblio.bib}

\end{document}